\documentclass[11pt]{article}
\usepackage{geometry}
\usepackage{bvrshorthand}
\usepackage{xcolor}
\usepackage{colonequals}
\usepackage[hidelinks]{hyperref}
\usepackage{stmaryrd}
\usepackage{amssymb}
\usepackage{amsmath}
\usepackage{amsthm}
\usepackage{amsfonts}
\usepackage{tikz}
\usetikzlibrary{patterns,arrows.meta}
\usepackage{mathtools}
\usepackage{enumitem}
\usepackage{float}
\newcommand{\doublepi}{$\Pi$\kern-0.61em{I}}
\newcommand{\doublepin}{\text{\doublepi}\;_n}

\newcommand{\doublepinp}{\text{\doublepi}\;_{n+1}}
\newcommand{\doublepinm}{\text{\doublepi}\;_N}
\newcommand{\thpbracket}[1]{\llparenthesis #1 \rrparenthesis}
\newcommand \uD {{\mathbf D}}
\newcommand \un {{\mathbf n}}

\title{Why S exists: the Haag-Ruelle theory}

\author{Aditya Hebbar$^{\text{out, ex}}$, Thomas Pochart$^{\text{in}}$, Balt C. van Rees$^{\text{in}}$}

\date{%
    $^\text{out}$Department of Theoretical Physics, Tata Institute of Fundamental Research,\\ Homi Bhabha Road, Mumbai 400005, India\\%
    $^\text{ex}$The Abdus Salam ICTP, Strada Costiera 11, 34151, Trieste, Italy\\%
    $^\text{in}$CPHT, CNRS, Ecole Polytechnique, Institut Polytechnique de Paris, Palaiseau, France%
}

\begin{document}

\maketitle

\begin{abstract}
Any Wightman quantum field theory with an isolated single-particle representation in its spectrum also contains the associated Fock spaces of multi-particle `in' and `out' states. 
The S-matrix can then be defined as the overlap between these states, and therefore exists in any such theory. We review this fundamental result in mathematical physics, known as the Haag-Ruelle theory. 
Our presentation is mathematically rigorous and entirely self-contained, requiring only basic knowledge of quantum field theory and functional analysis. 
Compared to earlier literature we streamline some proofs, consider arbitrary spacetime dimension $d$, and provide a useful spacetime picture: operators creating the particles are approximately
localized in spacetime regions that move off to infinity along the directions defined by the particles' velocities. We also show that the S operator is a contraction and use that to deduce that 2-to-2 
partial waves are $L^\infty$ functions bounded by 1.
\end{abstract}

\newpage

\setcounter{tocdepth}{2}
\tableofcontents

\newpage

\section{Introduction}
Take a Wightman quantum field theory with an isolated single-particle state in its spectrum. Does the 2-to-2 amplitude of this particle obey the Froissart-Martin bound, or for that matter any of the bounds obtained using the non-perturbative S-matrix bootstrap? The answer to this question appears to be yes, but there are many logical steps to take, with interesting physics to discover along the way.

In this note we review the first step in this logical chain: the fact that the S-matrix exists in the first place, even though the Wightman axioms stipulate only the existence of ordinary off-shell correlation functions. This is a fundamental result from the 1960s, known as the Haag-Ruelle theory \cite{Haag:1958vt,ruelle1962asymptotic}, which more precisely proves the inevitable existence of multi-particle `in' and `out' states in any Wightman field theory that has an isolated single-particle state. A direct corollary is the existence of the S-matrix, defined simply as the overlaps between `in' and `out' states. The Haag-Ruelle theory also provides the foundation for the rigorous justification of the LSZ prescriptions, as we will review in upcoming work \cite{LSZPAPER}. It is thereby the canonical starting point for any formal investigation of the non-perturbative properties of scattering amplitudes.

Here is how we will proceed. After introducing our conventions and free fields in section \ref{sec:freefields}, our derivation begins in earnest in section \ref{sec:interactingfields}. Here we first review the Wightman axioms. Then, we introduce the familiar-looking creation and annihilation operators as:
\begin{equation}
    \label{def:atp_intro}
\begin{split}
    a_t(\up) &= i e^{i E_\up t} \lrptl_t \phi(t,\up)\,,\\
    a^\dagg_t(\up) &= - i e^{- i E_\up t} \lrptl_t \phi(t,-\up)\,.
\end{split}
\end{equation}
If $\phi(x)$ were a free field then these operators would not depend on $t$, since their $t$-derivative produces the Klein-Gordon equation acting on $\phi(t, \up)$. In an interacting theory $a_t(\up)$ and $a^\dagg_t(\up)$ are $t$-dependent, but if we convolve them with suitable wave packets then their $t\to \pm \infty$ limits exist such that they become equivalent to their familiar free-field counterparts. The exact statement is formulated in the main theorem \ref{thm:haagruelle}.

The wavepackets need to be special in two ways. First, we need to solve the `single-particle problem', which is the problem of ensuring that $a_t^\dagg(\up)|\Omega\>$ lies entirely in the single-particle subspace and does not correspond to an uncontrollable multi-particle state. Solving this turns out to be straightforward: we just need to choose test functions $f(x)$ whose support in momentum space is such that $\phi(f) |\Omega\rangle$ can only lie in the single-particle subspace. 

The second condition ensures that the particles become widely separated and therefore non-interacting at large $|t|$. To do so we require the test functions to have disjoint \emph{velocity cones}, effectively ensuring that no two particles can have the same velocity.

With these wave packets, the proof of the main theorem relies on two key ingredients. The first is cluster decomposition of the connected correlation functions, which we prove in section \ref{sec:truncatedWigtman}. The version of cluster decomposition proven in \cite{streater2000pct} is unfortunately not sufficiently strong for our purposes. We thus review the original proof by Ruelle \cite{ruelle1962asymptotic} which is rather lengthy and technical. It is at this stage that the locality axiom is needed, and the mass gap also plays a role.

A second key ingredient is the large $|t|$ behavior of solutions to the massive Klein-Gordon equation, which is needed to understand the factors $e^{\pm i E_\up t}$ in the above equation. This will be a more straightforward exercise. We will provide bounds both inside and outside the velocity cones in section \ref{sec:solutionskg}.

With these ingredients in place we can finally prove the Haag-Ruelle theorem \ref{thm:haagruelle} in section \ref{sec:proof}. Besides the main statement we will also prove five additional claims, concerning among other things the convergence rate and Lorentz covariance. Some of these will also be essential for our upcoming review of the LSZ prescriptions \cite{LSZPAPER}.

The proof of the main theorem is formulated largely in momentum space, and it is not straightforward to see what is going on in position space. In section \ref{sec:spacetimepicture} we therefore provide a spacetime picture where we re-cast the essential arguments in position space and recover the idea of particles being created at widely separated positions and very early (or very late) times.

In section \ref{sec:conclusion} we interpret our main result as the isometric embedding of the free-field Fock space $\cF$ in the Hilbert space completion $\cH_\text{in}$ of the Haag-Ruelle states constructed in theorem \ref{thm:haagruelle}. We can then finally define the S-matrix as a linear contraction map:
\begin{equation}
    S: \cF \to \cF\,.
\end{equation}
We discuss how the $n$-to-$m$ particle scattering amplitudes are given in terms of very specific tempered distributions, and work out the consequences of (sub)unitarity and Lorentz covariance for 2-to-2 scattering amplitudes.

Finally, in section \ref{sec:discussion} we list various open questions, in particular concerning the extension of the Haag-Ruelle framework to massless particles in odd dimensions.

This review is intended for anyone with introductory knowledge of relativistic quantum field theory and functional analysis. As background we assume some familiarity with the usual spaces:
\begin{equation}
    \cS(\R^d), \qquad \cS'(\R^d), \qquad \cD(\R^d)\,,
\end{equation}
which are respectively the space of (complex-valued) Schwartz functions, tempered distributions, and compactly supported Schwartz functions on $\R^d$. We will also assume that the reader has encountered the Wightman axioms once before. All this material can be found, for example, in the first few pages of chapters 2 and 3 of Streater and Wightman \cite{streater2000pct}. Our results will make use of the Schwartz representation theorem, whose proof is omitted from many mathematical physics textbooks. We therefore provide one in appendix \ref{app:SchwartzRepThm}.

\subsubsection*{Historical context}
The foundational papers for this subject are actually those of the LSZ prescriptions \cite{Lehmann:1954rq, lehmann1957formulation}, which pre-date the Haag-Ruelle theory. LSZ simply postulated the asymptotic condition that the time dependent creation and annihilation operators defined in \eqref{def:atp_intro} converge (weakly) to free-field creation and annihilation operators:
\begin{equation}
    \label{eq:LSZ_asymptotic_condition}
	\lim_{t \rightarrow \mp \infty} \langle \Phi | a^{(\dagger)}_t(\up) | \Psi \rangle = \langle \Phi | a^{(\dagger)}_{\substack{\text{in}\\\text{out}}}(\up) | \Psi \rangle
\end{equation}
where $|\Phi \rangle$, $|\Psi \rangle$ are arbitrary states in $\cH$. While they acknowledged that such an operator may be ill-defined, they claimed that all their results were nevertheless still valid.

A few years later, Haag in \cite{Haag:1958vt} all but proved a different version of the asymptotic condition. Very roughly speaking he demonstrated a strong convergence for states of the form
\begin{equation}
	\lim_{t \rightarrow \mp \infty}  a^{(\dagger)}_{t}(\up_1) \ldots a^{(\dagger)}_t(\up_n)| \Omega \rangle = a^{(\dagger)}_{\substack{\text{in}\\\text{out}}}(\up_1) \ldots a^{(\dagger)}_{\substack{\text{in}\\\text{out}}}(\up_n) | \Omega \rangle\,,
\end{equation}
which notably start from the vacuum $|\Omega\>$ and not an arbitrary state $|\Psi\>$. The precise statement corresponds to our main theorem \ref{thm:haagruelle}. Our proof strategy is also essentially that of \cite{Haag:1958vt}, although we do streamline it in places.

Haag, however, did not provide a proof of the two key ingredients of cluster decomposition and the large $|t|$ behavior of solutions to the Klein-Gordon equation that we mentioned above. Both of these results were proven later still, namely by Ruelle in \cite{ruelle1962asymptotic}. This work forms the basis of sections \ref{sec:truncatedWigtman} and \ref{sec:solutionskg}.

Finally, it was Hepp \cite{hepp1965connection} who used Ruelle's fast falloff result to give a proof of the LSZ prescriptions themselves, by noting that the asymptotic condition \eqref{eq:LSZ_asymptotic_condition} holds at least for the Haag-Ruelle states. This result of Hepp is claim (iv) of theorem \ref{thm:haagruelle}, and Hepp's work also forms the foundation for the spacetime picture we provide in section \ref{sec:spacetimepicture}.

Other reviews of the Haag-Ruelle theory exist as chapters in textbooks on mathematical quantum field theory, see for example \cite{Jost:1965yxu,Reed:1979Scattering,araki1999mathematical,bogolubov2012general,glimm2012quantum}. We however felt it was useful to have a comprehensive and pedagogical stand-alone reference.

\section{Free fields}
\label{sec:freefields}
We briefly discuss free field theory in order to introduce notation and conventions. All our fields will be real scalars for simplicity.

First some general conventions. The Minkowski spacetime metric is
\begin{equation}
\eta_{\mu\nu} = \text{diag}(+,-,-,\ldots)
\end{equation}
and Fourier transforms are defined as
\begin{equation}
f(p) = \int d^d x\, e^{i p \cdot x} f(x)
\end{equation}
such that the inverse Fourier transform reads:
\begin{equation}
f(x) = \int d_d p \, e^{- i p \cdot x} f(p)
\end{equation}
with the shorthand
\begin{equation}
    d_d p \colonequals \frac{d^d p}{(2\pi)^d}\,.
\end{equation}
We decompose a vector into time and space components as $x = (t,\ux)$. The spatial Fourier transform is defined as
\begin{equation}
    f(t,\up) = \int d^{d-1} \ux \, e^{-i \up \cdot \ux} f(t,\ux)
\end{equation}
where we take the Euclidean inner product in the exponent and then
\begin{equation}
    f(p) = \int dt \, e^{i p^0 t} f(t,\up)
\end{equation}
with $p = (p^0,\up)$. In our technically incorrect notation the three different functions $f(x) = f(t,\ux)$, $f(p) = f(p^0,\up)$ and $f(t,\up)$ are distinguished by their argument.

We will consider a free field $\varphi(x)$. The equation of motion is
\begin{equation}
    (\square + m^2) \varphi(x) = 0
\end{equation}
We solve it in momentum space as
\begin{equation}
    \label{freefieldmomentumspace}
    \varphi(p) = 2 \pi \delta(p^2 - m^2) (\theta(p^0) a(\up) + \theta(-p^0) a^\dagg(-\up))
\end{equation}
meaning
\begin{equation}
    \varphi(t,\up) = \frac{1}{2 E_\up} \lr[{e^{-i E_\up t} a(\up) + e^{i E_\up t} a^\dagg(-\up)}]
\end{equation}
with
\begin{equation}
    E_\up \colonequals \sqrt{\up^2 + m^2}\,.
\end{equation}
We chose $\varphi(x)$ to be real in position space, so $\varphi(x) = \varphi^\dagg(x)$, which implies that $[\varphi(p)]^\dagg =  \varphi(-p)$. The usual convention is however that $[a(\up)]^\dagg = a^\dagg(\up)$, which requires the extra minus sign in equation \eqref{freefieldmomentumspace}. We can extract the annihilation and creation operators as\footnote{Compared to Peskin and Schroeder~\cite{peskin1995introduction} we have $a(\up)^\text{here} = \sqrt{2 E_\up} a(\up)^\text{PS}$ and similarly for $a^\dagg(\up)$.}
\begin{equation}
    \label{creationannihilation}    
    \begin{split}
    a(\up) &= i e^{i E_\up t} \lrptl_t \varphi(t,\up)\\
    a^\dagg(\up) &= - i e^{- i E_\up t} \lrptl_t \varphi(t,-\up) 
    \end{split} 
\end{equation}
with $a \lrD b \colonequals a D b - (D a) b$ for a differential operator $D$. Notice that we can pick any $t$ because
\begin{equation}
\label{t_independence_free_field}
\begin{split}
    \partial_t a(\up) &= i e^{i E_\up t} \partial^2_t \varphi(t,\up) - i \lr({\partial^2_t e^{i E_\up t}})  \varphi(t,\up)\\
    &= i e^{i E_\up t} \lr({\partial^2_t + E_\up^2  })\varphi(t,\up) = 0
\end{split}
\end{equation}
and similarly $\partial_t a^\dagg(\up) = 0$. The Hilbert space is then a Fock space which is informally speaking spanned by
\begin{equation}
    \label{fockspaceinformal}
    |\up_1, \ldots \up_n\> \colonequals a^\dagg(\up_1) \ldots a^\dagg(\up_n) |0\>\,,
\end{equation}
with a relativistic normalization such that
\begin{equation}
    \vev{\up | \uq} =  2 E_\up (2\pi)^{d-1} \delta^{(d-1)}(\up - \uq)\,.
\end{equation}
More precisely, one takes the Hilbert space completion of the normalizable states obtained by integrating the state-valued distributions $ |\up_1, \ldots \up_n\>$ against wavepackets, see section \ref{subsec:freefieldFockspace} for details.

Let us spell out our conventions concerning inner products and conjugation. We begin with the usual pairing in position space:
\begin{equation}
    \label{varphiagainsttest}
    \varphi(f) \colonequals \int d^d x f(x) \varphi(x) = \int d_d p f(-p) \varphi(p)
\end{equation}
Note that $[\varphi(f)]^\dagg  = \varphi(f^\dagg)$ where $f^\dagg(x) \colonequals f^*(x)$, hence $f^\dagg(p) = f^*(-p)$. We now introduce
\begin{equation}
    \label{aadaggersmearedfee}
\begin{split}
    a(f) &\colonequals 2 \pi \int d_d p  \, \delta(p^2 - m^2) \theta(p^0) a(\up) f(p)\, \\
    a^\dagg(f) &\colonequals 2 \pi \int d_d p \, \delta(p^2 - m^2) \theta(p^0) a^\dagg(\up) f(p)
\end{split}
\end{equation}
such that, by equation \eqref{freefieldmomentumspace},
\begin{equation}
    \varphi(f) = a(\check f) + a^\dagg(f)\,,
\end{equation}
where $\check f(p) \colonequals f(-p)$. Note that this leads to the following two-point functions
\begin{equation}
    \vev{\varphi(f) \varphi(g)} = \vev{a(\check f) a^\dagg(g)} = 2 \pi \int d_d p \, \delta(p^2 - m^2) \theta(p^0) f(-p) g(p) \label{eq:2ptfctssc}
\end{equation}
and hence the norms\footnote{The operation $\check f^\dagg(p) = f^*(p)$ is a conjugation in momentum space but not in position space $\check f^\dagg (x) = f^*(-x)$. Our conventions may appear somewhat cumbersome here, but they are the most economical for what follows.}
\begin{equation}
    \label{norms}
    || \varphi(f) |0\rangle ||^2 = || a^\dagg(f) |0\rangle ||^2  = \vev{a(\check f^\dagger) a^\dagg(f)} = 2 \pi \int d_d p \, \delta(p^2 - m^2) \theta(p^0) |f(p)|^2
\end{equation}
come out positive. We can also substitute equation \eqref{creationannihilation} and write, again for any $t \in \R$,
\begin{equation}
\label{aadaggersmearedfromfreefield}
\begin{split}
    a(f) &= +i \int d_d p  \, f(p) \frac{e^{+i E_\up t} \lrptl_t  e^{- i p^0 t}}{2 E_\up} \varphi(p)\, \\
    a^\dagg(f) &= - i \int d_d p \, f(p) \frac{e^{- i E_\up t} \lrptl_t  e^{+i p^0 t}}{2 E_\up} \varphi(-p) 
\end{split}    
\end{equation}
which will be useful later when we discuss the interacting theory below.

Equation \eqref{aadaggersmearedfee} implies that $a^\dagg(f)|0\> \neq 0$ only if the support of $f(p)$ intersects the physical single-particle momenta with $p^0 > 0$ and $p^2 = m^2$. On the other hand, $\varphi(p)|0\>$ is only non-zero if $-p$ is physical, so in particular only if $p^0 < 0$. To see this in more detail, consider the energy-momentum operator $P^\mu$. Combining
\begin{equation}
    [P^\mu, a^\dagg(\up)] = \begin{pmatrix}E_\up \\ \up \end{pmatrix} a^\dagg(\up)\,,
\end{equation}
and equation \eqref{freefieldmomentumspace}, we find
\begin{align}
    [P^\mu, \varphi(p)] &= % 2 \pi \delta(p^2 - m^2) \p{ - \theta(p^0) \begin{pmatrix}E_\up \\ \up \end{pmatrix}  a(\up) + \theta(-p^0)  \begin{pmatrix}E_\up \\ - \up \end{pmatrix}  a^\dagg(-\up) } = 
    - p^\mu \varphi(p)\,,
\end{align}
and so the operator $\varphi(p)$ injects a four-momentum $-p$.

For later use we also remark that
\begin{equation}
    [P^\mu, \varphi(x)] % = \left[P^\mu, \int d_d p \, e^{-i p x} \varphi(p)\right] 
    = -i \frac{\partial}{\partial x_\mu} \varphi(x)\,,
\end{equation}
so  $\varphi(x) = \exp(x^\mu \partial_\mu) \varphi(0) = e^{i P \cdot x} \varphi(0) e^{-i P \cdot x}$, and a Fourier transform of the form
\begin{equation}
    \begin{split}
        &\int d^d a\, e^{i p \cdot a}\vev{\varphi(x_1)\ldots \varphi(x_l) \varphi(x_{l+1} + a)\ldots \varphi(x_n+ a)} = \\
        &\int d^d a\, e^{i p \cdot a}\vev{\varphi(x_1)\ldots \varphi(x_l) e^{iP \cdot a}\varphi(x_{l+1})\ldots \varphi(x_n)} 
    \end{split}
\end{equation}
will thus be supported in the region where $-p^\mu$ lies in the physical spectrum.

\begin{definition}
    For $p \in \R^d$ and $\mu \geq 0$, we define the following regions:
    \begin{align}
    V_\mu &: p^2 > \mu^2 \nn \\
    V^\pm_\mu &: p^2 > \mu^2 \text{ and } \pm p^0 > 0 \\
    \bar V_\mu &: p^2 \geq \mu^2 \nn  \\
    \bar V^\pm_\mu &: p^2 \geq \mu^2 \text{ and } \pm p^0 \geq 0 \nn
    \end{align}
    When $\mu = 0$ we omit the subscript, $V^+ \colonequals V^+_0$ and so on.
\end{definition}

The support of a function is commonly defined to be a closed set, as per the following definition.

\begin{definition}
    For $f(p) \in \cS(\R^d)$ we define its \emph{support} as
    \begin{equation}
        \supp(f(p)) \colonequals \overline{\{p \, :\, f(p) \neq 0 \}}
    \end{equation}
    and its \emph{spatial support} as
    \begin{equation}
        \label{def:spatialsupport}
        \supp\!_{\up}(f(p)) \colonequals \overline{\{ \up \in \R^{d-1} \,: \, \exists p^0 \in \R  \text{ s.t. } f(p^0, \up) \neq 0\}}\,.
    \end{equation}
\end{definition}

\section{Interacting fields}
\label{sec:interactingfields}
We will consider quantum field theory in the framework of the Wightman axioms and also assume a mass gap. We will only consider particles and fields of spin $0$.

\subsection{Wightman-type axioms}
First, we postulate a separable Hilbert space $\cH$. States are unit rays of $\cH$. The state space features a unitary representation of the identity component of the $d$-dimensional Poincar\'e group.

\begin{axiom}[Spectrum] \label{axiom:spectrum}
There exist two distinguished representations: a Poincar\'e-invariant vacuum state $|\Omega\>$ and an isolated single-particle scalar representation with mass $m > 0$ whose states are spanned by wavepackets built from $|\up\>$ as above. All states $|\Phi\>$ orthogonal to these obey a further mass gap $M > m$:
\begin{equation} \label{spectrummu}
    \vev{\Psi|e^{-i P x}|\Phi} = \int d_d p\,  e^{-i p x} \mu_{\Psi,\Phi}(p)
\end{equation}
for any $\< \Psi|$ and with $\mu_{\Psi,\Phi}(p)$ a finite measure with support in $\bar V^+_M$.
\end{axiom}

Since $0 < m < M$ we have postulated two mass gaps. The usual Wightman axioms do not require either gap, but they are essential for what follows. Note that we could not have assumed anything less than a single-particle representation at $m$, since it is the smallest non-trivial unitary representation of the Poincar\'e group.

\begin{axiom}[Fields] \label{axiom:fields}
There exists a \emph{field} $\phi(x)$ as an operator-valued distribution. More precisely, $\phi$ is a linear map from $\cS(\R^d)$ to linear operators on $\cH$. The common domain of definition of these operators, $D$, is dense in $\mathcal{H}$. This domain contains in particular the states
\begin{equation}
    \label{npointstates}
    \phi(f_1) \ldots \phi(f_n) |\Omega\>\,,
\end{equation} 
for any finite $n$ and with $f_1, \ldots f_n$ elements of $\cS(\R^d)$. Furthermore, the map $\cS(\R^d)^n \to \cH$ defined by equation \eqref{npointstates} is strongly continuous. The fields are also self-adjoint, so $[\phi(f)]^\dagg = \phi(f^\dagg)$.
\end{axiom}

Just as for the free field, we will write
\begin{equation}
    \phi(f) = \int d^d x \, \phi(x) f(x)
\end{equation}
where $\phi(x)$ is now an operator-valued distribution on $\R^d$. The algebra of operators extends naturally to the matrix elements of such distributions; for example, the Schwartz kernel theorem implies that, for $|\Phi\> \in D$ or $|\Psi\> \in D$, the two-point function
\begin{equation}
    \vev{\Psi| \phi(x) \phi(y)|\Phi}
\end{equation}
corresponds to a tempered distribution on $\R^{2d}$.

\begin{axiom}[Properties of the fields]
    \label{axiom:fieldproperties}
    The fields are relativistically covariant scalars, in the sense that
    \begin{equation}
        \phi(g \cdot x) = U_g \phi(x) U_g^{-1}
    \end{equation}
    where $g$ is an element of the Poincar\'e group and $U_g$ its representation in $\cH$. Locality (or causality) is encoded as:
    \begin{equation}
        [\phi(x), \phi(y)] = 0 \qquad \text{ if }\qquad (x-y)^2 < 0\,.
    \end{equation}
    Finally $\phi(x)$ overlaps with the single-particle states in the sense that
    \begin{equation}
        \label{single-particle-overlap}
        \vev{\up | \phi(x) | \Omega} \neq 0\,.
    \end{equation}
\end{axiom}
One normally also supposes that states of the form $\phi(f_1)\ldots \phi(f_n)|\Omega\>$ span a dense subset in $\cH$, 
and then it is said that the vacuum is \emph{cyclic}. Such an assumption is not necessary for our purposes. On the other hand, if one is willing to assume cyclicity of the vacuum then equation \eqref{single-particle-overlap} can be dropped from axiom \ref{axiom:fieldproperties} without affecting the main results of the Haag-Ruelle theory; we explain this briefly in section \ref{sec:discussion}.

\subsection{Asymptotic states}
Equation \eqref{single-particle-overlap} tells us that states of the form $\phi(f)|\Omega\>$ have non-zero components in the single-particle subspace spanned by the vectors $|\up\>$. That however does not mean that $\phi(f)|\Omega\>$ is orthogonal to all the other states. This is the `single-particle problem' and our first order of business is to resolve it.

Now recall axiom \ref{axiom:spectrum} which states that the four-momentum of non-single-particle states is either $0$ (for the vacuum) or somewhere inside $\bar V^+_M$ for some $M > m$.
\begin{definition}
Suppose $f(x) \in \cS(\R^d)$. It is an element of $\widetilde{\cD_1(\R^d)}$, the set of \emph{single-particle test functions}, if the support of its Fourier transform $f(p)$ is both compact and such that
\begin{equation}
    \label{D1support}
    \supp(f(p)) \cap \left(\{0\} \cup \bar V_M \right) = \emptyset\,.
\end{equation}
We will equivalently write $f(p) \in \cD_1(\R^d)$, and further write $f(p) \in \cD_1^\pm(\R^d)$ if the support is restricted to $p^0 > 0$ or $p^0 < 0$, respectively.
\end{definition}

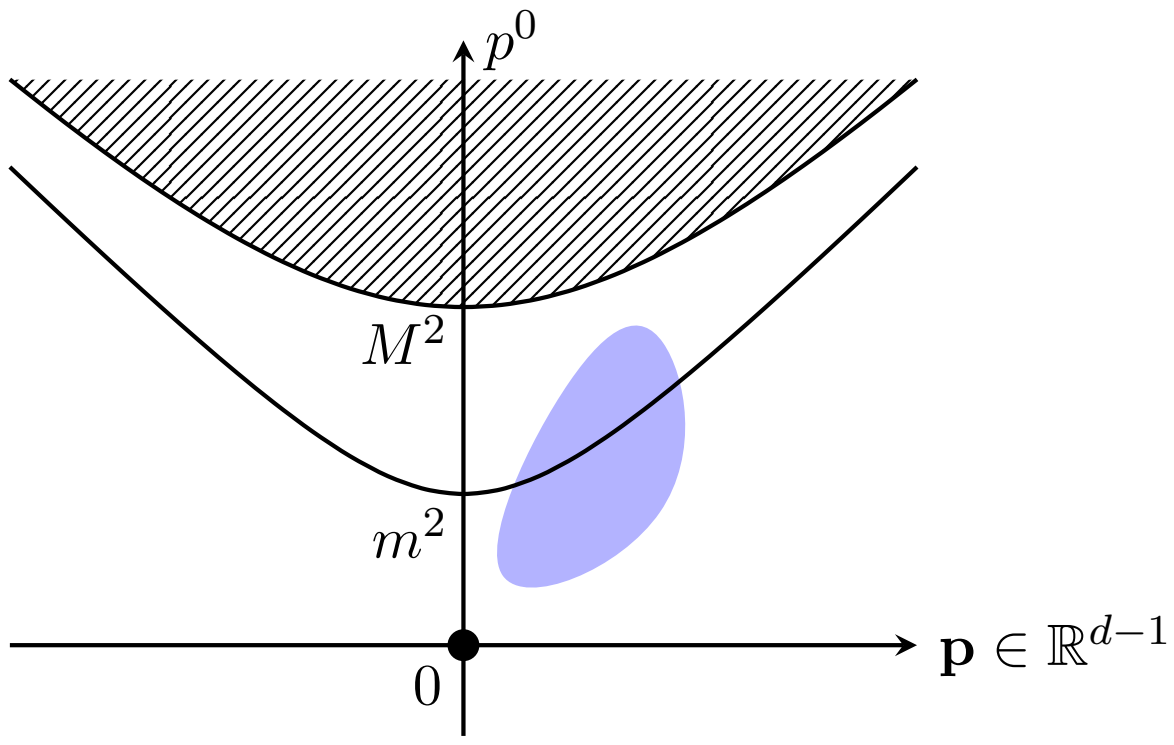
\begin{figure}[ht]
    \centering
\begin{tikzpicture}[>=stealth, thick, scale=2]
    
     \fill[blue!30] plot[smooth cycle, tension=0.9] coordinates {(0.13,0.23) (0.35,0.87) (0.66,1) (0.65,0.44)}; 

    % Shaded region above M^2
    \fill[pattern=north east lines] 
        (-1.5, 1.87) -- 
        plot[domain=-1.5:1.5, smooth] (\x, {sqrt(\x*\x+1.25)}) -- 
        (1.5, 1.87) -- 
        cycle;

    % Axes
    \draw[->] (-1.5,0) -- (1.5,0) node[right]  {$\up \in \mathbb{R}^{d-1}$};
    \draw[->] (0,-0.3) -- (0,2) node[right] {$p^0$};
    
    % m^2 and M^2 curves
    \draw[domain=-1.5:1.5,smooth,thick,black] plot (\x,{sqrt(\x*\x+0.25)});
    \node[left] at (0.02,0.37) {$m^2$};
    \draw[domain=-1.5:1.5,smooth,thick,black] plot (\x,{sqrt(\x*\x+1.25)});
    \node[left] at (0.02,1.01) {$M^2$};

    \fill (0,0) circle (1.5pt) node[below left] {$0$};

\end{tikzpicture}
    \caption{\label{fig:D1S1drawn}In blue, a possible support region for a function $f(p) \in \cD_1(\R^d)$. As per axiom \ref{axiom:spectrum}, the physical values of $p^2$ are either $0$ for the vacuum, $m^2$ for the single-particle representation, or at or above $M^2$ for any other state.}
\end{figure}

See figure \ref{fig:D1S1drawn} for a possible support region of a function $f(p) \in \cD_1(\R^d)$.

\begin{proposition}
\label{prop:singleparticlesubspace}
For any $f(p) \in \cD_1(\R^d)$, the state 
\begin{equation}
    \label{D1singleparticlestate}
    \phi(f) |\Omega \rangle = \int d_d p \, f(p) \phi(-p) |\Omega \rangle
\end{equation}
lies entirely in the single-particle subspace spanned by the vector-valued distributions $|\up\rangle$. It therefore obeys
\begin{equation}
    \int d_d p \, (p^2 - m^2) f(p) \phi(-p)|\Omega \rangle = 0
\end{equation}
\end{proposition}

\begin{proof}
The first statement is due to momentum conservation: equation \eqref{D1support} states that we simply do not inject the right amount of momentum to create anything besides the single-particle states. To prove the second statement we simply measure the overlap of the given state with any $|\uq\rangle$ in the single-particle subspace. Lorentz invariance dictates that
\begin{equation}
    \label{eq:one_particle_phi_vac}
    \vev{\uq|\phi(-p)|\Omega} = (2\pi)^d \sqrt{Z} \, \delta^{(d)}(q - p)
\end{equation}
for some $\sqrt{Z} \neq 0$ by equation \eqref{single-particle-overlap} and with $q = (E_\uq, \uq)$. Since $q^2 = m^2$, the overlap must vanish for all $|\uq \rangle$.
\end{proof}

In what follows, we assume a normalization of $\phi(x)$ such that $\sqrt{Z} = 1$. 

We now construct creation and annihilation operators exactly as in equation \eqref{aadaggersmearedfromfreefield}, but with the free field $\varphi(x)$ replaced by the interacting field $\phi(x)$ and with the test functions restricted to the single-particle ones.

\begin{definition}
For $f(p) \in \cD_1(\R^d)$ let
\begin{equation}
    \label{aadaggersmearedalternative}
    \begin{split}
        a_t(f) &\colonequals + i \int d_d p\,  f(p) \frac{e^{+i E_\up t} \lrptl_t e^{- i p^0 t}}{2E_\up}  \phi(p)\,,\\
        a^\dagger_t(f) &\colonequals - i \int d_d p\,  f(p) \frac{e^{-i E_\up t} \lrptl_t e^{+ i p^0 t} }{2E_\up} \phi(-p)\,.
    \end{split}
\end{equation}    
\end{definition}

In the free field case the operators so defined were independent of $t$ by virtue of the Klein-Gordon equation, see equation \eqref{t_independence_free_field}. In interacting theories this is no longer the case, but we will show that the large $|t|$ limit allows one to recover the free-field behavior.

Henceforth we will write $a^{(\dagger)}_t(f)$ to denote either $a_t(f)$ or $a_t^{\dagger}(f)$.

\begin{proposition}
    \label{prop:asdenselydefined}
    For $f(p) \in \cD_1(\R^d)$ the $a^{(\dagger)}_t(f)$ are densely defined operators. The states
    \begin{equation}
        \label{asactingonomega}
        a^{(\dagger)}_{t_1}(f_1) \ldots a^{(\dagger)}_{t_n}(f_n) |\Omega\> \in \cH
    \end{equation}
    are $C^\infty$ in $t_1, \ldots t_n$.
\end{proposition}

It is sometimes stated that the operators $a_t^{(\dagger)}(f)$ themselves are $C^\infty$ in $t$, but this requires defining the derivative of an unbounded operator. The above statement suffices for our purposes.

\begin{proof}
These statements directly follow from axiom \ref{axiom:fields}. For any $t$ the operators \eqref{aadaggersmearedalternative} are of the form $\phi(g)$ for some $g \in \cS(\R^d)$, whose domain of definition was postulated to be large enough for equation \eqref{asactingonomega} to hold. Smoothness in $t$ follows because we can use linearity to write, for example,
\begin{equation}
    \frac{1}{\epsilon} \left( a^\dagger_{t+ \epsilon}(f) - a^\dagger_t(f)\right)
\end{equation}
as a single operator with $a^\dagger_t(f_\epsilon)$ where
\begin{equation}
    f_\epsilon(p) = \frac{e^{-i (E_\up -p^0) \epsilon} - 1}{\epsilon} f(p)\,,
\end{equation}
When acting on a state as in equation \eqref{asactingonomega}, the strong continuity postulated in axiom \ref{axiom:fields} allows one to take the limit $\epsilon \to 0$ inside the pairing of $\phi$ and $f$, and all the $f_\epsilon(p)$ as well as the limiting function $f_0(p) = - i (E_\up - p^0) f(p)$ are elements of $\cD_1(\R^d)$.
\end{proof}
The main result is now captured in the following theorem, called the Haag-Ruelle theorem \cite{Haag:1958vt,ruelle1962asymptotic}:
\begin{theorem}
    \label{thm:haagruelle}
    Suppose $f_1(p), \ldots f_n(p) \in \cD_1^+(\R^d)$ have completely disjoint spatial support, so $\supp_\up(f_i(p)) \cap \supp_\up(f_j(p)) = \emptyset$ for all $i \neq j$. The states
    \begin{equation}
        \label{haagruelle}
        |\Phi_t\> = a^{(\dagger)}_{t}(f_1) \ldots a^{(\dagger)}_{t}(f_n) |\Omega\>
    \end{equation}
    converge strongly as $t \to \pm \infty$. The inner products between the limiting states reduce to those of the free field. Also:
    \begin{itemize}
        \item[(i)] The convergence rate is faster than any power of $|t|^{-1}$;
        \item[(ii)] the limiting states only depend on the $f(p)$ reduced to the mass shell $p^2 = m^2$ with $p^0 > 0$; 
        \item[(iii)] the limit is independent of the chosen Lorentz frame;
        \item[(iv)] the limits can be taken independently and sequentially provided time ordering is respected, so for example
        \begin{equation}
            \lim_{\substack{t_1 \ldots t_m \to \infty\\ t_1 \le \dots \le t_m}} a_{t_1}^{(\dagg)}(f_1) \ldots a_{t_m}^{(\dagg)}(f_m) \lim_{\substack{t_{m+1} \ldots t_n \to \infty\\ t_{m+1} \le \dots \le t_n}} a_{t_{m+1}}^{(\dagg)}(f_{m+1}) \ldots a_{t_n}^{(\dagg)}(f_n) |\Omega\> = \lim_{t \to \infty} |\Phi_t\>\,,
        \end{equation}
        with a convergence rate faster than any power of $|t_1|^{-1}$.
        \item[(v)] for $d\ge4$, the main statement as well as (ii) and (iii) hold without assuming the spatial supports of the $f_i$'s to be completely disjoint.
    \end{itemize} 
\end{theorem}

We denote the limiting states as $t \to \pm \infty$ as $|\Phi_\OUT\>$ and $|\Phi_\IN\>$, respectively. Their overlaps
\begin{equation}
    \label{Snaive}
    \vev{\Phi_\OUT|\Phi_\IN}
\end{equation}
will constitute the S-matrix.

\subsection{Intuitive derivation}
\label{subsec:intuitive}
Let us first rewrite the time-dependent annihilation and creation operators $a_t(f)$ and $a_t^\dagger(f)$ of equation \eqref{aadaggersmearedalternative} in a more intuitive form. We can do so by introducing:
\begin{equation}
    \begin{split}
    \phi(f)(x) &\colonequals \int d^d y \, f(y - x) \phi(y)\,,\\
    \chi(x) &\colonequals \int d_{d-1} \up\,  \frac{B(\up)}{2E_\up} e^{-i E_\up t + i \up \cdot \ux} \,,        
    \end{split}
\end{equation}
where $B(\up) \in \cD(\R^{d-1})$. Now choose $B(\up)$ to be a bump function, equal to 1 over a sufficiently large domain such that
\begin{equation}
    f(p) B(\up) = f(p)\,,
\end{equation}
which is always possible for $f(p) \in \cD_1(\R^d)$. With this choice we can rewrite $a_t^\dagger(f)$ and $a_t(f)$ of equation \eqref{aadaggersmearedalternative} as
\begin{equation}
    \label{adaggerintuitive}
    \begin{aligned}
    a^\dagger_t(f) &= - i  \int d^{d-1} \ux\, \chi(t,\ux) \, \lrptl_t  \, \phi(f)(t,\ux)\\
        a_t(f) &=  i  \int d^{d-1} \ux\, \chi^*(t,\ux) \, \lrptl_t  \, \phi(\check f)(t,\ux).
\end{aligned}
\end{equation}
 We see that the creation and annihilation operators are defined in terms of the \emph{translated smeared Wightman field} of the form $\phi(f)(x)$ and a \emph{smooth classical solution of the Klein-Gordon equation} $\chi(x)$. 

In section \ref{sec:truncatedWigtman} we will show that the operators $\phi(f)(x)$ obey a form of \emph{cluster decomposition}, meaning that \underline{the correlation functions of the $\phi(f)(x)$ factorize at large spacelike separations}. Although this result is intuitive and well-known, the proof itself is fairly complicated because our test functions $f(x)$ are not compactly supported in position space.

In section \ref{sec:solutionskg} we will discuss the behavior of smooth classical solutions of the Klein-Gordon equation $\chi(x)$. The upshot will be that, for compactly supported $B(\up)$, the field propagates almost entirely in the so-called \emph{velocity cone} $\cC_B \subset \R^d$, which is the set of directions associated to the support of $B(\up)$:
\begin{equation}
    \cC_B \colonequals \{ \alpha \, (E_\up, \up) : \up \in \supp(B(\up)) \text{ and } \alpha \in \R\}\,.
\end{equation}
The crucial result will be that \underline{$\chi(t,\ux)$ falls off rapidly in $|t|$ for $(t,\ux)$ outside the velocity cone}.

With the above results in hand we can argue for a key statement of the main theorem, namely that all the norms of the form
\begin{equation}
    \label{Phitnormintuitive}
    || a_t^{(\dagger)}(f_1)\ldots a_t^{(\dagger)}(f_n) |\Omega\> ||^2\,
\end{equation}
become those of the free field as $t \to \pm \infty$. The reasoning goes as follows.

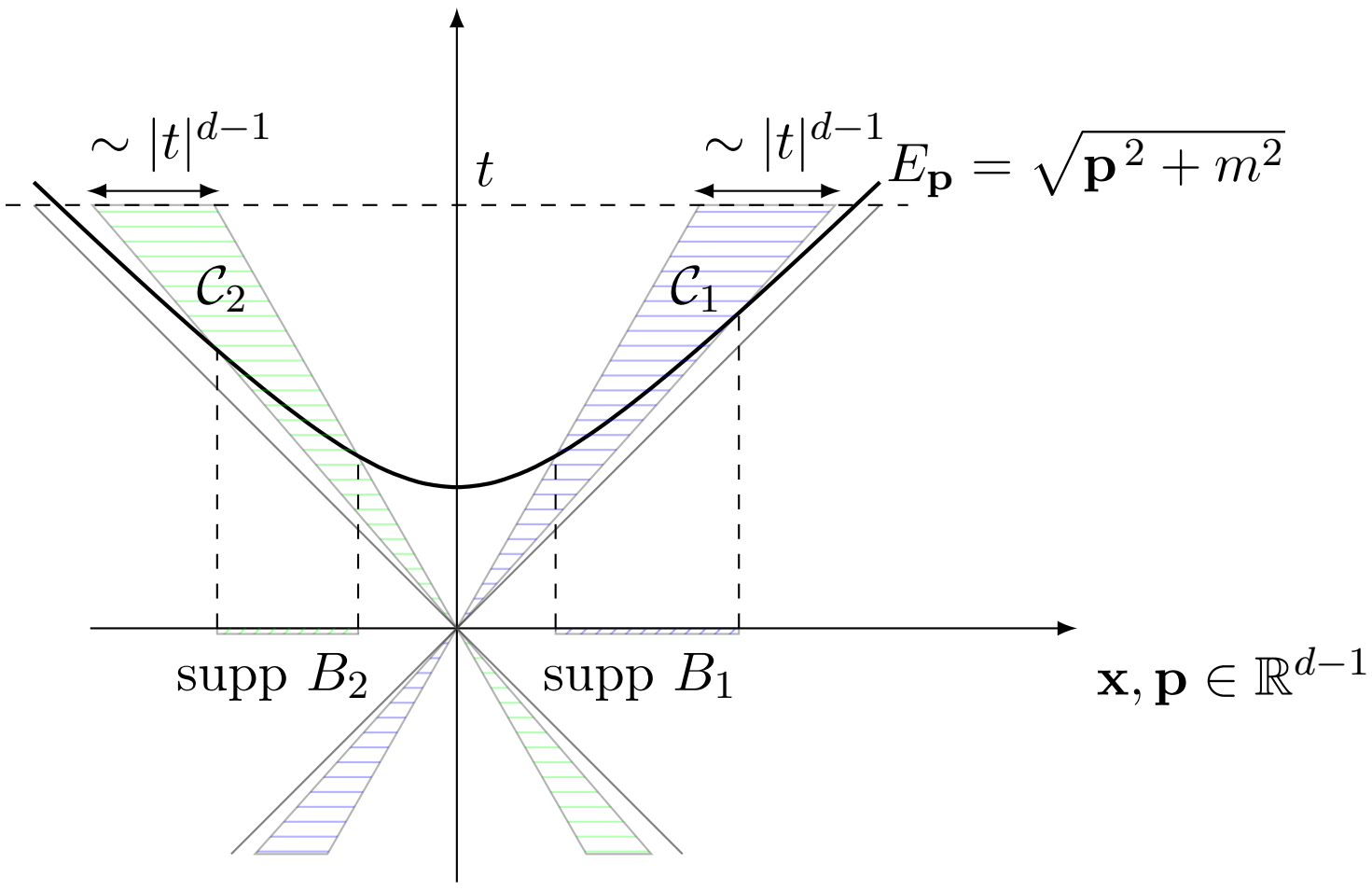
\begin{figure}
\begin{center}
    \begin{tikzpicture}[scale=2,>=latex]
        \centering
        % Axes
        \draw[->] (-1.3,0) -- (2.2,0) node[below right] {$\ux, \up \in \mathbb{R}^{d-1}$};
        \draw[->] (0,-0.9) -- (0,2.2);

        % E_p
        \draw[domain=-1.5:1.5,smooth,thick,black] plot (\x,{sqrt(\x*\x+0.25)});

        % velocity cone
        \draw[thin,gray] (-0.8,-0.8) -- (1.5,1.5);
        \draw[thin,gray] (0.8,-0.8) -- (-1.5,1.5);

        % Label de la courbe
        \node[black] at (2.23,1.65) {$E_{\up} = \sqrt{\up^{\,2} + m^2}$};

        % supp 1
        \draw[pattern=north east lines, pattern color=blue, opacity=0.3] (0.35,0) -- (1,0) -- (1,-0.02) -- (0.35,-0.02) -- cycle;
        \node[below] at (0.65,-0.02) {$\operatorname{supp}\,B_1$};

        % supp 2
        \draw[pattern=north east lines, pattern color=green, opacity=0.3] (-0.35,0) -- (-0.85,0) -- (-0.85,-0.02) -- (-0.35,-0.02) -- cycle;
        \node[below] at (-0.65,-0.02) {$\operatorname{supp}\,B_2$};

        %Dash lines supp 1
        \draw[dashed] (0.35,0) -- (0.35,{sqrt(0.35*0.35+0.25)});
        \draw[dashed] (1,0) -- (1,{sqrt(1+0.25)});

        %Dash lines supp 2
        \draw[dashed] (-0.85,0) -- (-0.85,{sqrt(0.85*0.85+0.25)});
        \draw[dashed] (-0.35,0) -- (-0.35,{sqrt(0.35*0.35+0.25)});       

        % C1
        \draw[pattern=horizontal lines, pattern color=blue, opacity=0.3] (0,0) -- ({1.5*0.35/sqrt(0.35*0.35+0.25)},1.5) -- ({1.5*1/sqrt(1+0.25)},1.5) -- cycle;
        \draw[pattern=horizontal lines, pattern color=blue, opacity=0.3] (0,0) -- ({-0.8*0.35/sqrt(0.35*0.35+0.25)},-0.8) -- ({-0.8*1/sqrt(1+0.25)},-0.8) -- cycle;

        % C2
        \draw[pattern=horizontal lines, pattern color=green, opacity=0.3] (0,0) -- ({-1.5*0.35/sqrt(0.35*0.35+0.25)},1.5) -- ({-1.5*0.85/sqrt(0.85*0.85+0.25)},1.5) -- cycle;
        \draw[pattern=horizontal lines, pattern color=green, opacity=0.3] (0,0) -- ({0.8*0.35/sqrt(0.35*0.35+0.25)},-0.8) -- ({0.8*0.85/sqrt(0.85*0.85+0.25)},-0.8) -- cycle;

        \node[below] at (0.85,1.35) {$\mathcal{C}_1$};
        \node[below] at (-0.83,1.35) {$\mathcal{C}_2$};

        %Velocity cone size estimates 
        \draw[dashed] (-1.6,1.5) -- (1.6,1.5);
        \node[above] at (0.1,1.5) {$t$};
        \draw[<->] ({-1.5*0.35/sqrt(0.35*0.35+0.25)+0.02},1.55) -- ({-1.5*0.85/sqrt(0.85*0.85+0.25)-0.02},1.55);
        \draw[<->] ({1.5*0.35/sqrt(0.35*0.35+0.25)-0.02},1.55) -- ({1.5*1/sqrt(1+0.25)+0.02},1.55);
        \node[above] at ({(1.5*0.35/sqrt(0.35*0.35+0.25)+1.5*1/sqrt(1+0.25))/2+0.1},1.55) {$\sim |t|^{d-1}$};
        \node[above] at ({(-1.5*0.35/sqrt(0.35*0.35+0.25)-1.5*0.85/sqrt(0.85*0.85+0.25))/2+0.1},1.55) {$\sim |t|^{d-1}$};

    \end{tikzpicture}
\end{center}
\caption{\label{fig:twovelocitycones}The velocity cones $\cC_1$ and $\cC_2$ in spacetime $(t,\ux)$ for two functions $B_1(\up)$ and $B_2(\up)$ with the indicated support in momentum space $\up$ along the horizontal axis. The size of the velocity cone grows at large $|t|$ like a ball in $d-1$ dimensions with radius $|t|$.}
\end{figure}

Equation \eqref{Phitnormintuitive} is a $2n$-point function of operators of the form given in equation \eqref{adaggerintuitive}, each evaluated at the same time $t$. As we make $|t|$ large, each of the $2n$ integrals over $\ux$ gets reduced only to the velocity cones (or more precisely the intersection of the velocity cones with the constant $t$ slice), because the $\chi(t,\ux)$ fall off rapidly elsewhere. But at large $|t|$ two non-overlapping velocity cones will gain a large spacelike separation (see figure \ref{fig:twovelocitycones}), and our cluster decomposition result then dictates a factorization of the $2n$-point Wightman function.

The precise factorization is easily deduced. By the assumption of the theorem, the $n$ operators in equation \eqref{Phitnormintuitive} all have non-overlapping velocity cones, so in computing the norm we only retain the pairwise overlaps between an operator and its adjoint. Therefore, at large $|t|$ the $2n$-point function factorizes into a product of $n$ 2-point functions, exactly as would be the case in a free theory.

To conclude the derivation, note that the 2-point functions are each of the form
\begin{equation}
\< \Omega | a^{(\dagger)}_t(f_1) a^{(\dagger)}_t(f_2)|\Omega\>\,,    
\end{equation}
and proposition \ref{prop:singleparticlesubspace} immediately implies that they are simply given by their corresponding free-field values. (This holds for any $t$ since, by the same proposition, $a^{(\dagger)}_t(f)|\Omega\>$ does not depend on time.) This is intuitively how the norms converge to the free-field values.

The argument for the other claims in the theorem, most importantly the convergence of the states themselves instead of just their norms, proceeds along similar lines. We have however opted to not bore the reader with more imprecise arguments, so we will dial the level of rigor back up and work our way towards a formal proof.

\section{Truncated Wightman functions}
\label{sec:truncatedWigtman}
The \emph{Wightman functions} are defined as
\begin{equation}
    \cW(x_1,\dots,x_n) \colonequals \< \Omega | \phi(x_1) \dots \phi(x_n) |\Omega\>\,.
\end{equation}
Technically they are distributions for test functions in $\R^{nd}$. In this section we will consider the definition and properties of \emph{truncated} Wightman functions, which we denote as
\begin{equation}
    \cW^T(x_1,\dots,x_n)\,.
\end{equation}
Truncated Wightman functions are similar to the connected correlation functions of Euclidean field theory, but more refined because they take into account the operator ordering. They obey a cluster decomposition property that we need to prove theorem \ref{thm:haagruelle}.

\subsection{Definition}
Consider the set $\doublepin$ of all unordered partitions of $\{x_1,\ldots, x_n\}$. A single such partition $\Pi \in \doublepin$ is a list of subsets of the $x_i$, so for $n=7$ the partition $\Pi = \{\{x_1,x_3,x_6\},\{x_2,x_4,x_5\},\{x_7\}\}$ would be a good example. We will denote the individual subsets as $\pi_\alpha$ and write $\Pi = \{\pi_1, \pi_2,\ldots \pi_{|\Pi|}\}$ with $|\Pi|$ the total number of subsets in $\Pi$. In the given example $|\Pi|= 3$ and $\pi_2 = \{x_2,x_4,x_5\}$.

For every partition we adopt the convention that the $x_i$ remain canonically ordered within each $\pi_\alpha$. Then we can write $\cW(\pi_2)$ which has the obvious interpretation; in the example it is the three-point Wightman function $\cW(x_2,x_4,x_5)$.

The truncated Wightman functions are then defined implicitly as
\begin{equation}
  \label{truncatedwightmandefinition}
  \cW(x_1,\ldots,x_n) = \sum_{\Pi \in \doublepin} \cW^T(\pi_1) \cW^T(\pi_2) \ldots \cW^T(\pi_{|\Pi|})\,.
\end{equation}
To give an idea, let us suppose the one-point Wightman function vanishes, $\cW(x_1) = 0$. In that case the one-, two- and three-point truncated Wightman functions are equal to their untruncated counterparts, whereas for four points we have:
\begin{multline}
    \label{truncatedwightman4pt}
    \cW^T(x_1,\ldots,x_4) = \\
    \cW(x_1,\ldots,x_4) - \cW(x_1,x_2) \cW(x_3,x_4) - \cW(x_1,x_3) \cW(x_2,x_4) - \cW(x_1,x_4) \cW(x_2,x_3)
\end{multline}

An explicit definition of the truncated Wightman functions can also be given:
\begin{equation}
    \label{truncatedwightmandefinitionexplicit}
    \cW^T(x_1,\ldots,x_n) = \sum_{\Pi \in \doublepin} (-1)^{|\Pi|-1} (|\Pi| - 1)! \, \cW(\pi_1) \cW(\pi_2) \ldots \cW(\pi_{|\Pi|})  
\end{equation}
The physicists' proof for the equivalence with equation \eqref{truncatedwightmandefinition} is obtained by considering that if $Z[J(x)]$ is a generating function for Wightman functions, then $\log Z[J(x)]$ is a generating function for the truncated Wightman functions. We will not need equation \eqref{truncatedwightmandefinitionexplicit} in what follows, but give a rigorous proof in appendix \ref{appendix:proof_of_truncated_Wightman_definition}.

\subsection{Spectral property}
According to axiom \ref{axiom:spectrum} there exists a mass gap in the theory, such that the spectrum of any state orthogonal to the vacuum $|\Omega\>$ lies entirely in $\bar V_m^+$, with $m > 0$ the single-particle mass. Therefore, for any integer $0 \leq l \leq n$, 
\begin{equation}
    \label{wightmanspectrum}
    % \forall l \in \llbracket 0; n \rrbracket, \: 
    \supp_p \left[ \int d^d a\, e^{i p a} \cW(x_1, \ldots, x_l, x_{l+1} + a, \ldots x_n + a) \right] \subseteq \{0\} \cup \bar V_m^-\,,
\end{equation}
where $\bar V_m^-$ appears because $\phi(p)$ creates a state of momentum $-p$.

For the truncated Wightman functions, on the other hand, the vacuum no longer contributes as an intermediate state.

\begin{theorem}\label{thm:truncatedsupport}
For any integer $0 < l < n$,
\begin{equation}
    \mathop{\mathrm{supp}}_p \left[ \int d^d a\, e^{i p a} \cW^T(x_1, \ldots, x_l, x_{l+1} + a, \ldots x_n + a) \right] \subseteq \bar V_m^-\,.
\end{equation}
\end{theorem} 

To see why the theorem is not obvious we can consider equation \eqref{truncatedwightman4pt} as an example with $n = 4$ and $l = 2$. The first two terms on the right-hand side taken together manifestly have the desired support property, but for the two remaining terms the situation is not immediately clear.
\begin{proof}
We proceed by induction in $n$. A slight generalization of our induction hypothesis includes the cases with $l = 0$ or $l = n$:
\begin{equation}
    \label{supportWln}
    \mathop{\mathrm{supp}}_p \left[ \int d^d a\, e^{i p a} \cW^T(x_1, \ldots, x_l, x_{l+1} + a, \ldots x_n + a) \right]
    \subseteq
    \begin{cases}
        \bar V_m^- \qquad \text{if }0 < l < n\\
        \{0\} \qquad \text{if }l = 0 \text{ or }l = n
    \end{cases}
\end{equation}
where the $\{0\}$ is just due to translation invariance. This statement is manifestly true for $n =1$ and $l = 0,1$. For the next step in the induction we begin with:
\begin{multline}
    \label{truncatednpoint}
    \cW^T(x_1, \ldots, x_l, x_{l+1} + a, \ldots x_n + a)  =\\ \cW(x_1, \ldots, x_l, x_{l+1} + a, \ldots x_n + a)  - \sum_{\substack{\Pi \in \doublepin \\ \Pi \neq \{\{1,2,\ldots n\}\}}} \cW^T(\ldots) \cW^T(\ldots) \ldots \cW^T(\ldots)
\end{multline}
where the sum is over all partitions except the trivial one $\{\{1,2,\ldots,n\}\}$.

Consider a single non-trivial partition $\Pi$. The corresponding term in the sum consists of a product of $|\Pi|$ truncated Wightman functions. Let us suppress the $x$ dependence and write each of these truncated Wightman functions as $W^T_\alpha(a)$, with $\alpha = 1, \ldots, |\Pi|$. From
\begin{equation}
    \label{fouriertransformproduct}
    \int d^d a \, e^{i p a} \prod_{\alpha=1}^{|\Pi|} W^T_\alpha(a) = \int d^d a\, e^{i p a} \prod_{\alpha=1}^{|\Pi|} \int d_d p_\alpha \, e^{- i p_\alpha a}\widetilde W^T_\alpha (p_\alpha)\,
\end{equation}
we see that the Fourier transform of the product is non-zero only if $p = \sum_\alpha p_\alpha$ with each $p_\alpha$ in the support of $\widetilde W^T_\alpha(p_\alpha)$. But each of the $\widetilde W^T_\alpha(p)$ is subject to the induction hypothesis, equation \eqref{supportWln}, and is therefore non-zero only if $p_\alpha \in \{0\} \cup \bar V_m^-$. But $\{0\} \cup \bar V_m^-$ is closed under addition, so the support in $p$ lies in the same set. 

Now consider the terms that contribute to the support at $p = 0$. For these terms $p_\alpha = 0$ for every $\alpha = 1, \ldots |\Pi|$. Equation \eqref{supportWln} then informs us that every $W_\alpha^T(a)$ must have either none or all of its arguments shifted by $a$.
The partitions $\Pi \in \doublepin$ obeying this property are in one-to-one correspondence with the subpartitions of $\{\{1,\ldots, l\}, \{l+1, \ldots, n\}\}$. Taken together, all these subpartitions sum to the product
\begin{equation}
    \cW(x_1, \ldots x_l) \cW(x_{l+1} + a, \ldots, x_n + a)\,.
\end{equation}
After a Fourier transform, this product precisely cancels the support at $\{0\}$ of the first term in equation \eqref{truncatednpoint}.
\end{proof}

\subsection{Cluster decomposition}

The statement and proof of the theorem below are due to Ruelle \cite{ruelle1962asymptotic} and summarized nicely in several places, for example \cite{Jost:1965yxu,glimm2012quantum}. We note that the textbook \cite{streater2000pct} also proves a version of cluster decomposition, but one which does not suffice for our purposes. Indeed, our test functions are assumed to be compactly supported in momentum space and therefore cannot be compactly supported in position space. One therefore needs to isolate and bound the tails where operators are not spacelike separated, and doing so complicates the analysis significantly.

\begin{theorem}
    \label{thm:clusterdecomposition}
    For any $n \in \N^+$, consider a test function on $n+1$ points $f(x_0,x_1,\ldots x_n) \in \cS(\R^{d (n + 1)})$ and $n$ spatial vectors $a_i = (0, \ua_i)$ with $i = 1,\ldots,n$. Then
    \begin{equation}
        \label{phiadefinition}
        \Phi_f(\ua_1, \ldots \ua_n) \colonequals \int d^d x_0 \, d^d x_1 \ldots d^d x_n \, f(x_0, x_1, \ldots x_n) \cW^T(x_0, x_1 + a_1, \ldots x_n + a_n)
    \end{equation}
    is an element of $\cS(\R^{(d-1)n})$.
\end{theorem}

Here we of course fix the spacetime point $x_0$ just to ensure rapid falloff for an otherwise translationally invariant object.

\begin{remark}
    Since $\cW^T(\cdot)$ is a tempered distribution we can immediately conclude that the convolution $\Phi = f * \cW^T$ is a polynomially bounded function. We can do substantially better by using locality and the mass gap condition. In the literature $\cW^T(\cdot)$ is sometimes called a tempered distribution of rapid decrease.
\end{remark}

To prove the Schwartz property we need to show $C^\infty$ and rapid falloff of both the function as well as any of its derivatives. But $\Phi_f(\ua_1,\ldots \ua_n)$ is manifestly $C^\infty$, and any derivative of $\Phi_f(\ua_1,\ldots \ua_n)$ again takes the form \eqref{phiadefinition}. It therefore suffices to show the rapid falloff of $\Phi_f(\ua_1,\ldots,\ua_n)$, which we will do below.

\newcommand{\ulua}{\underline a}
\newcommand{\ulx}{\underline x}
\newcommand{\ulp}{\underline p}

First some bookkeeping. Let us write $\ulua \colonequals (a_0, a_1, \ldots a_n)$ with $a_0 \colonequals 0$, and likewise $\ulx \colonequals (x_0, x_1, \ldots x_n)$. We also introduce the norms:
\begin{align}
    ||\ulua|| &\colonequals \sqrt{\sum_{i = 0}^n ||a_i||^2} = \sqrt{\sum_{i = 1}^n ||\ua_i||^2} \label{uluanormdef}\,, & 
    ||\ulx|| &\colonequals \sqrt{\sum_{i = 0}^n ||x_i||^2}\,,
\end{align}
so we can write the rapid falloff requirement in $\R^{n (d-1)}$ as:
\begin{equation}
    \label{rapiddfalloff}
    \forall k > 0 \, : \qquad \lim_{||\ulua|| \to \infty} ||\ulua||^k \Phi_f(\ua_1, \ldots \ua_n) = 0
\end{equation} 
Below we will separate the $n+1$ operators into two spacelike separated groups, say `left' and `right', in order to eventually use the locality axiom. We write $\Pi_2$ for such a partition of size 2, and we will denote the corresponding splits of $\ulx$ and $\ulua$ respectively as $\{\ulx_\text{left}, \ulx_\text{right}\}$ and $\{\ulua_\text{left}, \ulua_\text{right}\}$.

A priori it is not clear that we can consider only a single such partition, since different groups may become spacelike separated for different $\ulua$ when taking the limit \eqref{rapiddfalloff}. The following lemma tackles this issue.

\begin{lemma}
    \label{partitionexistence}
   For any $\ulua$ there exists a partition $\Pi_2(\ulua)$ of size 2 such that:
\begin{equation}
    \label{partitiondefiningequation}
    \delta(\ulua_\text{left},\ulua_\text{right}) \geq ||\ulua|| / n^{3/2}
\end{equation}
where $\delta(\ulua_\text{left},\ulua_\text{right})$ is the \emph{smallest} possible Euclidean distance between any point in $\ulua_\text{left}$ and any point in $\ulua_\text{right}$.
\end{lemma}
\begin{proof}
Suppose that for a given $\ulua$ no such partition exists. Then for every possible partition the above inequality is violated. So pick first a partition where $\ulua_\text{left}$ consists of a single point, say $\ua_k$. Then there must exist another point, say $\ua_l$, such that $||\ua_k - \ua_l || < ||\ulua|| / n^{3/2}$. We include this point in the `left' subset and apply the argument again to find a third point $\ua_m$. This point lies within a distance $ ||\ulua|| / n^{3/2}$ of either $\ua_k$ or $\ua_l$, and by the triangle inequality the \emph{maximum} distance between any pair of points in $\{\ua_k, \ua_l,\ua_m\}$ is then strictly less than $2 ||\ulua|| / n^{3/2}$. Iterating this procedure, we find that for $n+1$ points the maximum distance between any pair $\ua_i$ and $\ua_j$ is strictly less than $||\ulua||/n^{1/2}$. Setting $j = 0$ we in particular find that $||\ua_k|| = ||\ua_k - \ua_0||$ is strictly less than $||\ulua|| / n^{1/2}$ for any $k$. This leads us to conclude that
\begin{equation}
    \sum_{k = 1}^n ||\ua_k||^2 < ||\ulua||^2
\end{equation}
which contradicts the very definition of $||\ulua||$ given in equation \eqref{uluanormdef}.
\end{proof}
We now claim that we can replace the statement \eqref{rapiddfalloff} with the statement that:
\begin{equation}
    \label{rapiddfalloffrestricted}
    \forall \hat \Pi_2, \, \forall k > 0 \,:\qquad \lim_{\substack{||\ulua|| \to \infty\\\Pi_2(\ulua) = \hat \Pi_2}}  ||\ulua||^k \Phi_f(\ua_1, \ldots \ua_n) = 0\,.
\end{equation}
In words, we mean that the limit is taken within the subset of all $\ulua$ such that $\Pi_2(\ulua)$ defined as per the previous lemma is equal to a given partition $\hat \Pi_2$. To see the equivalence, consider a completely general sequence of configurations $\{\ulua^{(i)}\}_{i = 1, 2, 3, \ldots}$ such that $\lim_{i \to \infty} ||\ulua^{(i)}|| = \infty$. For every $\hat \Pi_2$ we define a subsequence by retaining those elements for which $\Pi_2(\ulua^{(i)}) = \hat \Pi_2$. There are only finitely many such subsequences, and lemma \ref{partitionexistence} proves that taken together they reconstitute the original sequence. The statement \eqref{rapiddfalloffrestricted} says that each subsequence has the same limit, which must then also be the limit of the original sequence. Therefore, equation \eqref{rapiddfalloff} follows from equation \eqref{rapiddfalloffrestricted}.

We have reduced the cluster decomposition theorem \ref{thm:clusterdecomposition} to proving equation \eqref{rapiddfalloffrestricted}. However, we are still not in a position to use the locality axiom: $f(x_0,x_1,\ldots x_n)$ does not need to be compactly supported and so the two groups in $\cW^T(x_0, x_1+a_1, \ldots x_n + a_n)$ might never get spacelike separated for all $x$, no matter how large we make $||\ulua||$. We will therefore split the integral into a region where $||\ulx||$ is small compared to $||\ulua||$ and we can use locality, and another region where $||\ulx||$ is large and the Schwartz property of $f(x_0,x_1, \ldots x_n)$ will ensure the required rapid falloff. A suitable transition point is provided by the following lemma.

\begin{lemma}
    Let $\Pi_2(\ulua)$ be defined as per lemma \ref{partitionexistence} and suppose that $||\ulx||^2 < ||\ulua||^2 / (4 n^3)$. Then all points $x_\mathrm{left} + a_\mathrm{left}$ in the `left' subset are spacelike separated from all points $x_\mathrm{right} + a_\mathrm{right}$ in the `right' subset.
\end{lemma}

Intuitively speaking points can only be non-spatially separated for large $||\ulua||$ if $x_\text{right}$ lies deep within a forward or backward lightcone emanating from $x_\text{left}$. But this implies large \emph{timelike} separation and hence a large Euclidean distance.

\begin{proof}
If any pair of points $(x_l + a_l, x_r + a_r) \in (\underline{x}_\text{left} + \underline{a}_\text{left},\underline{x}_\text{right} + \underline{a}_\text{right})$ is \emph{not} spacelike separated then:
\begin{equation}
\begin{split}
&
     (x_l - x_r + a_l - a_r)^2 \geq 0\\
    &\Rightarrow ||x_l - x_r||^2  \geq  ||\ux_l - \ux_r + \ua_l - \ua_r||^2 +  ||\ux_l - \ux_r||^2 \\
    &\Rightarrow  ||x_l - x_r||^2 \geq \frac{1}{2} \p{ ||\ux_l - \ux_r + \ua_l - \ua_r|| +  ||\ux_l - \ux_r|| }^2 \\
    &\Rightarrow ||x_l - x_r|| \geq \frac{1}{\sqrt{2}} ||\ua_l - \ua_r||  \geq \frac{1}{\sqrt{2}}\delta(\ulua_{\mathrm{left}}, \ulua_{\mathrm{right}}) \geq \frac{1}{\sqrt{2} n^{3/2}}||\ulua|| \\
    &\Rightarrow   ||x_l||  + ||x_r|| \geq \frac{1}{\sqrt{2} n^{3/2}} ||\ulua|| \\
    &\Rightarrow  ||x_l||^2 + ||x_r||^2 \geq \frac{1}{4 n^3} ||\ulua||^2 \\
    &\Rightarrow  ||\ulx||^2 \geq \frac{1}{4 n^3} ||\ulua||^2
\end{split}
\end{equation}
The lemma follows by negating the above reasoning.
\end{proof}

We will now split the integral around $||\ulx||^2 = ||\ulua||^2/(8n^3)$, safely inside the region $||\ulx||^2 < ||\ulua||^2/(4n^3)$ of the previous lemma. To do so properly we introduce a smoothed out step function $\tilde \theta : \R \to [0,1]$ which is $C^\infty$ and obeys:
\begin{align}
    \tilde \theta(t) = 0 & \qquad\text{if } t \leq -1 \nn \\
    \tilde \theta(t) > 0 & \qquad\text{if } -1 < t < 0 \label{smooththetaclusterdec} \\
    \tilde \theta(t) = 1 & \qquad\text{if } t \geq 0 \nn
\end{align}
Note that all derivatives of $\tilde{\theta}$ are bounded. Therefore, $f(\underline{x}) \, \tilde{\theta}(\|\ulx\|^2 - \|\ulua\|^2/(8n^3))$ is also Schwartz with respect to $\ulx$, which allows us to write $\Phi_f(\ua_1, \ldots \ua_n)$ as the sum of the following two terms:
\begin{align}
    &\int d^d x_0 \,  \ldots d^d x_n \, f(x_0, \ldots x_n) \cW^T(x_0 + a_0, \ldots x_n + a_n) \left[ \tilde \theta\!\left(||\ulx||^2  - ||\ulua||^2/(8n^3)\right) \right]\label{NSintegral} \\
    &\int d^d x_0 \, \ldots d^d x_n \, f(x_0, \ldots x_n) \cW^T(x_0 + a_0, \ldots x_n + a_n) \left[ 1 - \tilde \theta\!\left(||\ulx||^2  - ||\ulua||^2/(8n^3)\right) \right] \label{SSintegral}
\end{align}
The integral on the first line falls off rapidly by the following lemma. Note that this is just a general statement about distributions and does not require any field theoretic properties. In particular, for the purposes of the lemma $\ulx$ and $\ulua$ are just vectors in $\R^N$ for some $N$.

\begin{lemma}
Let $W \in \cS'(\R^N)$ and $F \in \cS(\R^N)$ and $c > 0$. Then we have, for every $k > 0$,
\begin{equation} \label{eqnfallofftails}
    \lim_{||\ulua|| \to \infty} ||\ulua||^k \int d \ulx \, F(\ulx) W(\ulx + \ulua) \, \tilde \theta\!\left(||\ulx||^2 - c^2 ||\ulua||^2\right) = 0
\end{equation}
\end{lemma}

Our proof uses the \emph{Schwartz representation theorem} which exposes a remarkable property of tempered distributions: they are always `just' a polynomial times a finite differential operator acting on a continuous and bounded function. This for example means that $W(\ulx)$ can essentially be written as:
\begin{equation}
    \label{SchwartzrepforW}
    W(\ulx) = (1+\ulx^2)^R (1-\Delta)^Q f_W(\ulx),
\end{equation}
for some $R,Q \in \N$ and for some bounded and continuous function $f_W(\ulx)$. The operator $\Delta$ is the Laplace operator on the $\ulx$. Its action should really be understood using the familiar `integration by parts' property of tempered distribution, meaning that the action of a tempered distribution on any Schwartz test function reduces to a finite combination of ordinary (Riemann) integrals of continuous functions with rapid falloff. A proof of the Schwartz representation theorem is given in appendix \ref{app:SchwartzRepThm}.

\begin{proof}
    The representation \eqref{SchwartzrepforW} means that the left-hand side of equation \eqref{eqnfallofftails} can be written as:
    \begin{align}
         \lim_{||\ulua|| \to \infty} ||\ulua||^k  \int d \ulx \, f_W(\ulx+\ulua) (1-\Delta)^Q \left[ (1+(\ulx+\ulua)^2)^R F(\ulx) \tilde \theta\!\left(||\ulx||^2 - c^2 ||\ulua||^2\right) \right]
    \end{align}
    After using the Leibniz rule to expand the derivatives, one gets a finite linear combination of terms of the form:
    \begin{equation}
        I(\ulua) = \int d \ulx\, \chi_{\ulua}(\ulx) w(\ulx + \ulua)
    \end{equation}
    where $\chi_{\ulua}(\ulx)$ contains $F(\ulx) \tilde \theta\!\left(||\ulx||^2 - c^2 ||\ulua||^2\right)$ and derivatives thereof, and $w(\ulx + \ulua)$ the remainder. Therefore $w$ is continuous and polynomially bounded, and $\chi_{\ulua}$ is  a Schwartz function supported on $||\ulx||^2 > c^2||\ulua||^2 -1$, which implies $||\ulx|| > c ||\ulua||/2$ for large enough $||\ulua||$. Note also that $\chi_{\ulua}(\ulx)$ depends on $\ulua$ only through the norm $||\ulua||$, and that
    \begin{equation}
        \label{subsubchi}
        \forall n > 0 \, : \qquad \sup_{\ulua} \sup_{\ulx} (||\ulx||^n | \chi_{\ulua}(\ulx) | ) < \infty\,,
    \end{equation}
    even when derivatives of $\tilde \theta(\cdot)$ with respect to $\ulx$ are involved, since those are continuous and compactly supported.

    We now bound $I(\ulua)$ as follows, using $\#$ to denote a constant (not always the same) that is independent of $\ulua$ and $\ulx$.
    \begin{align}
        |I(\ulua)| &\leq \phantom{\#} \int_{||\ulx|| > c||\ulua||/2} d \ulx\,  |\chi_{\ulua}(\ulx)| |w(\ulx + \ulua)| \nn\\
            &\leq \# \int_{||\ulx|| > c||\ulua||/2} d \ulx\,  |\chi_{\ulua}(\ulx)| (1 + ||\ulx + \ulua||^M) & &\text{($w$ is polynomially bounded)} \nn \\
            &\leq \# \int_{||\ulx|| > c||\ulua||/2} d \ulx\,  |\chi_{\ulua}(\ulx)| (1 + (||\ulx|| + ||\ulua||)^M ) & &\text{(triangle inequality)} \nn \\
            &\leq \# \int_{||\ulx|| > c||\ulua||/2} d \ulx\,  |\chi_{\ulua}(\ulx)| (1 + ||\ulx||^M) & &\text{(since $||\ulx|| > c||\ulua||/2$)} \nn \\
            &\leq \# \int_{||\ulx|| > c||\ulua||/2} d \ulx\,  ||\ulx||^{-n} (1 + ||\ulx||^M) & &\text{(equation \eqref{subsubchi})} \nn \\ 
            &\leq \# ||\ulua||^{M + N - n}
    \end{align}
    where we assumed $M > 0$ without loss of generality. We are free to pick any positive $n$, and for $n > k + M + N$ we obtain the lemma.
\end{proof}

After all these mathematical preliminaries we are left with the integral on the second line, equation \eqref{SSintegral}. Here the points in the two groups are spacelike separated, so by the locality axiom we may freely replace $\cW^T(\ulx + \ulua)$ with either $\cW^T(\ulx_\text{left} + \ulua_\text{left},\,\ulx_\text{right} + \ulua_\text{right})$ or with $\cW^T(\ulx_\text{right} + \ulua_\text{right},\,\ulx_\text{left} + \ulua_\text{left})$. After doing so we may immediately drop the term with the regularized theta function because it has rapid falloff, again by the above lemma. Therefore, up to terms that fall off rapidly our integral is equal to either one of:
\begin{equation}
\label{maybedangerousterms}
\begin{split}
    &\int d\ulx\, f(\ulx)\cW^T(\ulx_\text{left} + \ulua_\text{left},\,\ulx_\text{right} + \ulua_\text{right})\\
    &\int d\ulx\, f(\ulx)\cW^T(\ulx_\text{right} + \ulua_\text{right},\,\ulx_\text{left} + \ulua_\text{left})
\end{split}
\end{equation}

Now we use the mass gap axiom, or more precisely the implied spectral support of the truncated Wightman functions as stated in theorem \ref{thm:truncatedsupport}. The first line of \eqref{maybedangerousterms} is non-vanishing only if $f(p_0,\ldots p_n)$ has support in the region where the total momentum in the `left' group is in $\bar V_m^-$. The second line however requires this total momentum to be in $\bar V_m^+$. Since $m > 0$ (strictly!) these are completely disjoint closed sets, so these two conditions are mutually incompatible. Indeed, we can always split $f(p_0,\ldots p_n)$ smoothly into two parts, both Schwartz functions, with one part whose support is disjoint from $\bar V_m^-$ and one part whose support is disjoint from $\bar V_m^+$. We use the first expression in equation \eqref{maybedangerousterms} to show that the first part vanishes, and likewise for the second part. Neither of these two parts can then violate the rapid falloff condition, so the cluster decomposition theorem follows.

\section{Solutions to the Klein-Gordon equation}
\label{sec:solutionskg}
Two solutions to the (classical) Klein-Gordon equation
\begin{equation}
    (\square + m^2) \chi(x) = 0 
\end{equation}
are the Pauli-Jordan ``functions''
\begin{equation}
\label{paulijordan}
\begin{split}
    D^{(\pm)}(x) &\colonequals \pm 2 \pi i \int d_d p \, \theta(\pm p^0) \delta(p^2 - m^2) e^{-i p x} = \pm i \int \frac{d_{d-1} \up}{2 E_{\up}} e^{\mp i E_{\up} t + i \up \ux}\\
                &= \left. \pm i \frac{m^{d-2}}{(2 \pi)^{d/2}} z^{1-\frac{d}{2}}K_{\frac{d}{2}-1}(z) \right|_{\displaystyle{z = m \sqrt{\ux^2 - (t\mp i\epsilon)^2}}}
\end{split}
\end{equation}
The integral on the first line can be made well-defined with the replacement $t \to t \mp i \epsilon$ and taking the limit $\epsilon \downarrow 0$ outside the integral. The passage from the first to the second line is a standard exercise, so we explain it telegraphically. By Lorentz invariance the $D^{(\pm)}(x)$ are functions only of $x^2$, and in terms of this variable the Klein-Gordon equation becomes a simple second-order linear ordinary differential equation. It therefore has two independent solutions, but only the one we wrote matches the falloff properties of $D^{(\pm)}(x)$ for large imaginary $t$.

We will be interested in the late-time behavior of a classical solution to the Klein-Gordon equation with smooth initial data which is compactly supported in momentum space. For any $f(\up) \in \cD(\R^{d-1})$, let
\begin{equation}
    \label{chisolnkg}
    \chi(x) = \int \frac{d_{d-1} \up}{2 E_\up} \, e^{- i E_\up t + i \up \ux} f(\up) = - i \int d^{d-1} \uy \, D^{(+)}(t,\ux - \uy) f(\uy)
\end{equation}
For simplicity we consider only solutions with positive energies, but our results will apply equally to a linear combination with both positive and negative energies. Note that $\chi(x)$ is $C^\infty$ on $\R^d$.

\subsection{Uniform decay}
Let us begin with a universal bound on the solutions to the Klein-Gordon equations at large $|t|$.

\begin{proposition}\label{prop:KGEslowdecay}
    There exists a $C > 0$ such that
    \begin{align}
        |\chi(t,\ux)| \leq \frac{C}{|t|^{\frac{d-1}{2}}}\,,
    \end{align}
    for every $|t| > 1$ and $\ux \in \R^{d-1}$.
\end{proposition}

\begin{proof}
\textit{(For this proof, we set $m=1$.)}

The Pauli-Jordan functions themselves do not obey this condition because of the lightcone singularity at $z = 0$. We therefore introduce a new function \cite{bogolubov2012general}
\begin{equation}
    A(t,\ux) \colonequals \int d_{d-1}\up \, e^{i\up\ux} \frac{e^{-i E_\up t}}{2E_\up (1+E_\up)^{d-1} }
\end{equation}
and write
\begin{align}
    \left| \chi (t, \ux) \right| &= \left| \int \frac{d_{d-1} \up}{2 E_\up} e^{- i E_\up t + i \up \ux} f(\up) \right| \nn \\
    &= \left| \int d^{d-1} \uy A(t,\ux - \uy) \left[\int d_{d-1}\up \, e^{i\up\uy} (1+E_\up)^{d-1} f(\up) \right] \right| \nn \\
    &\leq \sup_\ux |A(t,\ux)| \,\, \int d^{d-1}\uy \left|\int d_{d-1}\up \, e^{i\up\uy} (1+E_\up)^{d-1} f(\up) \right| \label{chiintermediateboundslow}
\end{align}
The term inside the absolute value sign, as a function of $\uy$, is an element of $\cS(\R^{d-1})$, and therefore the remaining integral is finite and independent of $t$ and $\ux$. It thus remains to bound $A(t,\ux)$, which is a straightforward calculation.

First we use the identity $\int_0^\infty e^{-\omega \alpha} \alpha^{d-2} d \alpha =\Gamma(d-1)/\omega^{d-1}$ to write:
\begin{equation*}
    A(t,\ux) = \frac{1}{\Gamma(d-1)} \int d_{d-1} \up \int_0^\infty d \alpha\, \alpha^{d-2} e^{-\alpha} \frac{e^{-i[E_\up(t-i\alpha)-\up\ux]}}{2E_\up}
\end{equation*}
and therefore, using the absolute convergence of the integrals in question,
\begin{equation}
    A(t,\ux) = \frac{1}{\Gamma(d-1)(2\pi)^{d/2}} \int_0^\infty d \alpha\, \alpha^{d-2} e^{-\alpha} \left. z^{1-\frac{d}{2}}K_{\frac{d}{2}-1}(z) \right|_{\displaystyle{z = \sqrt{\ux^2 - (t - i\alpha)^2}}}\,.
\end{equation}
For the next step we need bounds on the Bessel function $K_\nu(z)$. We recall that it is smooth everywhere in the punctured $z$ plane (modulo the branch cut on the negative real axis) and has the asymptotic behaviors
\begin{equation}
    K_\nu(z) \underset{z \to 0}{=} O(z^{-\nu}), \qquad K_\nu(z) \underset{|z| \to \infty}{=} \sqrt{\frac{\pi}{2}}\frac{e^{-z}}{\sqrt{z}} \left( 1 + O\left(\frac{1}{z}\right)\right)\,.
\end{equation}
If $\nu = 0$, which corresponds to $d = 2$, the $z \to 0$ behavior becomes logarithmic instead, but in any case we can write:
\begin{equation}
    |K_\nu(z)| \leq \begin{cases}
        C_\nu \left( |z|^{-1/2} + |z|^{-\nu} \right)e^{-\Re(z)} & \qquad \nu > 1/2\\
        C_\nu |z|^{-1/2} e^{-\Re(z)} & \qquad \nu \leq 1/2\,.
    \end{cases}
\end{equation}
Furthermore, for $\Re(z) > 0$ and $s > 0$, it is straightforward to show that
\begin{equation}
    |z|^{-s} e^{-\Re(z)} \leq |\Im(z^2)|^{-s} (2\Re(z))^s e^{-\Re(z)} \leq C_s |\Im(z^2)|^{-s}\,.
\end{equation}
Therefore, for the values of $z$ of interest and for $d > 3$ we can write
\begin{equation}
    |z^{1-\frac{d}{2}}K_{\frac{d}{2}-1}(z) | \leq C_{d} \left( \frac{1}{|\Im(z^2)|^{\frac{d-1}{2}}} + \frac{1}{|\Im(z^2)|^{d-2}} \right)\,,
\end{equation}
whereas for $d \leq 3$ the second term is simply absent. Keeping this second term for now, we substitute to find:
\begin{align*}
        |A(t,\ux)| 
        &\leq C_d \int_0^\infty d \alpha\, \alpha^{d-2} e^{-\alpha} \left. \left( \frac{1}{|\Im(z^2)|^{\frac{d-1}{2}}} + \frac{1}{|\Im(z^2)|^{d-2}} \right) \right|_{\displaystyle{z = \sqrt{\ux^2 - (t - i\alpha)^2}}} \\
        &\leq C_d \int_0^\infty d \alpha\, \alpha^{d-2} e^{-\alpha} \left( \frac{1}{|2 \alpha t|^{\frac{d-1}{2}}} + \frac{1}{|2 \alpha t|^{d-2}} \right) \\
        &\leq \tilde C_d \left( \frac{1}{|t|^{\frac{d-1}{2}}} + \frac{1}{|t|^{d-2}} \right)\,.
\end{align*}
The statement of the proposition follows for both $d \leq 3$, where the second term is actually absent, and for $d > 3$ because we assumed that $|t| > 1$.
\end{proof}

\subsection{Fast decay outside of the velocity cone}\label{subseq:fastdecay}
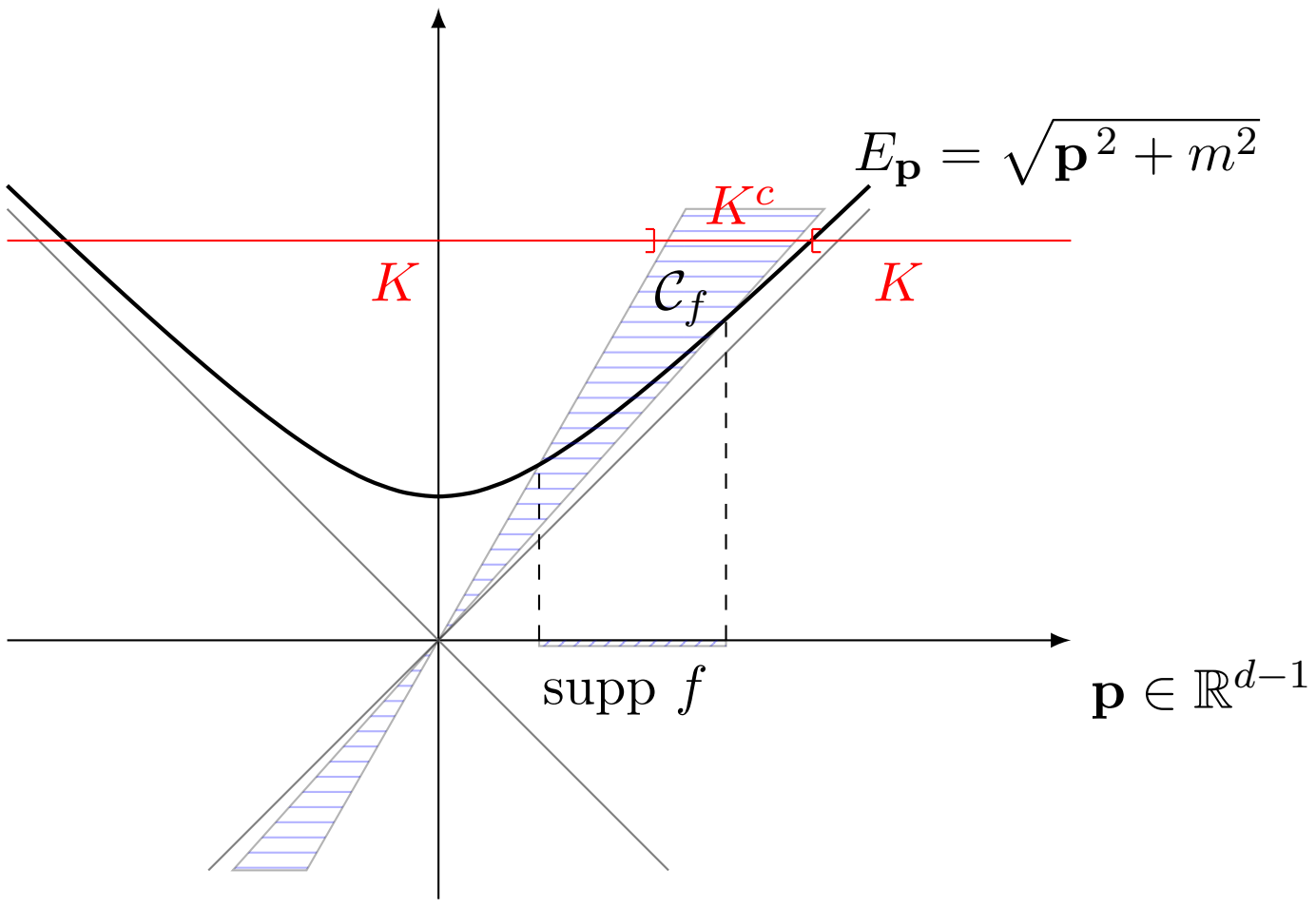
\begin{figure}
    \centering
    \begin{tikzpicture}[scale=2,>=latex]
        % \figureing
        % Axes
        \draw[->] (-1.5,0) -- (2.2,0) node[below right] {$\up \in \mathbb{R}^{d-1}$};
        \draw[->] (0,-0.9) -- (0,2.2);

        % E_p
        \draw[domain=-1.5:1.5,smooth,thick,black] plot (\x,{sqrt(\x*\x+0.25)});

        % velocity cone
        \draw[thin,gray] (-0.8,-0.8) -- (1.5,1.5);
        \draw[thin,gray] (0.8,-0.8) -- (-1.5,1.5);

        % Label de la courbe
        \node[black] at (2.15,1.7) {$E_{\up} = \sqrt{\up^{\,2} + m^2}$};

        % supp f
        \draw[pattern=north east lines, pattern color=blue, opacity=0.3] (0.35,0) -- (1,0) -- (1,-0.02) -- (0.35,-0.02) -- cycle;
        \node[below] at (0.65,-0.02) {$\operatorname{supp}\,f$};

        %Dash lines supp f
        \draw[dashed] (0.35,0) -- (0.35,{sqrt(0.35*0.35+0.25)});
        \draw[dashed] (1,0) -- (1,{sqrt(1+0.25)});

        % Cf
        \draw[pattern=horizontal lines, pattern color=blue, opacity=0.3] (0,0) -- ({1.5*0.35/sqrt(0.35*0.35+0.25)},1.5) -- ({1.5*1/sqrt(1+0.25)},1.5) -- cycle;
        \draw[pattern=horizontal lines, pattern color=blue, opacity=0.3] (0,0) -- ({-0.8*0.35/sqrt(0.35*0.35+0.25)},-0.8) -- ({-0.8*1/sqrt(1+0.25)},-0.8) -- cycle;
        \node[below] at (0.85,1.35) {$\mathcal{C}_f$};

        % K and Kc
        \draw[thin,red] (-1.5,1.39) -- (2.2,1.39);
        \draw[thin,red] (0.75,1.35) -- (0.75,1.43);
        \draw[thin,red] (0.72,1.35) -- (0.75,1.35);
        \draw[thin,red] (0.72,1.43) -- (0.75,1.43);
        \draw[thin,red] (1.3,1.35) -- (1.3,1.43);
        \draw[thin,red] (1.3,1.35) -- (1.33,1.35);
        \draw[thin,red] (1.3,1.43) -- (1.33,1.43);
        \node[above,red] at (1.05,1.38) {$K^c$};
        \node[below,red] at (-0.15,1.38) {$K$};
        \node[below,red] at (1.6,1.38) {$K$};

    \end{tikzpicture}
    \caption{\label{fig:velocitycone}Illustration of the velocity cone $\mathcal C_f$ given the support of $f(\up)$. The parts in red will be introduced in the proof of the auxiliary theorem below.}
\end{figure}

It turns out that the slow power-law decay of the previous proposition only holds inside the \emph{velocity cone} which is defined as follows.
\begin{definition}
    \label{def:velocitycone}
    We define the velocity cone of $f(\up)$ as the set of lines that are tangential to the on-shell four-momenta in its support:
    \begin{equation}
        \label{eq:velocitycone}
        \cC_f =\{  \alpha (E_{\up}, \up) \, | \, \up \in \supp(f(\up)) \text{ and } \alpha \in \R \} \subset \R^d
    \end{equation}
\end{definition}
See figure \ref{fig:velocitycone} for an illustration. Notice that $\cC_f$ is closed because $\supp(f(\up))$ is compact; also, since this set is invariant under rescalings it can be considered in position space despite having been defined in momentum space. In this subsection we will prove that the decay is much faster at large times when we consider points outside the velocity cones. It is useful to first consider the behavior of $\chi(x)$ along rays.

Given $n$ a $d$-dimensional Euclidean vector with $\|n\|^2 = 1$, we write 
\begin{equation}
    \label{gnsdefn}
    \chi(\tau n) = \int ds \, g_n(s) e^{-i \tau s}, \quad g_n(s) = \int \frac{d_{d-1} \up}{2 E_\up} \delta(s - n \cdot p) f(\up)\,,
\end{equation}
where in the last expression $p = (E_\up,\up)$. Proposition \ref{prop:fastdecay} will show that $\chi(\tau n)$ falls off rapidly at large $|\tau|$, provided that $n$ points outside the velocity cone.

\begin{lemma}
    The function $g_n(s)$ is supported for $|s| \leq S$, where $S$ can be chosen independently of $n$. For $n$ outside $\cC_f$ it is $C^\infty$ in both $n$ and $s$.
\end{lemma}

\begin{proof}
    Notice that
    \begin{equation}
        |n \cdot p| = |n^0 E_\up - \un \cdot \up| \leq |E_\up| + |\up|
    \end{equation}
    and the right-hand side is independent of $n$ and bounded as $\up$ varies over the compact support of $f(\up)$. So if $|s| > \sup_{\up \in \supp(f)} |E_\up| + |\up| \equalscolon S$ then the argument of the Dirac delta can never be zero and $g_n(s) = 0$.
      
    Next, for $|s| \leq S$ we can resolve the delta function by choosing an adapted coordinate equal to $n \cdot p$. Since $\up \mapsto n \cdot p$ is smooth, the coordinate transformation can be chosen to be smooth and locally invertible as long as $\nabla_\up(n \cdot p) \neq 0$ for all $p = (E_\up, \up)$ with $\up \in \supp(f)$.\footnote{This can be shown explicitly on use of the distributional generalization of the coarea formula. We make this precise in appendix \ref{sec:coarea}.} Now note that:
    \begin{equation*}
        \nabla_\up(n \cdot p) = 0\quad \Leftrightarrow\quad  \frac{n^0}{\sqrt{m^2+\up^2}} \up = \un\quad \Leftrightarrow\quad  \frac{n^0}{\sqrt{m^2+\up^2}} p = n\quad \Rightarrow\quad n \propto p
    \end{equation*}
    and the last equation says that $n \in \cC_f$. So smoothness is ensured as long as we keep $n$ outside the velocity cone.
    \end{proof}

\begin{proposition}\label{prop:fastdecay}
    Let $N$ be any compact subset of the unit Euclidean sphere in $\R^d$ such that $N \cap \cC_f = \emptyset$. Then for any $k \in \N$ there exists $C_{k,N}$ such that, for every $n \in N$ and $\tau > 0$,
    \begin{equation*}
        |\chi(\tau n)| \le \frac{C_{k,N}}{|\tau|^k}\,.
    \end{equation*}
\end{proposition}

\begin{proof}
    \begin{align*}
        |\tau|^k |\chi(\tau n)| &= \left| \int ds \, g_n^{(k)}(s) e^{- i\tau s} \right| \le \int ds |g_n^{(k)}(s)|\,.
    \end{align*}
    By the previous lemma, we know $g_n^{(k)}(s)$ is supported for $|s| \leq S$ and smooth in both $s$ and $n$. So for $n \in N$ we know that 
    \begin{equation}
        \label{gnssup}
        \sup_{n \in N,\,s \in [-S,S]} |g_n^{(k)}(s)| < \infty\,.
    \end{equation}
    The above integral is bounded by $2S$ times this supremum.
\end{proof}

Below is the version of the preceding proposition that we will use in the following.

\begin{proposition}
    \label{prop:KGEfastdecayfinal}
    Given $N$ a compactum of the Euclidean unit sphere that does not intersect $\cC_f$, for any $k,k' \in \N$, there exists a constant $C_{k,k',N}$ such that for $(t,\ux)$ on rays supported by this compactum $N$ with $|t|>1$,
    \begin{equation*}
        |\chi(t,\ux)| \leq \frac{C_{k,k',N}}{|t|^k (1 + \ux^2)^{k'}} 
    \end{equation*}
\end{proposition}

\begin{proof}
    Follows directly from the previous proposition, with the substitution $\tau = \sqrt{t^2 + \ux^2}$ and
    \begin{equation}
        (t^2 + \ux^2)^{k/2 + k'} = (t^2 + \ux^2)^{k/2} (t^2 + \ux^2)^{k'} \geq  |t|^k (1 + \ux^2)^{k'}
    \end{equation}
    for $|t| > 1$.
\end{proof}

\section{Proof of the Ruelle theorem}
\label{sec:proof}
We have now acquired the two key ingredients mentioned in the intuitive derivation of section \ref{subsec:intuitive}: cluster decomposition and rapid falloff of a solution to the Klein-Gordon equation outside its velocity cone. In theorem \ref{thm:haagruelletechnical} we will combine these ingredients to show that truncated $n$-point functions of the $a_t^{(\dagg)}(f)$ fall off as $|t| \to \infty$ for any $n > 2$. Therefore in the large $|t|$ limit only the truncated two-point functions are non-zero, exactly as in a free theory.

\subsection{Auxiliary theorem}
The main technical difficulty of the Haag-Ruelle theory lies in proving the following theorem. The connection between the $t$-dependent test functions $f_j(\pm p_j) e^{\pm i E_{\up_j}t - i p^0_j t}$ and the $a_t^{(\dagg)}(f)$ of the main theorem will be elucidated below.

\begin{theorem}
    \label{thm:haagruelletechnical}
    Let $f_j(p) \in \cD_1^+(\R^d)$ for $j = 1,\ldots n$. Then, for $|t| > 1$:
    \begin{equation}
        \label{haagruelletechnical}
        \left| \int d_d p_1 \ldots d_d p_n \left( \prod_j f_j(\pm p_j) e^{\pm i E_{\up_j}t - i p^0_j t} \right) \cW^T(p_1,\ldots p_n) \right| \leq C |t|^{-A}
    \end{equation}
    where the sign $\pm$ is chosen independently for each $j$ and with the exponent:
    \begin{equation}
        A = \frac{1}{2}(d-1)(n-2)\,.
    \end{equation}
    If in addition the $f_j(p)$ have some disjoint spatial support in the sense that $\cap_j \, \supp_{\up}(f_j(p)) = \emptyset$, then equation \eqref{haagruelletechnical} holds for all $A \geq 0$.
\end{theorem}

Note that the expression on the left-hand side of equation \eqref{haagruelletechnical} is $C^\infty$ in $t$. The specific value of $A$ will arise because we need to integrate a product of $n$ solutions to the Klein-Gordon equation, which each fall off like $|t|^{- \frac{d-1}{2}}$ by proposition \ref{prop:KGEslowdecay}, over a region whose size grows like $|t|^{d-1}$. 

\begin{proof}
Time translation invariance allows us to set $\exp\left(i \sum_j p_j^0 t\right) = 1$. We then introduce a set of bump functions $B_j(\up_j) \in \cD(\R^{d-1})$, $j \in \{1,\ldots n\}$ which equal $1$ in some domain such that, for all $j \in \{1,\ldots n\}$ and all $p \in \R^d$,
\begin{equation}
    f_j(p) B_j(\up_j) = f_j(p)
\end{equation}
We will demand that the bump functions $B_j(\up)$ and $B_k(\up)$ have disjoint support if $f_j(p)$ and $f_k(p)$ have disjoint spatial support, and more generally that if $\cap_j \, \supp_{\up}(f_j(p)) = \emptyset$ then $\cap_j \, \supp(B_j(\up)) = \emptyset$.

Introducing now the smooth solutions to the Klein-Gordon equation\footnote{Note that $[B_j^+(t,\ux)]^* = B_j^-(t,\ux)$ for real $B_j(\up)$ in momentum space.}
\begin{equation}
    B_j^\pm(t,\ux) = \int d_{d-1}\up\, B_j( \up) e^{\pm( i E_{\up} t - i \up \cdot \ux)}\,,
\end{equation}
allows us to write our target expression as,
\begin{equation}
    \int d^{d-1} \ux_1 \ldots d^{d-1} \ux_n \, B_1^\pm(t,\ux_1) \ldots B_n^\pm(t,\ux_n) \rho(\ux_2 - \ux_1, \ux_3 - \ux_1, \ldots \ux_n - \ux_1)
\end{equation}
where, using $x_i = (0, \ux_i)$,
\begin{multline}
    \rho(\ux_2 - \ux_1, \ux_3 - \ux_1, \ldots \ux_n - \ux_1) \colonequals \int d^d y_1 \ldots d^d y_n f_1(\mp y_1) \ldots f_n(\mp y_n) \cW^T(y_1 + x_1, \ldots y_n + x_n)\,.
\end{multline}
Now by the cluster decomposition theorem \ref{thm:clusterdecomposition} $\rho(\cdot) \in \cS(\R^{(d-1)(n-1)})$.

To apply the bounds of propositions \ref{prop:KGEslowdecay} and \ref{prop:KGEfastdecayfinal} we split each integral over $\ux_j$ into a closed region $K_j(t)$ and its complement $K_j^c(t)$. Here we choose $K_j^c(t)$ to contain the intersection of the velocity cone with the given $t$ slice, which we denote as $\cC_{B_j}|_t$, plus a small domain around it that grows linearly with $|t|$. (In figure \ref{fig:velocitycone} $K_j^c(t)$ should strictly contain the intersection of the striped area with a constant $t$ slice, as indicated by the red lines.) We will assume that this domain is chosen such that its $(d-1)$-dimensional volume grows more or less like that of $\cC_{B_j}|_t$ itself, in the sense that:
\begin{equation}
    \vol(K_j^c(t)) < C_j |t|^{d-1}
\end{equation}
which can always be done.

Consider now first the contribution from the region where, say, $\ux_1 \in K_1(t)$. Then:
\begin{align}
    &\phantom{\leq}\left|\int_{K_1(t)} d^{d-1} \ux_1 \int d^{d-1} \ux_2 \ldots d^{d-1} \ux_n \, B_1^\pm(t,\ux_1) \ldots B_n^\pm(t,\ux_n) \rho(\ux_2 - \ux_1, \ux_3 - \ux_1, \ldots \ux_n - \ux_1)\nn\right|\\
    & \leq \int_{K_1(t)} d^{d-1} \ux_1 \int d^{d-1} \ux_2 \ldots d^{d-1} \ux_n \, |B_1^\pm(t,\ux_1)| \ldots |B_n^\pm(t,\ux_n)| |\rho(\ux_2 - \ux_1, \ux_3 - \ux_1, \ldots \ux_n - \ux_1)|\nn\\
    & \leq \frac{\#}{|t|^{k + (n-1)(d-1)/2}}\int_{K_1(t)} d^{d-1} \ux_1 \int d^{d-1} \ux_2 \ldots d^{d-1} \ux_n \, \frac{1}{(1 + ||\ux_1||^2)^{d}}  |\rho(\ux_2 - \ux_1, \ux_3 - \ux_1, \ldots \ux_n - \ux_1)|\nn\\
    & = \frac{\#}{|t|^{k + (n-1)(d-1)/2}}\int_{K_1(t)} d^{d-1} \ux_1 \int d^{d-1} \tilde \ux_2 \ldots d^{d-1} \tilde \ux_n \, \frac{1}{(1 + ||\ux_1||^2)^{d}}  |\rho(\tilde \ux_2, \tilde \ux_3, \ldots \tilde \ux_n)|\nn\\
    & \le \frac{\#}{|t|^{k + (n-1)(d-1)/2}}\int d^{d-1} \ux_1\, d^{d-1} \tilde \ux_2 \ldots d^{d-1} \tilde \ux_n \, \frac{1}{(1 + ||\ux_1||^2)^{d}}  |\rho(\tilde \ux_2, \tilde \ux_3, \ldots \tilde \ux_n)|\nn\\
    &\leq \frac{\#}{|t|^K}\,.
\label{fastfalloff}
\end{align}
for arbitrary $K$. Here the second line is elementary, the third uses proposition \ref{prop:KGEfastdecayfinal} for $\ux_1$ (with arbitrary $k$) and proposition \ref{prop:KGEslowdecay} for $\ux_2, \ldots \ux_n$, the fourth is simply the change of variables $\tilde \ux_2 = \ux_2 - \ux_1$, the fifth enlarges the integration domain so the integral becomes independent of $t$, and the last uses the Schwartz property of $\rho(\cdot)$ to declare that the integral is finite.

For the complementary region we can only use proposition \ref{prop:KGEslowdecay} to obtain the bound
\begin{align}
    &\phantom{\leq}\left|\int_{K_1^c(t)} d^{d-1} \ux_1 \int d^{d-1} \ux_2 \ldots  d^{d-1} \ux_n \, B_1^\pm(t,\ux_1) \ldots B_n^\pm(t,\ux_n) \rho(\ux_2 - \ux_1, \ux_3 - \ux_1, \ldots \ux_n - \ux_1)\nn\right|\\
    &\leq \frac{\#}{|t|^{n(d-1)/2}}\int_{K_1^c(t)} d^{d-1} \ux_1 \int d^{d-1} \tilde \ux_2 \ldots \int d^{d-1} \tilde \ux_n \, |\rho(\tilde \ux_2, \tilde \ux_3, \ldots \tilde \ux_n)| \nn\\
    &= \frac{\#}{|t|^{n(d-1)/2}} \vol(K_1^c(t)) \int d^{d-1} \tilde \ux_2 \ldots \int d^{d-1} \tilde \ux_n \, |\rho(\tilde \ux_2, \tilde \ux_3, \ldots \tilde \ux_n)| \nn\\
    &\leq \frac{\#}{|t|^{(n-2)(d-1)/2}}\,,
\label{slowfalloff}
\end{align}
which produces the power of $|t|$ stated in the theorem.

Now by symmetry the bound of equation \eqref{fastfalloff} actually applies as soon as any of the $\ux_j \in K_j(t)$, so in actuality the bound \eqref{slowfalloff} can only be saturated by the region where all the $\ux_j \in K_j^c(t)$. But suppose now that the velocity cones do not overlap, that is, the $f_j$ have disjoint spatial support. Then, upon choosing the $K_j^c(t)$'s tight enough, the corresponding cones only have the origin in common, so that for any $\ux_1 \in K_1^c(t)$ there has to be at least one $k \neq 1$ such that $\ux_1$ is far from the whole of $K_k^c(t)$: the distance between $\ux_1$ and $K_k^c(t)$ grows linearly with $t$, by the nature of the velocity cones. As a result, so does the maximal distance between $\ux_1$ and the $\ux_j \in K_j^c(t)$, $j \neq 1$. In that case we use the Schwartz property of $\rho(\cdot)$, which gives
\begin{equation}
    \label{SchwarzRhoInAuxTheorem}
    |\rho(\ux_2 - \ux_1, \ux_3 - \ux_1, \ldots, \ux_n - \ux_1)| \leq \# \left[ 1 + \max( ||\ux_2 - \ux_1||, ||\ux_3 - \ux_1||, \ldots ||\ux_n - \ux_1|| )\right]^{-k} \leq \# |t|^{-k}\,,
\end{equation}
to conclude that the contribution where each $\ux_j \in K_j^c(t)$ is bounded by a constant times $|t|^{-k}$, for any $k > 0$. This leads to the better bound:
\begin{align}
    &\phantom{\leq}\left|\int_{K_1^c(t)} d^{d-1} \ux_1 \ldots \int_{K_n^c(t)} d^{d-1} \ux_n \, B_1^\pm(t,\ux_1) \ldots B_n^\pm(t,\ux_n) \rho(\ux_2 - \ux_1, \ux_3 - \ux_1, \ldots \ux_n - \ux_1)\nn\right|\\
    &\leq \frac{\#}{|t|^{n(d-1)/2}}\int_{K_1^c(t)} d^{d-1} \ux_1 \int_{K_2^c(t)} d^{d-1} \ux_2 \ldots \int_{K_n^c(t)} d^{d-1} \ux_n \, |\rho(\ux_2 - \ux_1, \ux_3 - \ux_1, \ldots \ux_n - \ux_1)| \nn\\
    &\leq \frac{\#}{|t|^{k + n(d-1)/2}} \vol(K_1^c(t)) \vol(K_2^c(t)) \ldots \vol(K_n^c(t))\nn\\
    &\leq\frac{\#}{|t|^{k - n(d-1)/2}} = \frac{\#}{|t|^{K}}  
\end{align}
again for arbitrary $K \geq 0$ and large $|t|$.
\end{proof}

\subsection{Proof of the main statement}
We recall the main statement of the Haag-Ruelle theorem.

\renewcommand{\thetheorem}{3.3}
\begin{theorem}
    Suppose $f_1(p), \ldots f_n(p) \in \cD_1^+(\R^d)$ have completely disjoint spatial support, so $\supp_\up(f_i(p)) \cap \supp_\up(f_j(p)) = \emptyset$ for all $i \neq j$. The states
    \begin{equation}
        |\Phi_t\> = a^{(\dagger)}_{t}(f_1) \ldots a^{(\dagger)}_{t}(f_n) |\Omega\>
    \end{equation}
    converge strongly as $t \to \pm \infty$. The inner products between the limiting states reduce to those of the free field. \end{theorem}
\renewcommand{\thetheorem}{\thesection.\arabic{theorem}}

\begin{proof}
For the first part we will show that $||d|\Phi_t\>/dt||^2$ falls off fast enough at large $|t|$ in order for the definition
\begin{equation}
    \label{phifromintegral}
    |\Phi_{\pm \infty}\> = |\Phi_{t=0}\> + \int_0^{\pm \infty} dt\, \frac{d|\Phi_t\>}{dt} \,
\end{equation}
to make sense. For the second part we need to investigate the limit of inner products $\<\Phi_t^{(1)} | \Phi_t^{(2)}\>$ between two different $|\Phi_t\>$.

Now $||d|\Phi_t\>/dt||^2$ is a sum of terms of the form
\begin{equation}
    \label{normsquaredschematic}
    \langle\Omega| a_{t}^{(\dagger)}(f_1) \ldots \frac{d}{dt} a_{t}^{(\dagger)}(f_i) \ldots \frac{d}{dt} a_{t}^{(\dagger)}(f_j) \ldots a_{t}^{(\dagger)}(f_{N}) | \Omega\rangle\,,
\end{equation}
with $N = 2n$ operator insertions. Our notation here is somewhat imprecise, so let us emphasize that there are exactly two time derivatives in this expression. Note also that the velocity cones are pairwise identical. On the other hand, $\<\Phi_t^{(1)} | \Phi_t^{(2)}\>$ is given by a similar expression but without time derivatives, with $N$ now an arbitrary positive integer, and with velocity cones only assumed disjoint for the operators within $\Phi^{(1)}_t$ and within $\Phi^{(2)}_t$.\footnote{\label{footnotevelcone}We originally defined the velocity cone in terms of $\supp(f(\up))$ for $f(\up) \in \cD(\R^{d-1})$. Whenever we refer to the velocity cone for functions $f(p) \in \cD(\R^d)$, we have in mind the set obtained by replacing $\supp(f(\up))$ in equation \eqref{eq:velocitycone} with $\supp_\up(f(p))$.}

Now expand the above expression in truncated expectation values. This yields a sum over partitions $\Pi \in \doublepinm$, where we recall that we write $\Pi = \{\pi_1, \pi_2,\ldots, \pi_{|\Pi|}\}$ with $\pi_\alpha$ the different subsets associated to the partition.

From equation \eqref{aadaggersmearedalternative} we see that, independently of the presence of a time derivative or a dagger, each operator is of the form:
\begin{equation}
    \label{eq:compact_f_irrespective_of_derivative}
    \int d_d p \, f(\pm p) e^{\pm i E_\up t  - i p^0 t} \phi(p)
\end{equation}
for some $f(p) \in \cD_1^+(\R^d)$, with the upper sign for $a_t(f)$ and the lower sign for $a^\dagg_t(f)$. The one-point functions of such operators vanish, since $f(p)$ vanishes near $p = 0$. As shown in the auxiliary theorem \ref{thm:haagruelletechnical}, the truncated $k$-point functions of such operators fall off faster than any power of $|t|$, unless all the $k$ velocity cones overlap. Using the completely disjoint spatial support assumption stated in the theorem,  such a complete overlap within each truncated $k$-point function can only occur when $k = 2$ for each subset in the partition. It therefore remains to analyze these completely disconnected contributions.

First, note that the contribution from these partitions is time-independent, because each $a^{(\dagger)}_t(f)$ acts on the vacuum either on the left or on the right, and
\begin{equation}
    \frac{d}{dt} a^{(\dagger)}_t(f) |\Omega\> = 0\,.
\end{equation}
Hence, the time derivatives ensure that these partitions completely drop out from $||d|\Phi_t\>/dt||^2$. For $\<\Phi_t^{(1)} | \Phi_t^{(2)}\>$ the same partitions produce precisely the claimed free-field value of the norm, because the test functions are in $\cD_1^+(\R^d)$ and therefore only the single-particle states contribute.
\end{proof}

\subsection{Proof of the additional claims}
\begin{claim}[i]
The convergence rate is faster than any power of $|t|^{-1}$.
\end{claim}

\begin{proof}
Since $||d|\Phi_t\>/dt||^2$ falls off faster than any power in $1/|t|$, the convergence rate can be estimated easily as:
\begin{equation*}
    \| |\Phi_t\rangle - |\Phi_{+\infty}\rangle \| = \left\| \int_t^{+\infty} \frac{d|\Phi_s\rangle}{ds} ds \right\|  \le \frac{\#}{t^{K-1}}
\end{equation*}
with arbitrary $K$.
\end{proof}

\begin{claim}[ii]
    The limiting states only depend on the $f(p)$ reduced to the mass shell $p^2 = m^2$ with $p^0 > 0$.
\end{claim}

\begin{proof}
Take two states $|\Phi_t\rangle$ and $|\tilde \Phi_t\rangle$ which differ only by the use of two different test functions $f_1(p), \tilde f_1(p) \in \cD_1^+(\R^d)$. Suppose however that they are equal on the mass shell, i.e.\ $f_1(p) - \tilde f_1(p) = 0$ whenever $p^2 = m^2$. Now by linearity $a^{(\dagger)} _t(f_1) - a^{(\dagger)}_t(\tilde f_1) = a^{(\dagger)}_t(f_1 - \tilde f_1)$. Repeating the steps of the main proof we then see that $|||\Phi_t\rangle - |\tilde \Phi_t\rangle||^2$ converges to zero, as the partitions with only $k=2$ subsets also vanish because $a^{(\dagger)}_t(f_1 - \tilde f_1)$ annihilates the vacuum.
\end{proof}

\begin{claim}[iii]
The limit is independent of the chosen Lorentz frame.
\end{claim}

Let $g$ be an element of the identity component of the $d$-dimensional Poincar\'e group. Axiom \ref{axiom:fieldproperties} stipulates the transformation properties of the fields under the corresponding unitary operator $U_g$, namely
\begin{equation}
    \label{phifPoincare}
    U_g \, \phi(f) U^{-1}_g = \phi(f_g)\,,\qquad f_g(x) \colonequals f(g^{-1} \cdot x)\,.
\end{equation}
We can now ask whether the creation and annihilation operators transform similarly, so whether
\begin{equation}
    \label{aadaggmaybecovariant}
    U_g \,a^{(\dagg)}_t(f) U^{-1}_g \overset{?}{=} a^{(\dagg)}_t(f_g)\,.
\end{equation}
It is a simple exercise to verify this equation in the free theory by virtue of the Klein-Gordon equation $(p^2 - m^2) \varphi(p) = 0$. In the interacting theory this equation does not always hold, but the proof below shows that the mismatch vanishes at large $|t|$ and covariance is restored.

\begin{proof}
We start from the definition of equation \eqref{aadaggersmearedalternative}. By inspection, equation \eqref{aadaggmaybecovariant} is trivially true if $g$ is either a translation or a spatial rotation, which leaves us with Lorentz boosts. We will consider an infinitesimal boost generated by a Hermitian operator $K_\uq$, such that
\begin{equation}
    \label{infinitesimalboostphi}
    i [K_\uq, \phi(p)] = (\delta_\uq p)^\mu \partial_\mu \phi(p)
\end{equation}
where
\begin{equation}
    \delta_\uq \begin{pmatrix}p^0 \\ \up\end{pmatrix} \colonequals \begin{pmatrix}\up \cdot \uq\\  p^0 \, \uq\end{pmatrix}\,.
\end{equation}
Of course $\delta_\uq( p^2 )  = 0$, and in the following we also use that:
\begin{equation}
    \delta_\uq( E_{\up} ) = \frac{p^0}{E_{\up}} \uq \cdot \up\,.
\end{equation}
Adding a test function, we may write, following a simple integration by parts,
\begin{equation}
    i [K_\uq, \phi(f)] = \phi(\delta_\uq f)
\end{equation}
with $(\delta_\uq f)(p) \colonequals - (\delta_\uq p)^\mu \partial_\mu f(p)$. The infinitesimal version of equation \eqref{aadaggmaybecovariant} is therefore whether
\begin{equation}
    i [K_\uq, a^{(\dagg)}_t(f)] - a^{(\dagg)}_t(\delta_\uq f) \overset{?}{=} 0\,.
\end{equation}
We will now only consider $a_t(f)$, since the computation for $a^\dagg_t(f)$ is entirely analogous. Explicit computation using equation \eqref{aadaggersmearedalternative} yields the non-invariance:
\begin{equation}
    i [K_\uq, a_t(f)] -  a_t(\delta_\uq f) = a_t(\delta_{\text{b}} f) - i t \, a_t(\delta_{\text{vb}} f)\,,
\end{equation}
where the `bad' and `very bad' pieces read:
\begin{equation}
    \begin{split}
    (\delta_\text{b} f)(p) = (\uq \cdot \up)  \frac{(p^0 - E_{\up})}{E_{\up}^2}\, f(p)\,,\qquad
    (\delta_\text{vb} f)(p) = (\uq \cdot \up) \frac{(p^0 - E_{\up})}{E_{\up}} \, f(p)\,.
    \end{split}
\end{equation}
Consider the infinitesimal variation $i K_\uq |\Phi_t\>$ of a state $|\Phi_t\>$ as defined in the main theorem. Since we work to first order, it suffices to expand each $a_t^{(\dagg)}(f)$ operator separately. We thus find unwanted contributions of the form
\begin{equation}
    \label{deltaPhiunwanted}
    \ldots a_t^{(\dagg)} (\delta_\text{b} f) \ldots |\Omega\>\,, \qquad \ldots \p{- i t \, a_t^{(\dagg)} (\delta_\text{vb} f)} \ldots |\Omega\>\,,
\end{equation}
with the dots denoting other creation and annihilation operators. We now claim that all of these contributions vanish as $|t| \to \infty$.

First note that $\delta_\text{b} f(p)$ and $\delta_\text{vb} f(p)$ are elements of $\cD_1^+(\R^d)$ if $f(p) \in \cD_1^+(\R^d)$, so we can make use of all the results above. Now expand the squared norm of any state of the form \eqref{deltaPhiunwanted} in truncated expectation values. By the same argument as the previous claim, the additional factor $(p^0 - E_\up)$, present for creation and annihilation operators alike, in both $\delta_\text{b} f(p)$ and $\delta_\text{vb} f(p)$ ensures that the partitions with subsets containing only two points vanish. And, as in the proof of the main theorem, the other partitions fall off faster than any power of $|t|^{-1}$. Since the squared norms vanish rapidly at large $|t|$, so do the states as vectors in the Hilbert space. In particular, the extra power of $|t|$ in the state involving $a^{(\dagg)}_t(\delta_\text{vb} f)$ is not dangerous.
\end{proof}

For the next claim we will need the lemma below that guarantees that the $a_t^{(\dagger)}(f)$ operators can be applied to $|\Phi_{\textrm{in}/\textrm{out}}\rangle$.
\begin{lemma}
    \label{non_overlapping_Phi_ex_in_D}
    The action of the operators $\phi(f)$ can be extended to the limit states $| \Phi_{\IN/\OUT} \rangle$ via
    \begin{equation}
        \label{eq:phi_action_on_Dex}
    \phi(f_1) \ldots \phi(f_n) | \Phi_{\IN/\OUT} \rangle = \phi(f_1) \ldots \phi(f_n)|\Phi_{t}\rangle + \int_t^{\mp \infty} \phi(f_1) \ldots \phi(f_n) \frac{d}{ds} |\Phi_s\rangle ds
    \end{equation}
    where $t$ is some arbitrary finite time.
\end{lemma}
% We will present our proof only for $\phi(f_1) \ldots \phi(f_n)|\Phi_{\OUT}\rangle$, the `in' case being entirely analogous.

\begin{proof}
    We begin our proof by first showing that the prescription above makes sense, \emph{i.e} it converges.
    
    By the Cauchy-Schwarz inequality, 
    \begin{equation}
    \begin{aligned}
        \label{DexinD_Cauchy_Schwartz}
        \left\| \phi(f_1) \ldots \phi(f_n)\frac{d}{ds} |\Phi_s\rangle \right\|^2 &= \left\langle \frac{d}{ds} \Phi_s \right| [\phi(f_1) \ldots \phi(f_n)]^\dagger \phi(f_1) \ldots \phi(f_n)\left| \frac{d}{ds} \Phi_s \right\rangle \\
                                                                            &\le \left\| \frac{d}{ds} |\Phi_s\rangle \right\| \left\| [\phi(f_1) \ldots \phi(f_n)]^\dagger \phi(f_1) \ldots \phi(f_n) \left| \frac{d}{ds} \Phi_s \right\rangle \right\|
    \end{aligned}
\end{equation}
    As shown in claim (i), the first factor in the right-most expression decays faster than any power law in $|s|$. The second factor is polynomially bounded in $|s|$, which we will now argue in the next paragraph.
    
    The square of the second term is a Wightman function, so a tempered distribution in many variables, integrated against a test function which depends on $s$. By the continuity of Schwartz distributions it is bounded by a finite linear combination of the Schwartz semi-norms on the test function, see proposition \ref{prop:estimateuf}. These semi-norms are defined in equation \eqref{Schwartzseminorms}, and for the case at hand the important property is that they involve taking a supremum not only of the test function itself but also of its derivatives. According to equation \eqref{aadaggersmearedalternative} our test functions are defined in momentum space $(p^0, \up)$ and depend on $s$ only through factors of the form $e^{\pm i (E_{\up} - p^0) s}$. Taking a supremum after taking finitely many derivatives yields at most a polynomial in $s$.
    Thus, 
    \begin{equation}
       \left\| \phi(f_1) \ldots \phi(f_n) \frac{d}{ds} |\Phi_s\rangle \right\|^2 \leq \# {|s|}^{-K} \quad \forall \quad K
    \end{equation}
    and we conclude that the integral in \eqref{eq:phi_action_on_Dex} absolutely converges.

    Finally consider the uniqueness of the prescription. If $|\Phi_{\IN/\OUT}\>$ are in the dense domain $D$ where the action of polynomials of $\phi(f)$ is defined as per axiom \ref{axiom:fields}, then the left-hand side of equation \eqref{eq:phi_action_on_Dex} has an independent meaning and we need to show that it equals the right-hand side. In other words, we need to show that
    \begin{equation}
        \phi(f_1) \ldots \phi(f_n)\left( \lim_{t \to \infty} |\Phi_t\> \right) = \lim_{t \to \infty} \phi(f_1) \ldots \phi(f_n)|\Phi_t\>
    \end{equation}
    This property means, by definition, that the operator $\phi(f_1) \ldots \phi(f_n)$ is \emph{closable}. Fortunately this is the case, because its adjoint operator is also densely defined.\footnote{Let $v_n$ be a sequence of vectors, each in the domain $D(A)$ in a Hilbert space $\cH$ for some linear operator $A$. Suppose that $v_n \to 0$, but $A v_n \to y$ as $n \to \infty$. For any $w \in D(A^*)$, continuity of the inner product gives $(w, y) = \lim_{n \to \infty} (w, A v_n) = \lim_{n \to \infty} (A^* w, v_n) = 0$. If $D(A^*)$ is dense in $\cH$, it follows that $y = 0$, so $A$ is closable.}
 Indeed, for $n=1$ we have $[\phi(f)]^\dagger = \phi(f^\dagger)$, which is defined in $D$ because $f^\dagger(x) = f^*(x)$ is Schwartz whenever $f(x)$ is. For $n > 1$ the argument is similar, although we need to use that $\phi(f)$ leaves $D$ invariant.
\end{proof}

\begin{claim}[iv]
    The limits can be taken independently and sequentially provided time ordering is respected, so for example
    \begin{equation}
        \lim_{\substack{t_1 \ldots t_m \to \infty\\ t_1 \le \dots \le t_m}} a_{t_1}^{(\dagg)}(f_1) \ldots a_{t_m}^{(\dagg)}(f_m) \lim_{\substack{t_{m+1} \ldots t_n \to \infty\\ t_{m+1} \le \dots \le t_n}} a_{t_{m+1}}^{(\dagg)}(f_{m+1}) \ldots a_{t_n}^{(\dagg)}(f_n) |\Omega\> = \lim_{t \to \infty} |\Phi_t\>\,,
    \end{equation}
    with a convergence rate faster than any power of $|t_1|^{-1}$.
\end{claim}

\begin{proof}
    We will content ourselves with showing that
    \begin{equation}
        \lim_{\substack{t_1 \ldots t_m \to \infty\\ t_1 \le \dots \le t_m}}  a_{t_1}^{(\dagg)}(f_1) \ldots a_{t_m}^{(\dagg)}(f_m) |\Phi'_{\textrm{out}}\> = \lim_{t \to \infty} |\Phi_t\> 
    \end{equation}
    with
    \begin{equation}
        |\Phi'_{\textrm{out}}\> = \lim_{t_{m+1} \to \infty} a_{t_{m+1}}^{(\dagg)}(f_{m+1}) \ldots a_{t_{m+1}}^{(\dagg)}(f_n) |\Omega\>,
    \end{equation}
    since all other cases are entirely analogous. We will in addition show that the convergence is faster than any power.

    By the previous lemma, all the states in this proof are well-defined. The result is evidently true when $m=0$. We will then proceed using induction in $m$.
    
    To lighten the notation we introduce a notation as follows:
    \begin{equation*}
        a_{t_1}^{(\dagg)}(f_1) \ldots a_{t_m}^{(\dagg)}(f_m) |\Phi'_{\textrm{out}}\> \equalscolon | t_1 \dots t_m \infty \dots \infty \rangle\,,
    \end{equation*}
    using which our induction hypothesis reads
    \begin{equation}
        \lim_{\substack{t_1 \ldots t_{m} \to \infty\\ t_1 \le \dots \le t_{m}}} | t_1 \dots t_{m} \infty \dots \infty \rangle = | \infty \dots \infty \rangle
    \end{equation}
    with convergence faster than any power in $|t_1|$. Our goal is to show the same equation holds with $m$ replaced by $m+1$.

    By the triangle inequality,
    \begin{multline}
        \left\| | t_1 \dots t_{m} t_{m+1}\infty  \dots \infty \rangle - | \infty \dots \infty \rangle \right\| \leq \\
        \begin{aligned}
        &\quad \left\| | t_1 \dots t_{m} t_{m+1}\infty  \dots \infty\rangle - |t_1 \dots t_{m} t_{m+1} t_{m+1} \dots t_{m+1} \rangle \right\|  \\
        &+\left\| |t_1 \dots t_{m} \, t_{m+1} t_{m+1} \dots t_{m+1} \rangle - |t_1 \dots t_m \, \infty \infty \dots \infty \rangle \right\| \\ 
        &+\left\| |t_1 \dots t_{m} \infty \infty \dots \infty \rangle - | \infty \dots \infty \rangle \right\|\,.
        \end{aligned}
    \end{multline}
    Our induction hypothesis ensures that the norm on the last line goes to zero rapidly in the desired limit. It remains to show that the same is true for the other two norms.
    
    The expression on the line before last reads:
    \begin{align*}
        &\left\| |t_1 \dots t_{m} \infty \dots \infty \rangle - |t_1 \dots t_{m} t_{m+1} \dots t_{m+1} \rangle \right\| \\
        &= \left\| \int_{t_{m+1}}^{\infty} a_{t_1}^{(\dagg)}(f_1) \ldots a_{t_{m}}^{(\dagg)}(f_{m}) \frac{d}{ds} a_{s}^{(\dagg)}(f_{m+1}) \ldots a_{s}^{(\dagg)}(f_{n}) | \Omega\rangle ds \right\| \\
        &\le \int_{t_{m+1}}^{\infty} \left\| a_{t_1}^{(\dagg)}(f_1) \ldots a_{t_{m}}^{(\dagg)}(f_{m}) \frac{d}{ds} a_{s}^{(\dagg)}(f_{m+1}) \ldots a_{s}^{(\dagg)}(f_{n}) | \Omega\rangle \right\| ds
    \end{align*}
    To control this expression, we follow the same logic as in the proof of the previous lemma. By the Cauchy-Schwarz inequality, the integrand is bounded by the norm of $\frac{d}{ds} a_{s}^{(\dagg)}(f_{m+1}) \ldots a_{s}^{(\dagg)}(f_{n}) \Omega\rangle$ which decays faster than any power law in $s$, times a term that grows at most polynomially in $s$ and $t_1, \ldots ,t_m$. Since $t_{m+1} \geq t_i \quad \forall \quad i \leq m$, we find that upon integrating in $s$,  $\left\| |t_1 \dots t_{m} \infty \dots \infty \rangle - |t_1 \dots t_{m} t_{m+1} \dots t_{m+1} \rangle \right\|$ decays faster than any power law in $1/t_{m+1}$.

    The final correction term is of almost the same form. We have
    \begin{align*}
        &\left\| | t_1 \dots t_{m} t_{m+1}\infty  \dots \infty\rangle - |t_1 \dots t_{m} t_{m+1} t_{m+1} \dots t_{m+1} \rangle \right\|\\
        &\le \left\| \int_{t_{m+1}}^{\infty} a_{t_1}^{(\dagg)}(f_1) \ldots a_{t_{m+1}}^{(\dagg)}(f_{m+1}) \frac{d}{ds} a_{s}^{(\dagg)}(f_{m+2}) \ldots a_{s}^{(\dagg)}(f_{n}) | \Omega\rangle ds \right\|
    \end{align*}
    We can again apply the logic of the previous lemma. Applying the Cauchy-Schwarz inequality results once more in a term that falls off rapidly in $1/s$, but now this term gets multiplied with a function that grows polynomially not only in $s$ and $t_1, \ldots, t_m$ but, because of the additional $a_{t_{m+1}}^{(\dagg)}(f_{m+1})$, also in $t_{m+1}$. That extra polynomial factor however does not offset the rapid decay in $t_{m+1}$ which is obtained after integrating over $s$.   
    \end{proof}

\begin{claim}[v]
    For $d\ge4$, the main statement as well as (ii) and (iii) hold without assuming the spatial supports of the $f_i$'s to be completely disjoint.
\end{claim}

\begin{proof}
    Without any assumption on the spatial supports of the test functions, the auxiliary theorem can only be used with $A = (n-2)(d-1)/2$ and so one must be more careful with the decay estimates.
    
    In particular, the falloff of $||d|\Phi_t\>/dt||^2$ is modified as follows. In the decomposition into truncated expectation values, the terms that contain only two-point functions still vanish because of the time derivatives. Any other term will either contain at least two three-point functions (because the total number of points is even) or contain at least one higher-point function. In either case this leads to decay like $1/|t|^{(d-1)}$. This means that $||d|\Phi_t\>/dt||$ falls off at least like $1/|t|^{(d-1)/2}$. The $t$-integral of $d|\Phi_t\>/dt$ converges if the decay is faster than $1/|t|^{1 + \epsilon}$ for some $\epsilon > 0$, which is the case if $d > 3$ strictly.
\end{proof}

It must be noted that the failure of the above proof does not mean that the underlying construction fails: $\int_0^\infty \frac{d|\Phi_t\>}{dt} dt$ could still converge absolutely if our upper bound is suboptimal, or perhaps it converges only conditionally.

Unfortunately, claim (iv) has not been proven to hold in case of overlapping wave functions. In fact the proof of
lemma \ref{non_overlapping_Phi_ex_in_D} does not hold if incoming/outgoing states have overlapping velocities. To see this,
consider equation \eqref{DexinD_Cauchy_Schwartz} in the proof of the lemma. For overlapping velocities
we have 
\begin{equation}
\left\| \frac{d}{ds} |\Phi_s\rangle \right\| \leq \frac{\#}{|s|^{(d-1)/2}}
\end{equation}
rather than a faster than any power law falloff. Whereas we only know that $\left\| \phi(f)^\dagger \phi(f) \left| \frac{d}{ds} \Phi_s \right\rangle \right\|$ is tempered
and hence bounded by some power of $|s|$. Without knowing this power, we cannot conclude that a definition of the form 
\eqref{eq:phi_action_on_Dex} makes sense for $|\Phi_\text{in/out}\rangle$ with overlapping wave functions.

\section{The spacetime picture}
\label{sec:spacetimepicture}
Theorem \ref{thm:haagruelle} shows the existence of the `in' and `out' free-field Fock spaces in any quantum field theory with the mass gap assumptions as stated in axiom \ref{axiom:spectrum}. An essential ingredient in the derivations was the compact support in momentum space of the wave packets $f(p)$, more precisely the assumption that they were elements of $\cD_1^+(\R^d)$. Such functions cannot be compactly supported in position space, which is somewhat contradictory to the idea that particles come in from (or move off to) `infinity' before (or after) the scattering event. There nevertheless exists an \emph{approximate} form of localization in position space, namely for each $t$ we can identify a region in spacetime $(x^0, \ux)$ that provides the essential contribution to the large $|t|$ behavior, with the contribution of its complement always falling off more rapidly than any power of $|t|$. The aim of this section is to identify such a region.

Given a test function $f(p) \in \cD_1^+(\R^d)$, we can introduce the new test functions:
\begin{equation}
    \label{def_f_t_pm}
    \begin{split}
        f_t^+(p) &\colonequals \left( \frac{E_\up - p^0}{2 E_\up}\right)  e^{i(p^0+E_\up)t} f(-p)\,, \\
        f_t^-(p) &\colonequals \left( \frac{E_\up + p^0}{2 E_\up} \right) e^{i(p^0-E_\up)t} f(p)\,,
    \end{split}
\end{equation}
in terms of which the operators $a_t(f)$ and $a^\dagger_t(f)$, defined originally in equation \eqref{aadaggersmearedalternative}, can be written as:
\begin{equation}
    \label{aadagger_t_f_t}
    \begin{split}
        a_t(f) &=  \int d_d p\,  f_t^+(-p) \, \phi(p)\,,\\
        a^\dagger_t(f) &= \int d_d p\,  f_t^-(-p)\, \phi(p)\,.
    \end{split}
\end{equation} 
We can then consider pairings with Wightman functions of the form:
\begin{equation}
    \label{eq:haagruelle_spacetime_expression}
\begin{aligned}
    \langle \Omega | a_{t_1}^{(\dagger)} (f_1) \dots a_{t_n}^{(\dagger)} (f_n) | \Omega \rangle
    &= \int d_d p_1 \dots d_d p_n f_{t_1}^\pm(-p_1) \dots f_{t_n}^\pm(-p_n) \langle \Omega | \phi(p_1) \dots \phi(p_n) | \Omega \rangle \\
    &= \int d^d x_1 \dots d^d x_n f_{t_1}^\pm(x_1) \dots f_{t_n}^\pm(x_n) \langle \Omega | \phi(x_1) \dots \phi(x_n) | \Omega \rangle \,.
\end{aligned}
\end{equation}
It is the latter position-space expression that we will work with below. Note that this expression is partially controlled at large $|t_i|$ by the Haag-Ruelle theorem. More precisely, we get an overlap between well-defined `in' and `out' states if we take the first $k$ of the $t_i$'s together to $+ \infty$ and the remaining $t_i$'s to $- \infty$, or vice versa, for any $k \in \{0, 1, \ldots n\}$. In fact, by claim (iv) of the main theorem of the previous section, we can take these limits sequentially, respecting the ordering condition of claim (iv). This however leaves the `mixed' limits where the operator ordering and $t_i$ ordering are not correlated. These are more difficult to understand. This is why, for example, in the standard LSZ prescription one replaces the Wightman function with a time-ordered one.

The aim of the remainder of this section is to prove that only a specific region in spacetime contributes at large $|t|$. As we will see, this is purely a property of the test functions $f_t(x)$ derived from an $f(p) \in \cD_1^+(\R^d)$.

\subsection{Families of solutions to the Klein-Gordon equation}\label{sss:familiesofKGsolution}
Let $f(p) \in \cD(\R^d)$. In the following we regard $p^0$ as a parameter and consider
\begin{equation}
    \label{defn:chi_p0_t_x}
    \chi^{\pm}_{p^0}(x) \colonequals \int \frac{d_{d-1}\up}{2E_\up} e^{\pm i(  E_\up t - \up \cdot \ux)} f(p^0, \up)
\end{equation}
as a family of solutions of the Klein-Gordon equation labeled by $p^0$. Notice that $\chi^\pm_{p^0}(x)$ are smooth and compactly supported in $p^0$.

\begin{lemma}
The family of functions $\chi^{\pm}_{p^0}(x)$ obeys the decay properties of propositions \ref{prop:KGEslowdecay}, \ref{prop:fastdecay} and \ref{prop:KGEfastdecayfinal} uniformly in $p^0$. That is, the constants $C$, $C_{k,N}$ and $C_{k,k', N}$ in these respective propositions can be chosen so the inequalities hold for all $p^0$ in the support of $\chi^\pm_{p^0}(x)$.
\end{lemma}

Note that for the equivalent of proposition \ref{prop:fastdecay} we need to define the velocity cone of the entire family $\chi_{p^0}^\pm(x)$ associated with an $f(p) \in \cD(\R^d)$. Just as in footnote \ref{footnotevelcone}, this velocity cone is defined using the spatial support of $f(p)$:
\begin{equation}
    \label{eq:newvelocitycone}
    \cC_f =\{  \alpha (E_{\up}, \up) \, | \, \up \in \supp\!_{\up}(f(p)) \text{ and } \alpha \in \R \} \subset \R^d\,.
\end{equation}
Recall that in the original definition \ref{def:velocitycone} we used $\supp(f(\up))$ to define the velocity $\cC_f$ associated to a lower-dimensional test function $f(\up) \in \cD(\R^{d-1})$. The notation $\cC_f$ is therefore slightly overloaded.

\begin{proof}
We will discuss how the proofs of the referenced propositions need to be modified if we replace $\chi^\pm(x)$, built from $f(\up)$, with $\chi^{\pm}_{p^0}(x)$, built from $f(p^0, \up)$, in order to show uniformity in $p^0$. In fact, we will do so only for $\chi_{p^0}^{-}(x)$, since the proofs for $\chi_{p^0}^{+}(x)$ are entirely analogous.

The dependence of the constant $C$ in proposition \ref{prop:KGEslowdecay} on $p^0$ enters through the estimate
\begin{equation}
    \int d^{d-1}\uy \left|\int d_{d-1}\up \, e^{i\up\uy} (1+E_\up)^{d-1} f(\up) \right|\,,
\end{equation}
which is just the relevant part of equation \eqref{chiintermediateboundslow}. If we now replace $f(\up)$ with $f(p^0, \up)$ then not much changes: the integral is uniformly bounded in $p^0$ since $f(p)$ is Schwartz.

To prove proposition \ref{prop:fastdecay} we introduced $g_n(s)$ in equation \eqref{gnsdefn}, which we recall was
\begin{equation}
    g_n(s) = \int \frac{d_{d-1}\up}{2 E_\up} \delta(s - n \cdot p) f(\up)\,.
\end{equation}
We showed that it is smooth in $(s,n)$ and compactly supported in $s$, provided we keep $n$ away from the velocity cone. If we now replace $f(\up)$ with $f(p^0, \up)$ then the modified function $g_{n,p^0}(s)$ is smooth in $(s,p^0, n)$ and compactly supported in both $s$ and $p^0$, again for $n$ away from the velocity cone defined through $\supp_\up(f(p))$. Therefore we can replace equation \eqref{gnssup} with
\begin{equation}
    \sup_{n \in N, p^0, s \in [-S, S]} |g_{n,p^0}^{(k)}(s)| < \infty\,,
\end{equation}
producing a $p^0$-independent constant $C_{k,N}$ in proposition \ref{prop:fastdecay}. 

Proposition \ref{prop:KGEfastdecayfinal} is just an immediate consequence of proposition \ref{prop:fastdecay}.
\end{proof}

Now we return to the test functions $f_t^\pm(p)$ defined in equation \eqref{def_f_t_pm}. Their position-space counterparts
\begin{equation}
     f_t^\pm(x) = \int d_d p \, e^{- i p x} f_t^\pm(p) \,,
\end{equation}
now obey the falloff conditions described in the following propositions.

\begin{proposition}\label{prop:slowdecaywithu}
    For all $q \in \N$, there exists a constant $C_q$ such that for $|t|>1$ and $x^0 \in \R, \, \ux \in \R^{d-1}$,
    \begin{equation}
        |f_t^\pm(x)| \le \frac{C_q}{(1+|x^0-t|)^q|t|^{\frac{d-1}{2}}}
    \end{equation}
\end{proposition} 
\begin{proposition}\label{prop:fastdecaywithu}
    Given $N$ a compactum of the Euclidean unit sphere that does not intersect $\cC_f$, for any $k,k',q \in \N$, there exists a constant $C_{q,k,k',N}$ such that for $(t,\ux)$ on rays supported by this compactum $N$ with $|t|>1$,
    \begin{equation}
        |f_t^\pm(x)| \le \frac{C_{q,k,k',N}}{(1+|x^0-t|)^q|t|^{k}(1+\ux^2)^{k'}}
    \end{equation}
\end{proposition}

\begin{proof}
Note that we can rewrite $f_t^\pm(x)$ as a Fourier transform (in $t$) of a function of the type 
considered in \eqref{defn:chi_p0_t_x}
\begin{equation}
    \begin{aligned}
    f_t^\pm(x) &= \int d_1 p^0 e^{\pm i p^0 (x^0 - t)} \chi_{p^0}^\pm(t,\ux)\,
    \end{aligned}
\end{equation}
by identifying $\chi^\pm_{p^0}(t,\ux)$ as 
\begin{equation}\label{eq:chitux}
    \chi^\pm_{p^0}(t,\ux) = \int \frac{d_{d-1}\up}{2E_\up} e^{\pm i (E_\up t -  \up \cdot \ux)}  (E_\up + p^0) f(p^0,\up) \,,
\end{equation}
and therefore we can estimate
\begin{equation}\label{powertimesftx}
    | (x^0 - t)^q f_t^\pm(x)| \leq \int d_1 p^0 | \partial_{p^0}^q \chi^\pm_{p^0}(t,\ux)| 
\end{equation}
The above lemma can be applied equally to $\chi^\pm_{p^0}(t,\ux)$ of equation \eqref{eq:chitux} but also to any of its derivatives with respect to $p^0$. Substituting the resulting bounds:
\begin{equation}
    |\partial_{p^0}^q\chi^\pm_{p^0}(t,\ux)| \leq \frac{C_q}{|t|^{\frac{d-1}{2}}}, \qquad |\partial_{p^0}^q\chi^\pm_{p^0}(t,\ux)| \leq \frac{C_{q, k, k', N}}{|t|^k (1 + \ux^2)^{k'}}
\end{equation}
in equation \eqref{powertimesftx} and using the compact support in $p^0$ yields almost exactly the requested bounds on $f_t^\pm(x)$; one only needs to replace $|x^0 - t|^q$ with $(1+ |x^0 - t|)^q$ but this is a trivial matter if the former result is acquired for all $q$.
\end{proof}

Note that $f_t^\pm(x)$ also solves the Klein-Gordon equation in $(t,\ux)$ for fixed $x^0 - t$, but we do not use that property here.

\subsection{Spacetime localization}
To summarize the previous propositions, we offer the following theorem. Recall that a pairing of a test function with a tempered distribution is not an ordinary integral; in order to capture a specific region, this time in position space, we need to introduce smooth bump functions $B(x) \in \cD(\R^d)$ which are non-negative and equal to 1 on a certain finite domain.

\begin{theorem}
\label{thm:spacetimelocalization}
Given $f(p) \in \cD_1^+(\R^d)$ with associated velocity cone $\cC_f$, let $B(y) \in \cD(\R^d)$ be a bump function which equals one in a neighborhood of
\begin{equation}
    \{  (y^0,  \uy) \in \R^d \, :\, \uy \in \cC_f|_{t = 1} \wedge y^0 = 1\}\,,
\end{equation}
then for any tempered distribution $T(x) \in \cS'(\R^d)$ and any $N > 0$ we have that
\begin{equation}
    \label{eq:ftx_spacetime_localization}
    \left|\int d^d x\, (1 - B(x/t))  f_t(x) T(x)\right| = O(|t|^{-N})
\end{equation}
\end{theorem}

\begin{figure}[H]
    \centering
    \begin{tikzpicture}[scale=2,>=latex]
        % \figureing
        % Axes
        \draw[->] (-1.5,0) -- (2.2,0) node[below right] {$\uy \in \mathbb{R}^{d-1}$};
        \draw[->] (0,-0.9) -- (0,1.8) node[above left] {$y^0$};
        
        % Cf
        \draw[pattern=horizontal lines, pattern color=blue, opacity=0.3] (0,0) -- ({1.5*0.35/sqrt(0.35*0.35+0.25)},1.5) -- ({1.5*1/sqrt(1+0.25)},1.5) -- cycle;
        \draw[pattern=horizontal lines, pattern color=blue, opacity=0.3] (0,0) -- ({-0.8*0.35/sqrt(0.35*0.35+0.25)},-0.8) -- ({-0.8*1/sqrt(1+0.25)},-0.8) -- cycle;
        \node[below] at (0.85,1.35) {$\mathcal{C}_f$};

        % B(y)=1
        \draw[thin, dashed] (0.79,1.39) -- (0,1.39) node[left] {$1$};
        \node[above,green!60!black] at (1.05,1.48) {$B(y)=1$};

        % B(y)=1 region
        \fill[green!60!black, opacity=0.5] plot[smooth cycle, tension=0.9] coordinates {(0.72,1.39) (1.0,1.46) (1.3,1.39) (1.0,1.32)}; 

    \end{tikzpicture}
    \caption{\label{fig:velocitycone_withbump}In green, a typical region where $B(y)$ equals $1$, so where $(1 - B(x/t))  f_t(x)$ equals $0$.}
\end{figure}
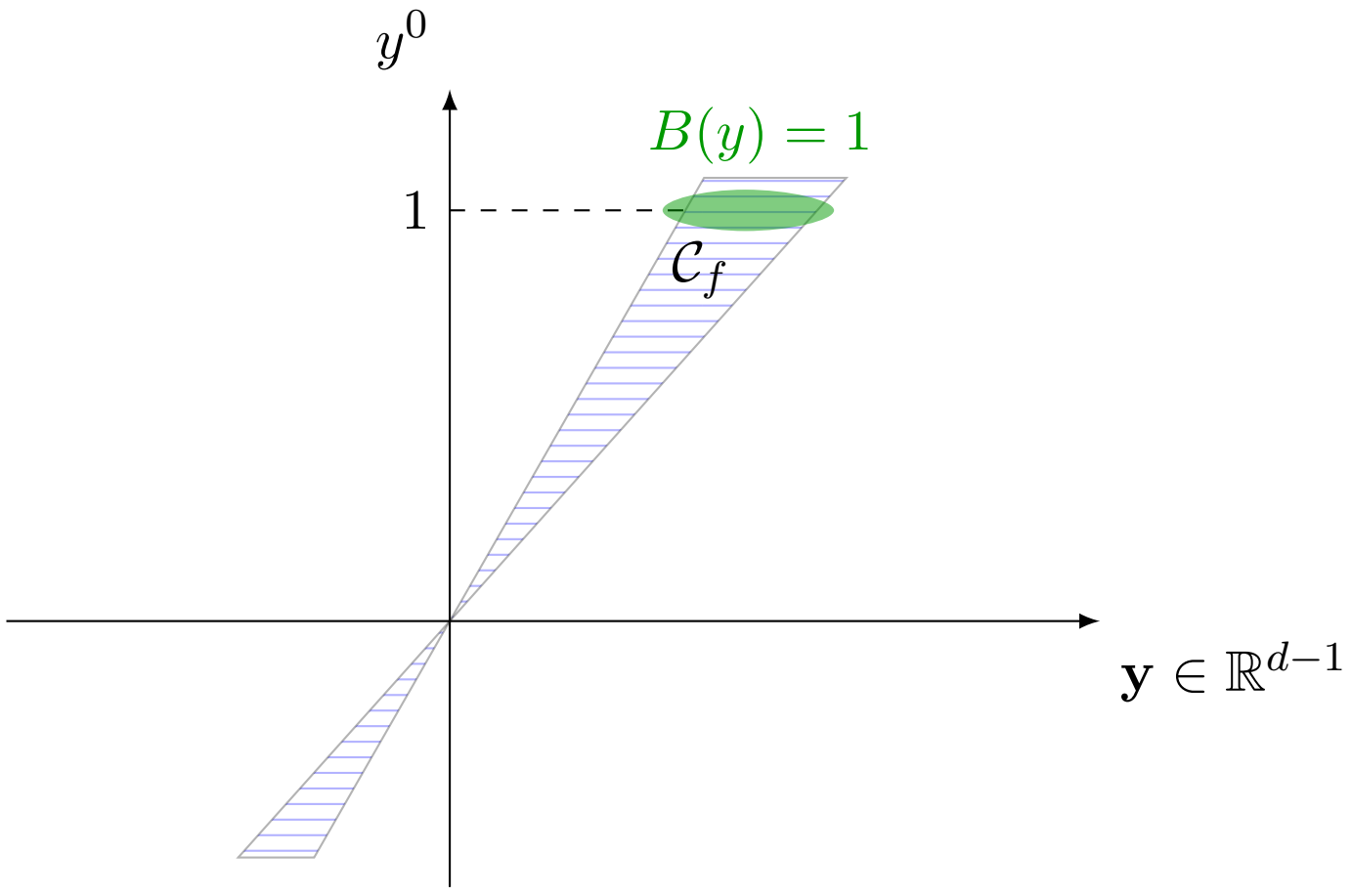

% See figure \ref{fig:velocitycone_withbump} for an illustration of the domain where the bump function should equal 1.

\begin{proof}
By the continuity property of tempered distributions, it suffices to bound the Schwartz semi-norms of the test function $(1 - B(x/t)) f_t(x)$ by $C |t|^{-N}$ for any $N > 0$. Let us consider instead
\begin{equation}
    B_c(x/t) f_t(x)
\end{equation}
for any $B_c(y)$ which is $C^\infty$, uniformly bounded and with uniformly bounded derivatives, and vanishing on the region given in the theorem. Our claim is that, up to a finite power of $t$, any finite derivative of $(1-B(x/t))f_t(x)$ can be written as a linear combination of functions of this more general form, and so it suffices to bound the supremum of the latter times a polynomial in $x$. Since $B_c(y)$ is bounded, we need to estimate
\begin{equation}
    \sup_x |P(x) f_t(x)|
\end{equation}
over the region where $1 - B(x/t)$ is not zero, with $P(x)$ a finite degree polynomial. In momentum space this polynomial becomes a derivative operator with respect to $p$ which, given the form \eqref{def_f_t_pm} of $f_t^\pm(p)$, does not meaningfully change $f_t(x)$ except for possibly a finite degree polynomial in $t$. But if
\begin{equation}
    \label{supf}
    \sup_{x\, \in\, \supp(1- B(x/t))} |f_t^\pm(x)| < \frac{C}{|t|^N}
\end{equation}
for any $N > 0$ then any finite degree polynomial will not matter, so the theorem follows if we manage to prove equation \eqref{supf}.

Now, without loss of generality we can suppose that $1- B(x/t)$ is zero in a small rectangle given by $|x^0/t - 1| < \epsilon$ and $\ux/t$ within an $\epsilon$ distance from $\cC_f|_{t=1}$, for some $\epsilon > 0$.  Rephrasing, this means that we need to estimate the supremum of $f_t(x)$ over all $x \in \R^d$ satisfying either $|x^0 - t| > \epsilon |t|$ or such that $(t,\ux)$ lies on a ray supported by a compactum of the Euclidean sphere that does not intersect $\cC_f$. Proposition \ref{prop:slowdecaywithu} ensures the upper bound on $|f_t(x)|$ proportional to $(1 + |x^0 - t|)^{-q} < (1 + \epsilon |t|)^{-q}$ for all $q > 0$ in the former case, and proposition \ref{prop:fastdecaywithu} provides an upper bound proportional to $|t|^{-k}$ for all $k > 0$ in the latter. In both cases we obtain rapid falloff as $|t| \to \infty$.
\end{proof}
\begin{remark}
Clearly the $x$-support of $B(x/t)$ grows linearly with $|t|$ in all directions. We can actually improve the localization in the $x^0$ direction without modifying the conclusion of the theorem. To do so, replace $B(x/t)$ with a family of bump functions $B_t(x)$, with derivatives bounded uniformly in $t$, whose support in $\ux$ grows linearly as before (so as to cover a neighborhood of $\cC_f|_t$), but whose support in $x^0$ (centered at $t$ as before) grows only like $|t|^\epsilon$ for some $\epsilon > 0$. The proof then uses that $(1+|x^0-t|)^{-q} < (1+|t|^\epsilon)^{-q}$ for all $q > 0$ where $|x^0 - t| > |t|^\epsilon$. 
\end{remark}

Although we wrote the theorem for a single test function, its meaning should be clear for the Haag-Ruelle theory: we can insert factors $B(x/t)$ in correlation functions of the operators $a_t^{(\dagger)}(f)$, as defined via the position-space expression \eqref{eq:haagruelle_spacetime_expression}, without affecting \emph{any} of our claims on the large $|t|$ limits.\footnote{This also applies to time derivatives, since acting with $\partial_t$ yields as extra contributions factors of the form $B(x/t) \partial_t f_t^\pm(x)$ and $\partial_t (B(x/t)) f_t^\pm (x) = -\frac{1}{t} (y \cdot \partial_y B)|_{y=x/t} f_t^\pm(x)$, which both behave at least as well as $B(x/t) f_t^\pm(x)$.} Doing so provides a manifest position-space localization of the integral to the support region of each $B(x/t)$. Since $B(x/t)$ is localized both in space and in time, we recover the intuitive picture that an `in' particle with a velocity ${\bf v} = \up / E_\up$ is created ``somewhere'' near the point $(t,\ux)$ with $\ux = {\bf v} t$ and $t \to - \infty$.

One may wonder if better localization is possible. Recall however that our test functions $f_t(x)$ are built upon functions $f(p) \in \cD_1^+(\R^d)$. But functions that are compactly supported in momentum space cannot be compactly supported in position space (where they are analytic). Therefore, \emph{exact} spacetime localization was never a real possibility. In other words, the Haag-Ruelle prescription is never equivalent to the naive idea of simply sending some operators to $x^0 = +\infty$ and others to $x^0 = -\infty$, perhaps along specific directions, to obtain an S-matrix element from a Wightman function. Indeed, physically speaking it would not at all be clear that every such operator would only create a particle or that this particle would move to the origin of spacetime.

Instead, the above theorem appears to be the most natural approximation of localization. Our theorem says that the Haag-Ruelle theory remains valid if we localize operators not to one specific direction, but to a region whose size grows with $|t|$. As $|t| \to \infty$ the region does move to infinity, in a direction given by the velocities found in the support of $f(p)$, but its size must simultaneously grow with $|t|$. In this way, as $|t| \to \infty$, it becomes infinitely big which allows for the necessary localization in momentum space.

\section{Why \texorpdfstring{$S$}{S} exists}
\label{sec:conclusion}
We have demonstrated that any interacting Wightman theory with the mass gap assumptions of axiom \ref{axiom:spectrum} has multi-particle `in' states spanning a subvectorspace of $\cH$. The Hilbert space completion of these states forms $\cH_\text{in}$, and of course we similarly have $\cH_\text{out}$.

In this section we first establish that $\cH_\text{in}$ and $\cH_\text{out}$ are both isometric to the free-field Fock space $\cF$, which is the statement of theorem \ref{them:H_in_is_Fock_space}. Using this result, we will show in subsection \ref{subsec:definingS} that the S-matrix can be naturally defined as a linear continuous map from $\cF$ to itself. Finally, in subsection \ref{subsec:S2to2} we will study the 2-to-2 S-matrix to exemplify the constraints of relativistic covariance and (sub)unitarity.

\subsection{The free-field Fock space}
\label{subsec:freefieldFockspace}
We introduced the free-field Fock space $\cF$ informally in section \ref{sec:freefields}, around equation \eqref{fockspaceinformal}. In this subsection we first give a precise definition of $\cF$ and establish some of its essential properties. This will provide the necessary background for establishing the isometry between $\cH_\text{in}$ and $\cF$ afterwards.

We define $\cF$ to be the Hilbert space completion of the vector space obtained by acting with polynomials of the free-field creation operators $a^\dagger(f)$ on the vacuum $|0\>$. It naturally splits into the direct sum of $n$-particle sectors
\begin{equation}\label{nparticlesubspace}
    \cF = \bigoplus\limits_{n=0}^{\infty} \cF^{(n)}
\end{equation}
where $\cF^{(n)}$ is the closed linear span of the $a^\dagger (f_1) \dots a^\dagger (f_n) |0\rangle$, with the $f_i$'s arbitrarily chosen Schwartz test functions. The scalar product in $\cF^{(n)}$, introduced informally in equations \eqref{eq:2ptfctssc} and \eqref{norms}, reads:
\begin{equation}
    \langle 0| a(\check f_n^\dagger) \dots a(\check f_1^\dagger) a^\dagger (g_1) \dots a^\dagger (g_n) |0\rangle = \frac{1}{n!} \sum\limits_{\sigma, \sigma' \in \cS_n} \big(f_{\sigma(1)}, g_{\sigma'(1)}\big) \ldots \big(f_{\sigma(n)}, g_{\sigma'(n)}\big) \,,
    \end{equation}
with the inner product between test functions defined as
\begin{equation}
    \big( f, g \big) \colonequals \int \frac{d_{d-1} \up}{2 E_{\up}} f^*(E_{\up}, \up) g(E_{\up}, \up)\,.
\end{equation}

This inner product identifies $\cF^{(n)}$ as an $L^2$ space, a point worth elaborating upon. The precise claim is that there is an isometric linear map we call $\iota$ from $\cF^{(n)}$ into the space $\Sym^n L^2\left(\R^{d-1}, \frac{d_{d-1} \up}{2 E_\up}\right)$, which is the $L^2$ space on $\R^{n(d-1)}$ with the relativistic $n$-particle measure
\begin{equation}
    \frac{d_{d-1} \up_1 \ldots d_{d-1} \up_n}{2^n E_{\up_1} \ldots E_{\up_n}}\,,
\end{equation}
restricted to elements that are completely symmetric in the $n$ arguments $\up_1, \ldots \up_n$. To explicate this, note for example that for $n=1$ the map is defined as
\begin{equation}
    \label{eq:omega_tilde_from_fn_to_l2}
    \iota\left[ a^{\dagger}(f) |0 \> \right] \colonequals \Big( \up \mapsto f(E_\up, \up) \Big) \in L^2\left(\R^{d-1}, \frac{d_{d-1} \up}{2 E_\up}\right)
\end{equation}
and for $n > 1$ we would apply the same operation to each creation operator, plus a symmetrization over the $n$ identical particles.

\begin{proposition}
For every $n \geq 1$, the map
\begin{equation}
    \iota : \cF^{(n)} \to \Sym^n L^2\left(\R^{d-1}, \frac{d_{d-1} \up}{2 E_\up}\right)
\end{equation}
is surjective.
\end{proposition}

\begin{proof}
Consider first the case $n=1$, so we consider the image of $\iota$ as defined in equation \eqref{eq:omega_tilde_from_fn_to_l2}. By the Wightman axioms, the test function $f(p)$ can be any element of $\cS(\R^d)$, and the mass-shell restricted map $\up \mapsto f(E_\up, \up)$ can be any element of $\cS(\R^{d-1})$. Then, we know that $\cS(\R^{d-1})$ is dense in $L^2(\R^{d-1})$ with the Lebesgue measure, because the Hermite functions form an orthonormal basis of $L^2(\R)$ and we can take simple products to get to $L^2(\R^{d-1})$. Finally, the factor $\sqrt{2 E_\up}$ is a Schwartz multiplier\footnote{A Schwartz multiplier $g$ is a function such that multiplication of any Schwartz function with $g$ yields another Schwartz function. This means that $g$ needs to be $C^\infty$, polynomially bounded, and have polynomially bounded derivatives.}, so if we have an orthonormal basis of Schwartz functions $h_k(\up)$ with $k \in \N$ for $L^2(\R^{d-1})$, then the functions $\sqrt{2 E_\up} h_k(\up)$ are still Schwartz and form an orthonormal basis of $L^2\left(\R^{d-1}, \frac{d_{d-1} \up}{2 E_\up}\right)$.

For $\cF^{(n)}$ with $n > 1$ we can simply take appropriately symmetrized products of the aforementioned single-particle basis to find a basis of $\Sym^n L^2\left(\R^{d-1}, \frac{d_{d-1} \up}{2 E_\up}\right)$ whose elements are all in the image of $\iota$.
\end{proof}

Since $\iota$ is also an isometry, the proposition directly implies that $\cF^{(n)}$ is unitarily isomorphic to $\Sym^n L^2\left(\R^{d-1}, \frac{d_{d-1} \up}{2 E_\up}\right)$ via $\iota$.

To explicitly construct the inverse of $\iota$ we need to address the subtlety that (in our conventions) the operators $a^\dagg(f)$ are defined using $d$-dimensional test functions $f$ whereas our $L^2$ space is over $\R^{d-1}$. Of course this is easily resolved: pick any $h \in \cD(\R)$ such that $h(m^2) = 1$ and define a map    
\begin{equation}
    \label{iotainverse}
    f(\up) \mapsto a^\dagger(f_h) |0\rangle, \quad f_h(p) :=  f(\up) h(p^2)
\end{equation}
for any $f(\up) \in \cS(\R^{d-1}) \subset L^2\left(\R^{d-1}, \frac{d_{d-1} \up}{2 E_\up}\right)$. The choice of $h$ is of course unimportant since, as in the proof of claim (ii) of theorem \ref{thm:haagruelle}, only the values of the test functions on the mass shell $p^2 = m^2$ are important. The map \eqref{iotainverse} is again an isometry, so it extends to the whole of $L^2\left(\R^{d-1}, \frac{d_{d-1} \up}{2 E_\up}\right)$ and in this way it provides the explicit inverse for $\iota$ on the single-particle states. An analogous construction goes through for any $n > 1$.

\subsection{The isometric embedding of \texorpdfstring{$\cF$}{F} into \texorpdfstring{$\cH$}{H}}
In the previous subsection we gathered some facts about the free-field Fock space $\cF$. In this subsection we return to the interacting Wightman theory.
Note that $\cF$ includes multi-particle states with overlapping velocity cones, which we explicitly excluded when formulating the Haag-Ruelle theorem. We will now show that these states are nevertheless part of $\cH_\text{in}$.

The main theorem \ref{thm:haagruelle} established that the Haag-Ruelle states are isometric to those in a subspace $\cF_{\mathrm{dis}} \subset \cF$ which is spanned by finite linear combinations of states with finitely many particles, created using wave functions in $\cD_1^+(\R^d)$ and with pairwise disjoint velocity cones. 

Let $\Omega_\text{in}$ be the corresponding linear map from $\cF_{\mathrm{dis}}$ into $\cH$, so the Haag-Ruelle states are given by $\Omega_\text{in} \cF_{\mathrm{dis}}$.\footnote{Of course, the ordering of the creation operators in a state of the form $a^\dagger (f_1) \dots a^\dagger (f_n) |0\rangle \in \cF$ does not matter. It follows from theorem \ref{thm:haagruelle} that different orderings also yield the same vector in $\cH$, so $\Omega_\text{in}$ is indeed a well-defined map.} Since $\Omega_\text{in}$ is norm-preserving and therefore bounded, it extends to the Hilbert space completion of $\cF_{\mathrm{dis}}$ and its image is the Hilbert space completion of the Haag-Ruelle states, which we defined to be $\cH_\text{in}$. 
\begin{theorem}
    \label{them:H_in_is_Fock_space}
The Hilbert space $\cH_{\text{in}}$ is isometrically isomorphic to the free-field Fock space $\cF$ 
via the map $\Omega_{\text{in}}$. 
\end{theorem}

\begin{proof}
By the decomposition \eqref{nparticlesubspace}, it suffices to show that each $n$-particle subspace $\cF^{(n)}_\text{dis}$ is dense in $\cF^{(n)}$. Since $\cF_\text{dis}^{(n)}$ is a subvectorspace of $\cF^{(n)}$, we can also apply $\iota$ to its elements and isometrically map them to elements of $\Sym^n L^2\left(\R^{d-1}, \frac{d_{d-1} \up}{2 E_\up}\right)$. In this language our objective is to show that the image
\begin{equation}
    \iota\left[ \cF_\text{dis}^{(n)} \right]
\end{equation}
is \emph{dense} in $\iota\left[ \cF^{(n)}\right]$, which we have established is just $\Sym^n L^2\left(\R^{d-1}, \frac{d_{d-1} \up}{2 E_\up}\right)$. We will do so for $n=2$, the general case being analogous.

Let us first discuss what elements of $ \iota\left[ \cF_\text{dis}^{(2)} \right]$ look like. In $\cF_\text{dis}^{(2)}$ each of the two test functions is an element of $\cD_1^+(\R^d)$, and since the passage to the mass shell is again surjective each test function can yield any element of $\cD(\R^{d-1})$. For $n> 1$ test functions we further need to take into account the symmetrization and the disjoint support. This means that for $n=2$ the space $ \iota\left[ \cF_\text{dis}^{(2)} \right]$ is spanned by
\begin{equation}
    \label{generalimageofOmegatilde}
    \Big((\up, \uq)\mapsto f(\up) g(\uq) + f(\uq) g(\up) \Big) \in \Sym^2 L^2\left(\R^{d-1}, \frac{d_{d-1} \up}{2 E_\up}\right)
\end{equation}
with $\up,\uq \in \R^{d-1}$ and with $f, g \in \cD(\R^{d-1})$ such that they have disjoint support.

Let us take an element $h(\up,\uq)$ in $\Sym^2 L^2\left(\R^{d-1}, \frac{d_{d-1} \up}{2 E_\up}\right)$ orthogonal to all elements of the form \eqref{generalimageofOmegatilde}. We will show that $h(\up,\uq)$ must be zero. To do so, consider $U, V$ any 
two \emph{disjoint} open sets of $\R^{d-1}$. Then by assumption
\begin{equation}
    \int_U d_{d-1}\up \int_V d_{d-1}\uq\, h(\up,\uq) f(\up) g(\uq) = 0
\end{equation}
for any $f \in \cD(U)$ and $g \in \cD(V)$. (The symmetrization \eqref{generalimageofOmegatilde} drops out because $h(\up,\uq)$ is also symmetric, and we used that $(2 E_\up)^{-1}$ is a Schwartz multiplier to absorb the measure factors in $f$ and $g$.) But $\cD(U) \times \cD(V)$ spans a dense subset of $L^2(U \times V)$, so it follows  that $h$ is $0$ almost everywhere on $U \times V$. This being true for arbitrary disjoint open sets $U$, $V$ of $\R^{d-1}$, one shows by taking an appropriate countable collection of such sets that $h$ is $0$ almost everywhere on $\R^{2(d-1)} \setminus \{ (\up,\up), \, \up \in \R^{d-1} \}$. But $\{ (\up,\up), \, \up \in \R^{d-1} \}$ is a subvectorspace of $\R^{2(d-1)}$ whose dimension is $d-1 < 2(d-1)$, and thus it has Lebesgue measure $0$. Therefore, $h$ is $0$ almost everywhere on $\R^{2(d-1)}$.
\end{proof}

\subsection{Defining \texorpdfstring{$S$}{S}}
\label{subsec:definingS}

The Haag-Ruelle theorem established the existence of the isometric maps
\begin{equation}
    \Omega_\text{in/out}: \cF_\text{dis} \to \cH
\end{equation}
whose images are, by construction, the `in' or `out' Haag-Ruelle states. Theorem \ref{them:H_in_is_Fock_space} then used the equivalence between $n$-particle Fock spaces and $L^2$ spaces to show that the completion of $\cF_\text{dis}$ inside $\cF$ is, in fact, all of $\cF$. Isometries like $\Omega_\text{in/out}$ then extend uniquely to $\cF$, so we can write:
\begin{equation}
    \Omega_\text{in/out}: \cF \to \cH_{\text{in/out}}\,,
\end{equation}
with $\cH_\text{in/out}$ the Hilbert space completion of the Haag-Ruelle states. The operators $\Omega_\text{in/out}$ so defined are called \emph{M\o{}ller operators}. They have adjoints $\Omega^\dagger_\text{in/out} : \cH \to \cF$ which are defined by the statement that for any $|\Psi\> \in \cH$ and $|\psi\> \in \cF$,
\begin{equation}
    \label{Omega_adjoint}
    \vev{ \psi | \Omega^\dagger_\text{ex} \Psi} = \vev{\Omega_\text{ex} \psi| \Psi}\,.
\end{equation}
Note that the first inner product is in $\cF$, but the second in $\cH$. We can therefore define
\begin{equation}
S = \Omega_\text{out}^{\dagg} \Omega_\text{in} : \cF \to \cF\,,
\end{equation}
and, as promised, we have shown that $S$ exists.

\begin{proposition}
    S is a contraction, i.e., $\| S |\psi \rangle \| \leq \| |\psi \rangle \|$ for any $|\psi \rangle$\,.
\end{proposition}

Of course, $S$ is normally supposed to be \emph{unitary}, meaning $\| S |\psi \rangle \| = \| S^\dagger |\psi \rangle \|= \| |\psi \rangle \|$. The reason we cannot prove this is because we have not shown that $\cH_\text{in} = \cH_\text{out}$ as sub-Hilbert spaces of $\cH$, which is known as \emph{weak asymptotic completeness}. We will have more to say about this below.

\begin{proof}
We claim that the operator $\cP_\text{out} \colonequals \Omega_\text{out} \Omega_\text{out}^\dagger$ is an orthogonal projector from $\cH$ onto $\cH_\text{out}$. Indeed, it acts as the identity operator on $\cH_\text{out}$ and from equation \eqref{Omega_adjoint} it follows that $\Omega^\dagger_\text{out}$ and therefore $\cP_\text{out}$ annihilates any states in the orthogonal complement $\cH_\text{out}^\perp$.\footnote{Similarly, $\cP_\text{in} \colonequals \Omega_\text{in} \Omega_\text{in}^\dagger$ is an orthogonal projector from $\cH$ onto $\cH_\text{in}$.}

For any state $| \psi \rangle \in \cF$ we now have,
\begin{equation}
    \begin{aligned}
\| S |\psi \rangle \|^2 &= \langle \psi | S^\dagger S |\psi \rangle \\
                      &= \langle \psi | \Omega_\text{in}^\dagger \Omega_\text{out} \Omega_\text{out}^\dagger \Omega_\text{in} |\psi \rangle \\
                     &= \langle \psi | \Omega_\text{in}^\dagger \cP_\text{out} \Omega_\text{in} |\psi \rangle \\
                    & \leq \| \Omega_\text{in} | \psi \rangle \|^2 \\
                    &= \| | \psi \rangle \|^2\,,
    \end{aligned}
\end{equation}
where we used that an orthogonal projector is a contraction and that $\Omega_\text{in}$ is an isometry. This of course implies that the $S$ operator is bounded
\begin{equation}
\| S \| = \sup_{|\psi\rangle \in \cF} \frac{ \| S | \psi \rangle \|}{\| |\psi \rangle \|} \leq 1.
\end{equation}
\end{proof}

Note that we defined $S$ as an operator on $\cF$, but we could just as well have defined it on $\cH$ by conjugating with one of $\Omega_\text{in/out}$ and its adjoint. In any case, the matrix elements of $S$ are indeed given by the familiar overlaps $\vev{\Psi_\OUT|\Phi_\IN}$, as follows from our prescription by taking two Fock space vectors $|\phi\>$ and $|\psi\>$ such that $|\Phi_\IN\> = \Omega_\text{in} |\phi\>$ and $\< \Psi_\OUT| = \<\psi| \Omega^\dagg_\text{out}$.

Next we consider, for a given $n$ and $m$, the \emph{$n$-to-$m$ S-matrix} which is obtained by restricting the domain of $S$ to the $n$-particle Fock space, and projecting its image to the $m$-particle Fock space:
\begin{equation}
    \label{ntomS}
    S_{n \to m} : \cF^{(n)} \to \cF^{(m)}
\end{equation}
Its matrix elements
\begin{equation}
    \< \psi_m | S | \psi_n \> 
\end{equation}
with $| \psi_n \rangle \in \cF^{(n)}$, and $| \psi_m \rangle \in \cF^{(m)}$, are the \emph{$n$-to-$m$ scattering amplitudes}.

\begin{proposition}\label{ntomistempered}
The $n$-to-$m$ S-matrix is entirely defined by an element of $\cS'(\R^{(d-1)(n+m)})$.
\end{proposition}

Note that it is not immediately clear that an inner product of the form $\vev{\Psi_\OUT|\Phi_\IN}$ is a tempered distribution, since the temperedness of $\vev{\Psi_{+t}|\Phi_{-t}}$ at all finite $t$ may not survive in the limit $t \to \infty$. This is why the proposition is non-trivial.

\begin{proof}
   
    By conjugating the $n$-to-$m$ S-matrix with $\iota$ we can define the maps
    \begin{equation}
        S_{n \to m} : \Sym^n L^2\left(\R^{d-1}, \frac{d_{d-1} \up}{2 E_\up}\right) \to \Sym^m L^2\left(\R^{d-1}, \frac{d_{d-1} \up}{2 E_\up}\right)
    \end{equation}
    These maps are in one-to-one correspondence with the original maps \eqref{ntomS} between Fock spaces, so we use the same symbol for both. Now we use the straightforward inclusions
    \begin{equation}
        \label{ScwartzInL2}
        \Sym^n \cS(\R^{d-1}) \subset\Sym^n L^2\left(\R^{d-1}, \frac{d_{d-1} \up}{2 E_\up}\right)
    \end{equation}
    and
    \begin{equation}
        \label{L2InScwartzPrime}
        \Sym^m L^2\left(\R^{d-1}, \frac{d_{d-1} \up}{2 E_\up}\right) \subset \Sym^m \cS'(\R^{d-1})
    \end{equation}
    along with the fact that $S_{n \to m}$ is a contraction, i.e.\ a bounded operator from $\Sym^n L^2\left(\R^{d-1}, \frac{d_{d-1} \up}{2 E_\up}\right)$ to $\Sym^m L^2\left(\R^{d-1}, \frac{d_{d-1} \up}{2 E_\up}\right)$ to deduce that $S_{n \to m}$ is \emph{also} a continuous map from $\Sym^n \cS(\R^{d-1})$ to $\Sym^m \cS'(\R^{d-1})$. This map again uniquely determines $S_{n \to m}$, because $\cS(\R^k)$ is dense in $L^2(\R^k)$ and the same is true for the corresponding symmetrized function spaces.
    
    Now we invoke the Schwartz kernel theorem \cite{reed1980functional}, which states that any continuous linear map from $\cS(\R^{n(d-1)})$ to $\cS'(\R^{m(d-1)})$ uniquely defines a tempered distribution on $(\R^{d-1})^{n+m}$. The same is true for the symmetrized spaces considered here, since a map defined on $\Sym^n \cS(\R^{d-1})$ can be extended to a map on $\cS(\R^{n(d-1)})$ by adding an appropriate symmetrization operator.
    \end{proof}
There are further restrictions on the tempered distribution defining $S_{n \to m}$. For one, it must of course respect the symmetry under permutations of the $n$ `in' particles and of the $m$ `out' particles. Second, it should be compatible with the Poincar\'e invariance of $S$, which says that
\begin{equation}
    \label{ScommutesWithU}
    U_g S U_g^{-1} = S\,.
\end{equation}
This invariance follows from the proof of claim (iii) of the main theorem, which shows that $\Omega_\text{in/out}$, i.e.\ the maps from $\cF$ to $\cH_\text{in/out}$, intertwine the action of the Poincar\'e group as defined axiomatically on $\cH$ and canonically on $\cF$. And last, we have to recall that we started with a bounded linear map between $L^2$ spaces, and certainly not every element of $\cS'(\R^{(d-1)(n+m)})$ corresponds to such a map.

We will now explore the consequences of these constraints for $S_{2 \to 2}$.

\subsection{The 2-to-2 S-matrix}
\label{subsec:S2to2}
Take two on-shell momenta $p_1$ and $p_2$ with positive energies and set $p_t = p_1 + p_2$. Note that $p_t^2 \geq 4m^2$ and $p_t^0 > 0$, which corresponds to a region in $\R^d$ we denote as $\bar V_{2m}^+$.

We now change coordinates from $p_1$, $p_2$ to $p_t$ and a unit vector $\nhat\in S^{d-2}$, as follows. Let $\Lambda(p_t)$ be the pure boost whose inverse brings us to the center of mass frame, meaning that $p_t = \Lambda(p_t) (\sqrt{p_t^2}, 0, 0, \ldots 0)$. Then we can write:
\begin{equation}
    p_1 = \Lambda(p_t) \begin{pmatrix}\sqrt{p_t^2}/2\\ k(p_t^2)\, \nhat \end{pmatrix} \,, \qquad p_2 = \Lambda(p_t) \begin{pmatrix}\sqrt{p_t^2}/2\\ - k(p_t^2)\, \nhat \end{pmatrix}\,,
\end{equation}
where $k(p_t^2) \colonequals \frac{1}{2} \sqrt{p_t^2 - 4 m^2} \geq 0$. Under this change of coordinates the measure changes as
\begin{equation}
    \label{twoparticlemeasure}
    \frac{d_{d-1} \up_1 d_{d-1} \up_2}{4 E_{\up_1} E_{\up_2}} = d\mu(p_t) \, d\nhat\,, \qquad d\mu(p_t) \colonequals \frac{1}{(2\pi)^{d-2}} \frac{k(p_t^2)^{d-3}}{4 \sqrt{p_t^2}} \theta(p_t^2 - 4m^2) \theta(p_t^0) d_d p_t % = \varrho(p_t) d^d p_t \,,
\end{equation}
with $d\nhat$ the measure for a round metric on $S^{d-2}$. The original $\Sym^2 L^2\left(\R^{d-1}, \frac{d_{d-1} \up}{2 E_\up}\right)$ splits into $L^2(\R^d, d\mu(p_t))$ times $L^2_\text{sym}(S^{d-2})$, with the symmetrization just corresponding to invariance under $\nhat \to - \nhat$. Next, we introduce the orthonormal hyperspherical harmonics $Y^\ell_m(\nhat)$ as a basis for $L^2(S^{d-2})$, with $m$ a discrete index labeling the harmonics of total spin $\ell$.  Since $Y^\ell_m(- \nhat) = (-1)^\ell Y^\ell_m(\nhat)$ the symmetrization demands that $\ell$ is even.\footnote{With appropriate modifications for spacetime dimensions $d \leq 3$ where there is no $m$ index. For $d=3$ we just have $\ell \in 2\Z$ and for $d=2$ we have $\ell = 0$ only.} After repeating the construction for the two outgoing particles we find that in the 2-to-2 sector the S-matrix operator can be written as a family of linear maps
\begin{equation}
    \label{Sllmma_as_map}S^{\ell\, \ell'}_{m\, m'} : L^2(\R^d, d\mu(p_t)) \to L^2(\R^d, d\mu(p_t))
\end{equation}
Note these must all be contractions, $\| S^{\ell\, \ell'}_{m\, m'} \| \le \| S_{2 \to 2} \| \le 1$.

\begin{proposition}
The map $S^{\ell\, \ell'}_{m\, m'}$ is determined uniquely by a tempered distribution 
\begin{equation}
    S^{\ell\, \ell'}_{m\, m'} (p_t, p_t') \in \cS'(\R^{2d})
\end{equation}
supported at $p_t, p_t' \in \bar V_{2m}^+$.
\end{proposition}

\begin{proof}
It is straightforward to verify that inclusions like equation \eqref{ScwartzInL2} and \eqref{L2InScwartzPrime} also work for $L^2(\R^d, d\mu(p_t))$, despite the fact that the step function in equation \eqref{twoparticlemeasure} is not a Schwartz multiplier. So we indeed obtain an element of $\cS'(\R^{2d})$ by a repetition of the proof of proposition \ref{ntomistempered}, and again it determines $S^{\ell\, \ell'}_{m\, m'}$ uniquely because $\cS(\R^d)$ is dense in $L^2(\R^d, d\mu(p_t))$. 

As for the support property, notice that any test function supported outside the given region is zero as an element of $L^2(\R^d, d\mu(p_t))$, and since $S^{\ell\, \ell'}_{m\, m'}$ is linear its image must also be zero. Therefore, $S^{\ell\, \ell'}_{m\, m'} (p_t, p_t')$ must vanish when $p_t' \notin \bar V_{2m}^+$. On the other hand, by the same logic the tempered distribution in $p_t$ defined as 
\begin{equation}
 \int d_d p_t'\, S^{\ell\, \ell'}_{m\, m'} (p_t, p_t') f(p_t') \in \cS'(\R^d)
\end{equation}
must be supported in $\bar V_{2m}^+$ for any Schwartz test function $f$. But this means simply that $S^{\ell\, \ell'}_{m\, m'} (p_t, p_t')$ must also vanish when $p_t \notin \bar V_{2m}^+$.
\end{proof}

Since the Poincar\'e transformations act within the $n$-particle subspace, $S_{2 \to 2}$ also obeys equation \eqref{ScommutesWithU}, so $U_g S_{2 \to 2} U_g^{-1} = S_{2 \to 2}$. For spacetime translations this leads immediately to
\begin{equation}
    e^{i (p_t - p_t') a} S^{\ell\, \ell'}_{m\, m'} (p_t, p_t') = S^{\ell\, \ell'}_{m\, m'} (p_t, p_t')
\end{equation}
for every $a \in \R^d$. This implies that
\begin{equation}\label{Fouriermultiplier}
    S^{\ell\, \ell'}_{m\, m'} (p_t, p_t') = \delta^{(d)} (p_t - p_t') S^{\ell\, \ell'}_{m\, m'} (p_t)\,
\end{equation}
with $S^{\ell\, \ell'}_{m\, m'} (p_t) \in \cS'(\R^d)$.
The kernels $S^{\ell\, \ell'}_{m\, m'} (p_t, p_t')$ are then called \emph{Fourier multipliers}, see for example \cite{Hormander1960, grafakos2014classical} for more details. 

Next we can consider boosts and rotations. It is a standard quantum field theory exercise\footnote{For $d = 3$, we drop the $m$ and $m'$ indices and let $\ell,\ell' \in 2 \Z$, whereas for $d=2$, $\ell = \ell' = 0$ anyway and the statement directly applies to the 2-to-2 scattering amplitude.} that this leads to the further restrictions that
\begin{equation}
    S^{\ell\, \ell'}_{m\, m'} (p_t) = \delta^{\ell\, \ell'} \delta_{m\, m'} S^\ell(p_t)
\end{equation}
and also that, for every Lorentz transformation $\Lambda$,
\begin{equation}
    \label{Linfty_boost}
    S^\ell(p_t) = S^\ell(\Lambda p_t)\,.
\end{equation}
so altogether we can pass to the center-of-mass variable $s = p_t^2$ and write, for every $\ell$,
\begin{equation}
    S^\ell(p_t) = \varsigma^\ell(p_t^2)
\end{equation}
with $\varsigma^\ell \in \cS'(\R)$ supported on $p_t^2 \geq 4m^2$.

For our last constraint we return to the viewpoint that $S^\ell(p_t)$ is also a contraction mapping between $L^2(\R^d, d\mu(p_t))$ as we wrote in equation \eqref{Sllmma_as_map}. Therefore, for every spin $\ell$ the object $\varsigma^\ell(p_t^2) f(p_t)$ is an element of $L^2(\R^d, d\mu(p_t))$ if $f(p_t)$ is, and furthermore we must have that
\begin{equation}\label{towardsLinfty}
    \int d\mu(p_t) \, |\varsigma^{\ell} (p_t^2) f(p_t)|^2 \leq \int d\mu(p_t)| f(p_t)|^2\,.
\end{equation}
This means that $\varsigma^\ell(p_t^2)$ must in fact be a measurable function, defined up to a set of measure zero, and with bounded essential supremum\footnote{This $L^\infty$ property of the partial wave amplitudes $\varsigma^\ell(s)$ was known in the 1960s since it was stated in \cite{Martin:1966zz}, but we are not sure where the above argument was presented for the first time.}
\begin{equation}
\label{partial_wave_unitarity}
\| \varsigma^\ell(s)\|_\infty \colonequals \operatorname*{ess\,sup}\ |\varsigma^\ell(s)| \leq 1\,.
\end{equation}
We recall that the $L^\infty$ norm or essential supremum of a measurable function $\sigma$ is defined by the smallest (or infimum) of those $C \in \R$ for which the preimage $|\sigma|^{-1}( \{ x\geq C \})$ has measure zero. The reason this property follows from equation \eqref{towardsLinfty} is by contradiction: if $|\varsigma^{\ell} (p_t^2)| > 1$ on a set of finite non-zero measure then we can use indicator functions on this set to find a violation of equation \eqref{towardsLinfty}.

Using the constraints of subunitarity and Lorentz invariance we have managed to show that the 2-to-2 S-matrix, initially merely defined as a contraction mapping between $L^2$ spaces, can be written entirely in terms of the functions $\varsigma^{\ell}(s)$ with $\ell \in 2\N$ and $s \geq 4m^2$, defined up to a set of measure zero and obeying the $L^\infty$ bound of equation \eqref{partial_wave_unitarity}.

This is not an exhaustive set of constraints, as not every such family can originate from a Wightman theory. Figuring out the space of consistent S-matrices is the purview of the S-matrix bootstrap program, which will be a story for another time.
\section{Discussion}
\label{sec:discussion}
The Haag-Ruelle theory demonstrates that every Wightman theory with a unique vacuum and an isolated single-particle state has multi-particle `in' and `out' states, and an associated S-matrix. On the one hand this is physically intuitive, but on the other it is quite remarkable that this result can be derived without assuming anything else about the states in $\cH$. In the following we will discuss several related topics and open questions.

\subsubsection*{Asymptotic completeness} Theorem \ref{them:H_in_is_Fock_space} shows that the Haag-Ruelle construction furnishes two Fock spaces, $\cH_\text{in}$ and $\cH_\text{out}$, built out of tensor products of 
the single particle state of mass $m$. Of course $\cH_\text{in} \subseteq \cH$ and $\cH_\text{out} \subseteq \cH$. Unitarity of the S-matrix requires that 
\begin{equation}
\cH_\text{in} = \cH_\text{out}\,,
\end{equation}
as sub-Hilbert spaces of $\cH$, a condition that is sometimes called weak asymptotic completeness.

\emph{Asymptotic completeness} is the stronger statement that
\begin{equation}
\cH_\text{in} = \cH_\text{out} = \cH\,.
\end{equation}
It has been proven in some models in $1+1d$ \cite{Lechner:2006kb}, but it does not follow from the Wightman axioms alone.\footnote{For example, take a two-point function defined by a K\"allen--Lehmann measure $\rho(\mu^2) = \delta(\mu^2 - m^2) + \sigma(\mu^2)$, with $\sigma$ continuous and supported on $[M^2,\infty)$, and construct the generalized free field by setting all truncated $n$-point functions to zero for $n \neq 2$. It satisfies all our axioms but is not asymptotically complete: the Haag-Ruelle states only span the Fock space of the mass $m$ particle.} Note that the next lightest state in a Fock space after the single-particle state is the two-particle state with both particles at rest, so total mass $2m$. Therefore, asymptotic completeness implies that $M = 2m$.

One can also consider the scenario with multiple isolated single-particle states with mass $m < m_i < 2 m$. In this case one solves the single-particle problem of proposition \ref{prop:singleparticlesubspace} for each particle separately and repeats the Haag-Ruelle construction with (mixed) products of the corresponding operators. The resulting incoming and outgoing Hilbert space is the tensor product of the Fock spaces:
\begin{equation}
\cH_\text{in/out}  = \bigotimes_i \mathcal F_i
\end{equation}
On physical grounds, we usually posit a finite number of single-particle states, which also avoids accumulation points that would arise in the case of an infinite number of single-particle states. In the latter scenario, we could always restrict to a finite subset of single-particle states to repeat the Haag-Ruelle construction. The resulting S-matrix would, however, only be a contraction and not unitary.

\subsubsection*{Cyclicity of the vacuum} As part of axiom \ref{axiom:fieldproperties}, we included the statement that the state $\phi(x)|\Omega\rangle$ has an overlap with
a one-particle state. Strictly speaking this is not necessary if one includes the cyclicity of the vacuum as part of the axioms. This is because cyclicity implies that some polynomial in the smeared fields $B \sim \phi(f_1) \phi(f_2) \ldots \phi(f_n)$ has non-zero overlap with the one-particle state. Defining $B(x) = e^{i P x} B e^{-i P x}$, one can re-run the entire Haag-Ruelle argument with this field $B(x)$. 

\subsubsection*{Local fields and cluster decomposition}
In a gapped theory, the local fields $\phi(x)$ are known to obey a strong form of cluster decomposition where the truncated correlation functions of operators compactly supported in position space decay \emph{exponentially} with spacelike separation \cite{Araki:1962zhd}. This is stronger than the Schwartz-type falloff we established in theorem \ref{thm:clusterdecomposition}.

In the Haag-Ruelle construction, one however needs compact support in momentum space, such that $\phi(f)$ only creates
single-particle states. This means that in position space these wave functions cannot be supported compactly, and for this reason we are in any case not expecting any faster falloff than the one established in theorem \ref{thm:clusterdecomposition}. Haag \cite{Haag:1958vt} already realized that this is the essential property for the Haag-Ruelle construction, and he motivated the rapid falloff by physical arguments. As we stated in the introduction, this form of cluster decomposition was proven definitively by Ruelle \cite{ruelle1962asymptotic}, although partial results had been obtained by others \cite{dellantonio_asymptotic_1959, ARAKI1960260,jostandhepp}.

An operator whose correlation functions obey cluster decomposition of the form proven in theorem \ref{thm:clusterdecomposition} for the Wightman field is sometimes called an ``almost-local" operator, and in fact the Haag-Ruelle construction works for any such operator. The polynomial field $B(x)$ introduced earlier is an example of an almost-local operator.

\subsubsection*{Temperedness}
Axiom \ref{axiom:fields} implies that the Wightman correlation functions are tempered distributions and therefore (roughly speaking) polynomially bounded in momentum space. In the past, however, more exotic scenarios have been considered.

Jaffe \cite{Jaffe:1967nb} considered the possibility of letting correlation functions grow super-polynomially in momentum space
\begin{equation}
\langle \phi(p) \ldots \rangle \sim e^{|p|^\alpha}
\end{equation}
He did so by considering a space of test functions which in momentum space have decay at large $|p|$ given by
\begin{equation}
f(p) \sim e^{-g(|p|)},
\end{equation}
stronger than usual decay of Schwartz functions in momentum space. This strong decay is dual to 
smoothness in position space. In particular, by the Paley-Wiener theorem, if $g(|p|) \gtrsim |p|$ 
then $f(x)$ is analytic and therefore cannot have compact support. 
The precise condition to have test functions with compact support in position space turns out to be \cite{Jaffe:1967nb}
\begin{equation}
    \label{eq:strict_localizability_condition}
\int_0^\infty \frac{g(|p|)}{1+|p|^2}\, d|p| < \infty \,.
\end{equation}
In this case, the fields are called ``strictly localizable fields" and one can impose locality in the form
\begin{equation}
[\phi(f_1),\phi(f_2)] = 0 
\end{equation}
if the compact supports of $f_1$ and $f_2$ in position space are space-like separated.

If the condition \eqref{eq:strict_localizability_condition} is violated, the fields are non-localizable since  
there are no test functions with compact supports and hence it is not possible to write a locality condition.
It was later pointed out that this case might be relevant for little string theory \cite{Kapustin:1999ci} and galileon field theory 
\cite{Keltner:2015xda}. In any case, Steinmann \cite{Steinmann:1970cm} formulated a scattering theory for the non-localizable case by assuming analyticity of correlation functions in momentum space.

\subsubsection*{Generalizations to spinning particles and global symmetries} 
In this review we only considered a neutral scalar particle for simplicity, but extending the 
Haag-Ruelle construction to particles with spin and/or charged under a global symmetry is largely straightforward.
One replaces the scalar field by a multi-component (and, for charged particles, possibly complex) 
Wightman field $\phi_\alpha (x)$ transforming in a representation of the global symmetry group and of the covering group of the Lorentz group.
The single-particle problem is then solved by the same restriction as in \eqref{D1support} on the momentum-space support of the test functions $f_\alpha(p)$, which now also carry an index. Equation \eqref{eq:one_particle_phi_vac} is modified by including a polarization tensor or spinor $u_\alpha (p, \sigma)$ where $\sigma$ is a 
little group representation index. For fermionic fields, the commutator in the statement of locality in axiom \ref{axiom:fieldproperties} becomes an anti-commutator and truncated Wightman functions 
are defined with permutation signs. The proof of cluster decomposition only uses locality and the mass gap, and the wave packets are built from the same scalar solutions of the Klein-Gordon equation for
 each component of the Wightman field, so the estimates of sections \ref{sec:truncatedWigtman} and \ref{sec:solutionskg} carry over essentially unchanged. One thus obtains asymptotic bosonic or fermionic Fock spaces carrying the 
 appropriate spin and symmetry representations (with particles and antiparticles as distinct species), and an S-matrix that commutes with the global symmetry.

 Note that when the operator is charged (or fermionic), it is 
technically no longer part of the algebra of observables. Of course, in the Wightman framework, one is allowed to use 
such `unobservable' fields to carry out the Haag-Ruelle construction and arrive at the S-matrix. However,
as a matter of principle, it is also possible to use only observable operators (and their correlation functions) to compute the cross section (rather than the S-matrix) \cite{araki_collision_1967}.

\subsubsection*{Anyons in $d=3$}
In $d = 3$ there exist massive particles called anyons that fall outside our axioms.
Their spin can take any real value, and their statistics is governed by a representation of the braid group rather than the permutation group \cite{Frohlich:1990zg,Mund:2008hn}.
They cannot be created from the vacuum by a local field as in axiom \ref{axiom:fieldproperties},
 because the topological charges they carry cannot be localized in bounded regions, only in spacelike cones \cite{Buchholz:1981fj}. 
Nevertheless, a Haag–Ruelle scattering theory for such particles was formulated in \cite{Frohlich:1990zg,Fredenhagen:1994dk}.
The space of scattering states is then no longer the Fock space $\mathcal{F}$ of section \ref{sec:conclusion}: instead, the $n$-particle states are (square-integrable) sections of a flat vector bundle over the space of non-coinciding velocities.
 The restriction to disjoint velocity supports, which for us was a technical assumption in $d<4$, thereby becomes essential.
 Chern–Simons theories coupled to massive matter provide the natural field-theoretic realizations of anyons and their 2-to-2 S-matrices indeed display unusual features, including a term localized on forward scattering and modified crossing rules \cite{Jain:2014nza,Inbasekar:2015tsa,Gabai:2022snc,Mehta:2022lgq}.
  It would be interesting to understand these features from an axiomatic starting point.

\subsubsection*{Massless particles} In the construction presented here, the double mass gap postulated in axiom \ref{axiom:spectrum} was instrumental. This is further highlighted by the challenges that arise in the massless case: to begin with, the provided solution to the single-particle problem now fails since any smooth bump function around $p^2 = 0$ would also pick up the multi-particle states (which should exist if any version of the Haag-Ruelle theory can be extended to the massless case). Furthermore, cluster decomposition does not yield Schwartz-class functions, and solutions to the massless Klein-Gordon equation also do not fall off as rapidly as before.

Nevertheless, a scattering theory for massless particles in $3+1d$ \cite{Buchholz:1974je,Buchholz:1976rs} was constructed using locality and Huygens' principle, instead of the estimates of theorem \ref{thm:clusterdecomposition} and of proposition \ref{prop:KGEfastdecayfinal} that were reliant on a mass gap. These results are probably straightforwardly generalizable to other even spacetime dimensions, but we are not aware of a proof that would work in odd dimensions. 

Other constructions available in the literature cover both the case of massive particles in the presence of massless ones \cite{dybalski2005haag} and that of massless bosons in $3+1$ dimensions \cite{duch2015massless}, at the price of additional regularity assumptions.

\subsubsection*{Gauge theories and infraparticles} In gauge theories with a mass gap, particles that are created from the vacuum by local gauge-invariant operators, such as the hadrons of QCD, fit into the framework of this review, with the generalizations to several particle species and to particles with spin.
The situation is different in gauge theories with massless gauge bosons, such as four-dimensional QED. By Gauss's law, the electric charge can be measured via the electric flux through an arbitrarily large sphere, so a charged state cannot be created from the vacuum by local fields as in axiom~\ref{axiom:fieldproperties}. 
The charged particle is inseparably accompanied by a cloud of soft photons, and the mass shell $p^2 = m^2$  is only the lower edge of a continuum instead of a single-particle representation as in axiom~\ref{axiom:spectrum}. Such particles are called infraparticles \cite{Schroer:1963gw}. Any smooth bump function around $p^2 = m^2$ also picks up these arbitrarily soft photons, and therefore  a Haag-Ruelle type construction is precluded. Proposals for a scattering theory either replace the Fock states by dressed asymptotic states, as suggested by Kulish and Faddeev \cite{Kulish:1970ut}, or abandon the S-matrix in favor of cross sections.

\subsubsection*{Generalizations to inclusive observables} As we discussed in section \ref{sec:spacetimepicture}, in the Haag-Ruelle construction the operator ordering is always correlated with the $t$-ordering, in the sense that one  never considers, for example, states of the form
\begin{equation}
    a^{(\dagg)}_\OUT(f_1) a^{(\dagg)}_\IN(f_2) a^{(\dagg)}_\OUT(f_3)|\Omega\>\,.
\end{equation}
The authors of \cite{Caron-Huot:2023vxl}, see also \cite{Caron-Huot:2025ymc}, however proposed that such states exist at least in the weak sense. The corresponding out-of-time-order asymptotic observables were then called `inclusive amplitudes'. It would be very interesting to see if these can also be rigorously defined, especially because they appear to be naturally related to standard scattering amplitudes by analytic continuation \cite{Caron-Huot:2023ikn}.

\subsubsection*{Scattering amplitudes from Euclidean correlators} The Haag-Ruelle theorem expresses scattering amplitudes in terms of Wightman functions smeared with Schwartz functions, without time ordering or analytic continuation in the momenta. Stripping the energy-momentum conserving delta function off a truncated Wightman function in momentum space yields a spectral density, and Euclidean correlators are Laplace transforms of the same spectral densities with respect to the energy variables.
This makes Haag-Ruelle theory well suited to situations where only Euclidean correlation functions are available. One example of this is Euclidean lattice field theory, where Barata and Fredenhagen \cite{Barata:1990rn} developed a Haag-Ruelle theory and showed that the resulting amplitudes can be approximated arbitrarily well by linear combinations of Euclidean correlators. More recently, Patella and Tantalo \cite{Patella:2024cto} derived approximation formulae of this type directly in the continuum and in infinite volume, in terms of spatially smeared Euclidean correlators evaluated at discrete time separations. This could offer a route to hadronic amplitudes  in lattice QCD that is complementary to Lüscher’s finite-volume method \cite{Luscher:1986pf,Luscher:1990ux,Luscher:1990ck}. One contribution to their error estimate is the cost of creating the asymptotic states at a finite time $t$. It is governed by the rate of convergence in theorem~\ref{thm:haagruelle} and therefore vanishes faster than any power of $|t|$ for non-overlapping velocities.

\subsubsection*{Towards the analytic S-matrix}
The Haag-Ruelle theory gives a manifest way to compute S-matrix elements from large $|t|$ limits of Wightman functions. The much more common prescription is however that of LSZ, which we will review rigorously in our upcoming work \cite{LSZPAPER}. This is however not sufficient to answer the question we posed in the introduction: does the 2-to-2 amplitude obey any of the bounds obtained using the non-perturbative S-matrix bootstrap like the Froissart-Martin bound? These results rely crucially on the \emph{analyticity} of scattering amplitudes.

To prove these analyticity results one starts from certain non-time-ordered reduction formulae. The idea is always to obtain the scattering amplitude by LSZ reducing correlation functions which, by virtue of causality, have some manifest analyticity properties in momentum space. By then using various analytic extension theorems like the edge-of-the-wedge theorem, one can squeeze out a domain of analyticity for the on-shell scattering amplitudes.

Analyticity (and polynomial boundedness at infinity) underlie all of the S-matrix bootstrap results, both old and new. Unfortunately, as mentioned already in \cite{Kruczenski:2022lot}, a real gap remains between the rigorously proven analyticity properties of scattering amplitudes, reviewed for example in \cite{Sommer:1970mr,bogolubov2012general}, and the expected analyticity domain obtained from e.g.~an analysis of Landau diagrams. It would be great to bridge this gap in the future.
\section*{Acknowledgments}
We are grateful for discussions with Simon Caron-Huot, Alexandre Fouquet, Filip Her\v{c}ek, Slava Rychkov and Sasha Zhiboedov. We also benefited from discussions with the participants of the ``S-matrix marathon'' workshop at the IAS in 2024, and thank the organizers for creating a stimulating environment. The authors would like to thank ICTP-SAIFR (FAPESP grant 2021/14335-0) where part of this work was done. AH and BvR also thank the Yukawa Institute for Theoretical Physics at Kyoto University, where part of this work was written during ``Progress of Theoretical Bootstrap''. The authors were also supported by Simons Foundation grant \#488659 for the Simons Collaboration on the non-perturbative bootstrap. We acknowledge funding from the European Union (ERC “QFTinAdS”, project number 101087025).
AH was also supported by “Exotic High Energy Phenomenology” (X-HEP), a project funded by the European Union Grant Agreement n. 101039756 (PI: J. Elias Miró).
AI LLMs were used to finetune some of the arguments in this paper.

\appendix

\section{Proof of equation \texorpdfstring{\eqref{truncatedwightmandefinitionexplicit}}{}}
\label{appendix:proof_of_truncated_Wightman_definition}
\begin{proof}
    We will use strong induction. Equations \eqref{truncatedwightmandefinition} and \eqref{truncatedwightmandefinitionexplicit} are clearly equivalent for the starting point $n=1$. We will now prove that the result holds for $n+1$ if it holds for $\{1, \dots, n\}$.

    For the remainder of this proof, we will ease the notations by denoting $\cW(\dots)$ by $\<\dots\>$ and $\cW^T(\dots)$ by $[\dots]$.
    \begin{align*}
        \< x_1 \dots x_{n+1} \> &= \sum\limits_{\Pi \in \doublepinp} [\pi_1] \dots [\pi_{|\Pi|}] \\
        &= \sum\limits_{\substack{\Pi \in \doublepinp \\ |\Pi| \ne 1}} [\pi_1] \dots [\pi_{|\Pi|}] + [x_1 \dots x_{n+1}] \\
        &= \sum\limits_{\substack{\Pi \in \doublepinp \\ |\Pi| \ne 1}} \prod\limits_{i=1}^{|\Pi|} \left( \sum\limits_{\Gamma_i \in \text{\doublepi}\:_{|\pi_i|}} (-1)^{|\Gamma_i|-1} (|\Gamma_i|-1)! \<\gamma_{i,1}\> \dots \<\gamma_{i,|\Gamma_i|}\> \right) + [x_1 \dots x_{n+1}] \\
        &= \sum\limits_{\substack{\Pi \in \doublepinp \\ |\Pi| \ne 1}} \sum\limits_{\Gamma_1 \in \text{\doublepi}\:_{|\pi_1|}} \dots \sum\limits_{\Gamma_{|\Pi|} \in \text{\doublepi}\:_{|\pi_{|\Pi|}|}} (-1)^{\sum\limits_{i=1}^{|\Pi|}|\Gamma_i|-|\Pi|} \left( \prod\limits_{i=1}^{|\Pi|} (|\Gamma_i|-1)! \right) \< \gamma_{1,1} \> \dots \< \gamma_{|\Pi|,|\Gamma_{|\Pi|}|} \> \\
        &\qquad+ [x_1 \dots x_{n+1}] 
    \end{align*}
    For any $\tilde{\Pi} \in \doublepinp$ with $|\tilde{\Pi}|\ne 1$, the $\< \gamma_{1,1} \> \dots \< \gamma_{|\Pi|,|\Gamma_{|\Pi|}|} \>$ that correspond to $\< \tilde{\pi}_1 \> \dots \< \tilde{\pi}_{|\tilde{\Pi}|}\>$ are exactly the ones such that $\sqcup_i \Gamma_i = \tilde{\Pi}$ (that is, the $\Gamma_i$ that refine $\Pi$ in a way that gives $\tilde{\Pi}$). Hence,
    \begin{align*}
        \< x_1 \dots x_{n+1} \> &= \sum\limits_{\substack{\tilde{\Pi} \in \doublepinp \\ |\tilde{\Pi}| \ne 1}} \left( \sum\limits_{\substack{\Pi,\Gamma_1,\dots,\Gamma_{|\Pi|} \\ \sqcup_i \Gamma_i = \tilde{\Pi}}} (-1)^{\sum\limits_{i=1}^{|\Pi|}|\Gamma_i|-|\Pi|} \left( \prod\limits_{i=1}^{|\Pi|} (|\Gamma_i|-1)! \right) \right) \< \tilde{\pi}_1 \> \dots \< \tilde{\pi}_{|\tilde{\Pi}|}\> +[x_1 \dots x_{n+1}] \\
        &= \sum\limits_{\substack{\tilde{\Pi} \in \doublepinp \\ |\tilde{\Pi}| \ne 1}} (-1)^{|\tilde{\Pi}|} \left( \sum\limits_{\substack{\Pi,\Gamma_1,\dots,\Gamma_{|\Pi|} \\ \sqcup_i \Gamma_i = \tilde{\Pi}}} (-1)^{|\Pi|} \left( \prod\limits_{i=1}^{|\Pi|} (|\Gamma_i|-1)! \right) \right) \< \tilde{\pi}_1 \> \dots \< \tilde{\pi}_{|\tilde{\Pi}|}\> +[x_1 \dots x_{n+1}] \\
        &= \sum\limits_{\substack{\tilde{\Pi} \in \doublepinp \\ |\tilde{\Pi}| \ne 1}} (-1)^{|\tilde{\Pi}|} (|\tilde{\Pi}|-1)! \< \tilde{\pi}_1 \> \dots \< \tilde{\pi}_{|\tilde{\Pi}|}\> +[x_1 \dots x_{n+1}]
    \end{align*}
    where the last equality was obtained by using the standard combinatorics result
    \begin{equation*}
        \forall N \in \mathbb{N}^* \setminus \{1\}, \: \sum\limits_{\Pi'\in\doublepinm} (-1)^{|\Pi'|} \prod\limits_{i=1}^{|\Pi'|} (|\pi'_i|-1)! = 0\,.
    \end{equation*}
    Finally, 
    \begin{align*}
        [x_1 \dots x_{n+1}] &= \< x_1 \dots x_{n+1} \> - \sum\limits_{\substack{\tilde{\Pi} \in \doublepinp \\ |\tilde{\Pi}| \ne 1}} (-1)^{|\tilde{\Pi}|} (|\tilde{\Pi}|-1)! \< \tilde{\pi}_1 \> \dots \< \tilde{\pi}_{|\tilde{\Pi}|}\> \\
        &= \sum\limits_{\tilde{\Pi}\in\doublepinp} (-1)^{|\tilde{\Pi}|-1} (|\tilde{\Pi}|-1)! \< \tilde{\pi}_1 \> \dots \< \tilde{\pi}_{|\tilde{\Pi}|}\> 
    \end{align*}
\end{proof}

\section{The Schwartz representation theorem}
\label{app:SchwartzRepThm}
\renewcommand{\ux}{x}
\renewcommand{\up}{p}
\renewcommand{\uD}{D}
In this appendix, we state and prove the Schwartz representation theorem. Note that everything occurs in $\R^n$, endowed with a Euclidean scalar product.
 
\subsection{Sobolev spaces}

\subsubsection{Distributional setting}

To begin with, let us start by recalling some notations. As we work with distributions on $\R^n$, we consider given a \textit{multi-index} $\alpha = (\alpha_1,\dots,\alpha_n) \in \mathbb{N}^n$
\begin{itemize}[noitemsep,topsep=0pt]
    \item The \textit{length} of $\alpha$, $|\alpha| = \alpha_1 + \dots + \alpha_n$
    \item For $\ux \in \mathbb{R}^n$, $\ux^\alpha = x_1^{\alpha_1} \dots x_n^{\alpha_n}$
    \item The differential operator $\partial^\alpha = \frac{\partial^{|\alpha|}}{\partial x_1^{\alpha_1}\dots\partial x_n^{\alpha_n}}$
\end{itemize}
From there, we call \textbf{Schwartz semi-norms} the quantities defined for $f \in \mathcal{C}^\infty (\mathbb{R}^n)$  and $(\alpha,\beta) \in (\N^n)^2$ by :
\begin{equation}
    \label{Schwartzseminorms}
    \| f \|_{\alpha,\beta} = \sup\limits_{x \in \mathbb{R}^n} |x^\alpha \partial^\beta f (x)|
\end{equation}
and for $k \in \N$, we'll also consider the
\begin{equation}
    \label{Sknorms}
    \| f \|^\cS_{k} = \max\limits_{|\alpha|+|\beta| \le k} \| f \|_{\alpha,\beta}\,.
\end{equation}
The \textbf{Schwartz space} (or \textit{space of rapidly decreasing test functions}) is then
\begin{equation*}
    \mathcal{S}(\mathbb{R}^n) = \{ f \in \mathcal{C}^\infty (\mathbb{R}^n) \: / \: \forall (\alpha,\beta) \in (\N^n)^2, \: \| f \|_{\alpha,\beta} < \infty \}
\end{equation*}
endowed with the semi-norms $\| \cdot \|_{\alpha,\beta}$, while a \textbf{tempered distribution} is a \textit{continuous linear functional} on the space of Schwartz functions; that is, the space of tempered distributions is $\mathcal{S}'(\mathbb{R}^n)$, the \textit{dual space} to that of Schwartz functions $\mathcal{S}(\mathbb{R}^n)$.

\subsubsection{Introducing Sobolev spaces}
Let us introduce a few more notations. In this appendix, we will denote:
\begin{itemize}[noitemsep,topsep=0pt]
    \item For $\ux \in \R^n$, $\thpbracket{\ux} = \sqrt{1+\ux^2}$ and its Fourier-space analogue $\thpbracket{\up}$. This is what in the mathematics literature is usually denoted as $\langle x \rangle$; we opted against this notation, for it is not an expectation value in the physical sense of the term.
    \item For $s \in \R$, $\thpbracket{\ux}^s \centerdot$ the map $\varpi(\ux) \mapsto \thpbracket{\ux}^s\varpi(\ux)$, that one may define in ``position space'' for $\varpi$ being a Schwartz function or a tempered distribution.
    \item $\thpbracket{\up}^s \centerdot$, the ``momentum space'' analogue of $\thpbracket{\ux}^s \centerdot$.
    \item $\thpbracket{\uD}^s \centerdot = \cF^{-1} \circ \thpbracket{\up}^s \centerdot \circ \cF$ (with $\cF$ the Fourier transform) the fractional derivative of order $s$, defined on both $\mathcal{S}(\mathbb{R}^n)$ and $\mathcal{S}'(\mathbb{R}^n)$.
\end{itemize}
\textbf{(Distributional) Sobolev spaces} are then the sets
\begin{equation*}
    H^s(\R^n) = \{ u \in \cS'(\R^n) \: / \: \thpbracket{D}^s u \in L^2(\R^n) \}
\end{equation*}
where $s \in \R$ is the order of the space, and it is understood that $\thpbracket{D}^s u \in L^2(\R^n) $ actually means
\begin{equation*}
    \exists f \in L^2(\R^n), \: \thpbracket{D}^s u = T_f
\end{equation*}
where $T_f \in \cS'(\R^n)$ is the tempered distribution defined as integration against $f$. We will keep using this notation abuse in the remainder of this appendix.

Topologically speaking, Sobolev spaces become Hilbert spaces when endowed with the norm:
\begin{equation*}
    \| u \|_{H^s(\R^n)} = \| \thpbracket{D}^su \|_{L^2(\R^n)} = (2\pi)^{-n/2} \| \thpbracket{\up}^s \cF u \|_{L^2(\R^n)}
\end{equation*}
It can be noted already that Sobolev spaces are \textit{ordered for $\subset$}:
\begin{equation*}
    \forall (s,s') \in \R^2, \: s' > s \Rightarrow H^{s'}(\R^n) \subset H^s(\R^n)
\end{equation*}

\begin{proof}
    For $u \in H^{s'}(\R^n)$, $\| \thpbracket{\up}^{s'} \cF u \|_{L^2(\R^n)} < \infty$; and since $\thpbracket{\up}^{s} \cF u = \thpbracket{\up}^{s-s'} \thpbracket{\up}^{s'} \cF u$ with $\thpbracket{\up}^{s-s'} \in ]0;1]$, $\| \thpbracket{\up}^{s} \cF u \|_{L^2(\R^n)} < \infty$.
\end{proof}

\subsection{A few technical lemmas}

\subsubsection{On tempered distributions and Schwartz functions}

\begin{proposition}[A multi-index Leibniz rule]
    Let $(f,g) \in \cS(\R^n)^2$, and $\alpha \in \N^n$.
    \begin{equation}
        \label{eq:multindexleibniz}
        \partial^\alpha (fg) = \sum\limits_{\beta \le \alpha} \begin{pmatrix}
            \alpha \\ \beta
        \end{pmatrix} \partial^{\beta} f \partial^{\alpha - \beta} g
    \end{equation}
    where $\beta \le \alpha$ means $\forall i \in \{1, \dots, n\}, \: \beta_i \le \alpha_i$ and $\begin{pmatrix} \alpha \\ \beta \end{pmatrix} = \prod\limits_{i=1}^n \begin{pmatrix} \alpha_i \\ \beta_i \end{pmatrix}$.
\end{proposition}

\begin{proof}
    We show how the proof works for $n=2$; it is analogous for general $n$.
    \begin{align*}
        \partial^\alpha (fg) &= \partial_1^{\alpha_1} \partial_2^{\alpha_2}(fg) \\
        &= \partial_1^{\alpha_1} \left[ \sum\limits_{\beta_2 \le \alpha_2} \begin{pmatrix} \alpha_2 \\ \beta_2 \end{pmatrix} \partial^{\beta_2} f \partial^{\alpha_2 - \beta_2} g \right] \\
        &= \sum\limits_{\beta_2 \le \alpha_2} \begin{pmatrix} \alpha_2 \\ \beta_2 \end{pmatrix} \sum\limits_{\beta_1 \le \alpha_1} \begin{pmatrix} \alpha_1 \\ \beta_1 \end{pmatrix} \partial^{\beta_2} \partial^{\beta_1} f \partial^{\alpha_2 - \beta_2} \partial^{\alpha_1 - \beta_1} g \\
        &= \sum\limits_{\beta \le \alpha} \begin{pmatrix} \alpha \\ \beta \end{pmatrix} \partial^{\beta} f \partial^{\alpha - \beta} g\,.
    \end{align*}
\end{proof}

\begin{proposition}[A consequence of the continuity of tempered distributions]
    \label{prop:estimateuf}
    \begin{equation*}
        \forall u \in \cS'(\R^n), \exists \tilde{C}_u \in \R_+^*, \exists k \in \N, \forall f \in \cS(\R^n), \: |u(f)| \le \tilde{C}_u \| f \|^\cS_k
    \end{equation*}
\end{proposition}

\begin{proof}
    Assume for the sake of contradiction the opposite:
    \begin{equation*}
        \exists u \in \cS'(\R^n), \forall C >0, \forall k \in \N, \exists f_{k,C} \in \cS(\R^n), \: |u(f_{k,C})| > C \| f_{k,C} \|^\cS_k
    \end{equation*}
    In particular,
    \begin{equation*}
        \forall k \in \N^*, \exists f_k \in \cS(\R^n), |u(f_k)| > k \| f_k \|^\cS_k\,.
    \end{equation*}
    Note that $\| f_k \|^\cS_k \ne 0$, since otherwise $\| f_k \|_\infty \le \| f_k \|_k^\cS = 0$ imposes $f_k \equiv 0$, which in turn yields $u(f_k) = 0$, contradicting the strict inequality $|u(f_k)| > k \| f_k \|^\cS_k$. Thus, we can define
    \begin{equation}
        (\tilde{f})_{k \in \N^*} \colonequals \left( \frac{f_k}{k \| f_k \|_k^\cS} \right)_{k \in \N^*}\,,  
    \end{equation}
    which obeys
    \begin{equation*}
        \forall k \in \N^*, \: |u(\tilde{f}_k)| > 1.
    \end{equation*}
    Yet, $(\tilde{f})_{k \in \N^*}$ goes to $0$ in $\cS(\R^n)$:
    \begin{equation*}
        \forall (\alpha, \beta) \in (\N^n)^2, \forall k \ge |\alpha|+|\beta|, \: \| \tilde{f}_k \|_{\alpha, \beta} \le \| \tilde{f}_k \|^\cS_{|\alpha|+|\beta|} \le \| \tilde{f}_k \|^\cS_k \underset{k \rightarrow \infty}{\longrightarrow} 0
    \end{equation*}
    Which contradicts the continuity of $u$ over $\cS(\R^n)$.
\end{proof}

\subsubsection{On $\thpbracket{\cdot}$}

\begin{proposition}
    One can write
    \begin{equation}
        \label{eq:partialbrakpoly}
        \forall q \in \R, \forall \gamma \in \N^n, \exists P_\gamma \in \R_{|\gamma|}[X_1,\dots,X_n], \: \partial^\gamma \thpbracket{\ux}^q = \thpbracket{\ux}^{q - 2|\gamma|} P_\gamma(\ux)
    \end{equation}
    which yields the estimate
    \begin{equation}
        \label{eq:partialbrakestimate}
        \forall q \in \R, \forall \gamma \in \N^n, \exists C_\gamma \in \R_+^*, \forall \ux \in \R^n, \: |\partial^\gamma \thpbracket{\ux}^q| \le C_\gamma \thpbracket{\ux}^{q - |\gamma|}
    \end{equation}
\end{proposition}

\begin{proof}
    We start by proving the existence of the polynomial by induction on $|\gamma|$ at fixed $q$.
    \begin{itemize}[noitemsep,topsep=0pt]
        \item Basis case: for $|\gamma|=0$, $\gamma = (0,\dots,0)$ and $P_0(\ux) = 1$ is a suitable choice.
        \item Induction case: assume there is some $N \in \N$ such that the result holds for all $\gamma \in \N^n$ with $|\gamma| = N$, and now consider a $\gamma \in \N^n$ such that $|\gamma| = N+1$. Then write $\gamma = \gamma_0 + (0, \dots, 0, 1, 0, \dots, 0)$ with $|\gamma_0| = N$ and the $1$ of $(0, \dots, 0, 1, 0, \dots, 0)$ in position $i$.
        \begin{align*}
            \partial^\gamma \thpbracket{\ux}^q &= \partial_i \partial^{\gamma_0} \thpbracket{\ux}^q \\
            &= \partial_i \left[ \thpbracket{x}^{q-2|\gamma_0|} P_{\gamma_0} (\ux) \right] \\
            &= (q - 2|\gamma_0|) \left(\partial_i \thpbracket{\ux}\right) \thpbracket{\ux}^{q - 2 |\gamma_0| -1} P_{\gamma_0}(\ux) + \thpbracket{x}^{q - 2 |\gamma_0|} \partial_i P_{\gamma_0}(\ux) \\
            &= \thpbracket{\ux}^{q-2|\gamma|} \underbrace{\left[ (q - 2|\gamma_0|) x_i P_{\gamma_0}(\ux) + \thpbracket{x}^2 \partial_i P_{\gamma_0}(\ux) \right]}_{P_\gamma(\ux)}
        \end{align*}
        which concludes the induction.
    \end{itemize}

    \noindent The estimates follow from the triangle inequality and $\thpbracket{\ux} \ge 1$.
\end{proof}

\begin{proposition}
    For $s \in \R$, $\thpbracket{x}^s\centerdot$ is a topological vector space isomorphism of $\cS(\R^n)$.
\end{proposition}

\begin{proof}
    \quad
    \begin{itemize}[noitemsep,topsep=0pt]
        \item $\thpbracket{\ux}^s\centerdot$, when defined on $\cS(\R^n)$, is valued in $\cS(\R^n)$. To see this, let $f \in \cS(\R^n)$, and consider some $(\alpha, \beta) \in (\N^n)^2$. For $\ux \in \R^n$, 
        \begin{align*}
            | \ux^\alpha \partial^\beta (\thpbracket{\ux}^s f)| &= \left| \ux^\alpha \sum\limits_{\gamma \le \beta} \begin{pmatrix} \beta \\ \gamma \end{pmatrix} \partial^\gamma \thpbracket{\ux}^s \partial^{\beta - \gamma} f \right| \\
            &\le \sum\limits_{\gamma \le \beta} \begin{pmatrix} \beta \\ \gamma \end{pmatrix} C_\gamma \left| \ux^\alpha \thpbracket{\ux}^{s - |\gamma|} \partial^{\beta - \gamma} f \right| \\
            &\le \sum\limits_{\gamma \le \beta} \begin{pmatrix} \beta \\ \gamma \end{pmatrix} C_\gamma \left| \ux^\alpha \thpbracket{\ux}^{2 | \lceil s \rceil |} \partial^{\beta - \gamma} f \right|
        \end{align*}
        Upon expanding the (multivariate) polynomial $\thpbracket{\ux}^{2|\lceil s \rceil|}$, one gets a finite linear combination of terms bounded by Schwartz seminorms of $f$. Since the latter are finite, it follows that so does $\| \thpbracket{\ux}^s f \|_{\alpha, \beta}$, which in turn shows that indeed $\thpbracket{\ux}^s f \in \cS(\R^n)$.
        \item $\thpbracket{\ux}^s\centerdot$ is linear by construction.
        \item $\thpbracket{\ux}^s\centerdot$ is bijective, since $\thpbracket{\ux}^{-s}\centerdot$ provides an inverse to it.
        \item $\thpbracket{\ux}^s\centerdot$ is continuous:
        
        \noindent Let us argue by sequential characterization of continuity in Fréchet spaces (the Schwartz space being such a space). Consider $(f_k) \in (\cS(\R^n)^\N)$ a sequence of Schwartz functions that converges to some $f \in \cS(\R^n)$ for the Schwartz seminorms, that is, 
        \begin{equation*}
            \forall (\alpha, \beta) \in (\N^n)^2, \: \| f - f_k \|_{\alpha, \beta} \underset{k \rightarrow \infty}{\longrightarrow} 0
        \end{equation*}
        Using the inequality shown in the first point, it appears that:
        \begin{equation*}
            | \ux^\alpha \partial^\beta (\thpbracket{\ux}^s (f - f_k))| \le \sum\limits_{\gamma \le \beta} \begin{pmatrix} \beta \\ \gamma \end{pmatrix} C_\gamma \left| \ux^\alpha \thpbracket{\ux}^{2 | \lceil s \rceil |} \partial^{\beta - \gamma} (f - f_k) \right|
        \end{equation*}
        and again, upon expanding the polynomial $\thpbracket{\ux}^{2|\lceil s \rceil|}$ on the rhs and taking the sup, it appears that $\| (\thpbracket{\ux}^s (f - f_k)) \|_{\alpha, \beta}$ is upper bounded by a finite linear combination of Schwartz seminorms of $f - f_k$, which all go to zero; thus, $\| (\thpbracket{\ux}^s (f - f_k)) \|_{\alpha, \beta} \underset{k \rightarrow \infty}{\longrightarrow} 0$.
        \item The argument of the previous point applies to show that the inverse mapping $\thpbracket{\ux}^{-s}\centerdot$ also is continuous.
    \end{itemize}
\end{proof}

\begin{proposition}
    For $s \in \R$, $\thpbracket{x}^s\centerdot$ is a topological vector space isomorphism of $\cS'(\R^n)$.
\end{proposition}

\begin{proof}
    \quad
    \begin{itemize}[noitemsep,topsep=0pt]
        \item $\thpbracket{\ux}^s\centerdot$, when defined on $\cS'(\R^n)$, is valued in $\cS'(\R^n)$ since
        \begin{equation*}
            \forall u \in \cS'(\R^n), \forall f \in \cS(\R^n), \: (\thpbracket{\ux}^s u)(f) = u(\thpbracket{\ux}^s f)
        \end{equation*} 
        and $\thpbracket{x}^s\centerdot$ is valued in $\cS(\R^n)$ when defined on $\cS(\R^n)$.
        \item $\thpbracket{\ux}^s\centerdot$ is linear by construction.
        \item $\thpbracket{\ux}^s\centerdot$ is bijective, since $\thpbracket{\ux}^{-s}\centerdot$ provides an inverse to it.
        \item $\thpbracket{\ux}^s\centerdot$ is continuous:
        
        \noindent Since the weak topology that $\cS'(\R^n)$ is endowed with is not metrizable, we characterize continuity with nets rather than sequences. Let $(u_k)$ be a net of tempered distributions that converges to some $u \in \cS'(\R^n)$ for the weak topology:
        \begin{equation*}
            \forall f \in \cS(\R^n), \: |u(f) - u_k(f)| \underset{k \rightarrow \infty}{\longrightarrow} 0
        \end{equation*}
        One has:
        \begin{align*}
            \forall f \in \cS(\R^n), \: | (\thpbracket{\ux}^s u)(f) -(\thpbracket{\ux}^s u_k)(f) | &= | u(\thpbracket{\ux}^s f) - u_k(\thpbracket{\ux}^s f) | 
            \underset{k \rightarrow \infty}{\longrightarrow} 0
        \end{align*}
        Which proves the desired continuity.

        \item The argument of the previous point applies to show that the inverse mapping $\thpbracket{\ux}^{-s}\centerdot$ also is continuous.
    \end{itemize}
\end{proof}

\subsubsection{On Sobolev spaces}
Generally speaking, Sobolev spaces\footnote{as defined here, that is, Sobolev-Bessel potential spaces with $p=2$} are not spaces of \textit{functions}, but of \textit{distributions}. Nonetheless, these distributions are ``never too far away from (very nice) functions'' in the sense that:

\begin{lemma}
    \label{lemma:hsdensity}
    For $s \in \R$, $H^s(\R^n) \cap \cS(\R^n)$ is dense in $H^s(\R^n)$.
\end{lemma}

\begin{proof}
    We proceed by sequential characterization of density. 

    Let $u \in H^s(\R^n)$. By definition of $H^s(\R^n)$, $\thpbracket{\up}^s u(\up) \in L^2(\R^n)$; since $\cS(\R^n)$ is dense in $L^2(\R^n)$, 
    \begin{equation*}
        \exists (v_k) \in \left( \cS(\R^n) \right)^\N, \: \| v_k(\up) - \thpbracket{\up}^s u(\up) \|_{L^2(\R^n)} \underset{k \rightarrow \infty}{\longrightarrow} 0
    \end{equation*}
    Consider now $(u_k) = (\thpbracket{\up}^{-s} v_k)$. By construction, $(u_k) \in \left( H^s(\R^n) \cap \cS(\R^n) \right)^\N$, and
    \begin{align*}
        \forall k \in \N, \: \| u_k - u \|_{H^s(\R^n)} &= (2\pi)^{-n/2} \| \thpbracket{\up}^s (u_k (\up) -u(\up)) \|_{L^2(\R^n)} \\
        &= (2\pi)^{-n/2} \| v_k(\up) - \thpbracket{\up}^s u(\up) \|_{L^2(\R^n)} \\
        &\underset{k \rightarrow \infty}{\longrightarrow} 0
    \end{align*}
\end{proof}

\noindent Furthermore, for high enough $s$, Sobolev spaces actually are spaces of functions:

\begin{theorem}[Sobolev embedding] Let $k \in \N$ and
    let $\cC^k_b(\R^n)$ be the set of $\cC^k$ scalar-valued functions on $\R^n$ whose derivatives of order up to $k$ are all bounded. Then, for $s > \frac{n}{2} + k$, $H^s(\R^n) \subset \cC^k_b(\R^n)$ and moreover
    \begin{equation*}
        \exists C_s^k \in \R_+^*, \forall \alpha \in \N^n \,/\, |\alpha| \le k,  \forall u \in H^s(\R^n), \| \partial^\alpha u \|_\infty \le C_s^k\|u\|_{H^s(\R^n)}\,.
    \end{equation*}
\end{theorem}

\begin{proof}
    We fix in the entire proof $k \in \N$.

    Let $\alpha \in \N^n$ with $|\alpha| \le k$ and let us first consider $u \in H^s(\R^n) \cap \cS(\R^n)$.
    \begin{align*}
        \forall \ux \in \R^n, \: |\partial^\alpha u(\ux)| &= \left| \int d_n \up e^{i \ux \cdot \up} \up^\alpha u(\up) \right| \\
        &\le \int d_n \up | \up^\alpha| |u(\up)| \\
        &\le \int d_n \up \thpbracket{\up}^{|\alpha|} |u(\up)| \\
        &\le \int d_n \up \thpbracket{\up}^{k} |u(\up)| \\
        &\le \int d_n \up \thpbracket{\up}^{k-s} \thpbracket{\up}^{s} |u(\up)| \\
        &\le \underbrace{\sqrt{\int d_n \up \thpbracket{\up}^{2(k-s)}}}_{C_s^k} \underbrace{\sqrt{\int d_n \up\thpbracket{\up}^{2s}|u(\up)|^2}}_{\|u\|_{H^s(\R^n)}}
    \end{align*}
    Note that $C_s^k$ is indeed independent of the specific choice of $\alpha$, and since $s > n/2+k$, it is finite. By taking the sup, 
    \begin{equation}
        \label{eq:soboembd4HS}
        \| \partial^\alpha u \|_\infty \le C^k_s \| u \|_{H^s(\R^n)}\,,
    \end{equation}
    which proves along the way that $u \in \cC^k_b(\R^n)$.

    Let us now generalize this result to $u \in H^s(\R^n)$. Lemma \ref{lemma:hsdensity} ensures the existence of a sequence $(u_\ell) \in \left( H^s(\R^n) \cap \cS(\R^n) \right)^\N$ that converges to $u$ in the $H^s(\R^n)$ norm:
    \begin{equation*}
        \| u_\ell - u \|_{H^s(\R^n)} \underset{\ell \rightarrow \infty}{\longrightarrow} 0
    \end{equation*}
    $(u_\ell)$ is a Cauchy sequence in $H^s(\R^n)$. Thus, by the previously established inequality (\ref{eq:soboembd4HS}), applied to every $|\alpha| \le k$, it is a Cauchy sequence in $\cC^k_b(\R^n)$ with the norm $\max_{|\alpha| \le k} \| \partial^\alpha \cdot \|_\infty$; since the latter is complete, $(u_\ell)$ admits some limit $\tilde{u}$ in $\cC^k_b(\R^n)$. Let us now show that $u$ and $\tilde{u}$ can be identified in the sense of distributions.
    \begin{itemize}[noitemsep,topsep=0pt]
        \item On the one hand,
        \begin{align*}
            \forall f \in \cS(\R^n), \forall \ell \in \N, \: |u_\ell(f) - \tilde{u}(f)| &= \left| \int d^n \ux (u_\ell - \tilde{u})(\ux) f(\ux) \right| \\
            &\le \| u_\ell - \tilde{u} \|_\infty \|f\|_1 \\
            &\underset{\ell \rightarrow \infty}{\longrightarrow} 0
        \end{align*}
        \noindent Thus $(u_\ell) \underset{\ell \rightarrow \infty}{\longrightarrow} \tilde{u}$ in the sense of distributions.
        \item On the other hand,
        \begin{align*}
            \forall f \in \cS(\R^n), \forall \ell \in \N, \: |u_\ell(f) - u(f)| &\le \|u_\ell -u \|_{H^s(\R^n)} \|f\|_{H^{-s}(\R^n)} \\
            &\underset{\ell \rightarrow \infty}{\longrightarrow} 0
        \end{align*}
        \noindent Thus $(u_\ell) \underset{\ell \rightarrow \infty}{\longrightarrow} u$ in the sense of distributions.
    \end{itemize}
    \noindent By uniqueness of the limit, $\tilde{u} = u$ (in the sense of distributions).
    \noindent Finally, by writing inequality (\ref{eq:soboembd4HS}) for $u_\ell$ and taking $\ell \to \infty$, it gets extended to $u$. 
\end{proof}

\begin{remark}
    The theorem was again stated with the abuse of notations introduced in the definition of Sobolev spaces, in which distributions associated to functions are identified to said functions; this, to avoid unnecessarily heavy notations such as the distribution set $T \cC^k_b(\R^n)$.
\end{remark}

\subsection{The Schwartz representation theorem}
We now turn to the main result of this appendix, which we will present in two forms.

\begin{theorem}[Schwartz representation - intersection form]
    \begin{equation*}
        \cS(\R^n) = \bigcap\limits_{s,r \in \R} \thpbracket{\ux}^r H^s(\R^n)
    \end{equation*}
\end{theorem}

\begin{proof}
    We proceed by double inclusion.

    $\boxed{\supset}$ First, note that using Sobolev embedding,
    \begin{equation*}
        \bigcap\limits_{s,r \in \R} \thpbracket{\ux}^r H^s(\R^n) \subset \bigcap\limits_{k \in \N} H^{\frac{n}{2} + k + 1}(\R^n) \subset \bigcap\limits_{k \in \N} \cC^k_b(\R^n) = \cC^\infty_b(\R^n)
    \end{equation*}
    Thus, we are indeed dealing with a set of functions.

    Let us now show that given a $f \in \bigcap\limits_{s,r \in \R} \thpbracket{\ux}^r H^s(\R^n)$, for $(\alpha,\beta) \in (\N^n)^2$, $\| f \|_{\alpha,\beta} < \infty$.

    \noindent Since $f \in \thpbracket{\ux}^{-|\alpha|} H^{\frac{n}{2} + |\beta| +1} (\R^n)$, there exists $g \in H^{\frac{n}{2} + |\beta| +1} (\R^n) \subset \cC^{|\beta|}_b(\R^n)$ such that $f(\ux) = \thpbracket{\ux}^{-|\alpha|} g(\ux)$.

    \begin{align*}
        \forall \ux \in \R^n, \: |\ux^\alpha \partial^\beta f(\ux)| &\le \thpbracket{\ux}^{|\alpha|}|\partial^\beta (\thpbracket{\ux}^{-|\alpha|}g(\ux))| \\
        &\le \thpbracket{\ux}^{|\alpha|} \sum\limits_{\gamma \le \beta} \begin{pmatrix} \beta \\ \gamma \end{pmatrix} C_\gamma \left| \thpbracket{\ux}^{-|\alpha| - |\gamma|} \partial^{\beta - \gamma} g (\ux)\right| \textrm{using } (\ref{eq:multindexleibniz}), (\ref{eq:partialbrakestimate}) \\
        &\le \sum\limits_{\gamma \le \beta} \begin{pmatrix} \beta \\ \gamma \end{pmatrix} C_\gamma \left| \thpbracket{\ux}^{- |\gamma|} \partial^{\beta - \gamma} g (\ux) \right| \\
        &\le \sum\limits_{\gamma \le \beta} \begin{pmatrix} \beta \\ \gamma \end{pmatrix} C_\gamma \left| \partial^{\beta - \gamma} g (\ux) \right| \\
        &\le \underbrace{\sum\limits_{\gamma \le \beta} \begin{pmatrix} \beta \\ \gamma \end{pmatrix} C_\gamma \| \partial^{\beta - \gamma} g \|_\infty }_{<\infty \textrm{ since } g \in \cC^{|\beta|}_b(\R^n)}
    \end{align*}

    \noindent Taking the sup concludes that $\| f \|_{\alpha,\beta} < \infty$, and hence (since this holds for any $\alpha,\beta$) that $f \in \cS(\R^n)$.

    $\boxed{\subset}$ Let $f(\ux) \in \cS(\R^n)$ (in this part of the proof, we will be a bit sloppy in notations and abusively denote by $f(\ux)$ a function and by $f(\up)$ its Fourier transform). Let us show that for $(r,s) \in \R^2$, $f(\ux) \in \thpbracket{\ux}^r H^s(\R^n)$.

    \noindent Since $\thpbracket{\ux}^{-r}\centerdot$ is an isomorphism of $\cS(\R^n)$, $g(\ux) = \thpbracket{\ux}^{-r}f(\ux) \in \cS(\R^n)$.

    \noindent Now, the Fourier transform also is an isomorphism of $\cS(\R^n)$; hence, $g(\up) \in \cS(\R^n)$ as well.

    \noindent As $\thpbracket{\up}^{s+n+1}$ is again an isomorphism of $\cS(\R^n)$, $\thpbracket{\up}^{s+n+1} g(\up) \in \cS(\R^n)$.

    \noindent It follows that $\| \thpbracket{\up}^{s+n+1} g(\up) \|_\infty < \infty$; hence, the inequality
    \begin{equation*}
        \forall \up \in \R^n, \: | \thpbracket{\up}^s g(\up) |^2 \le \frac{\| \thpbracket{\up}^{s+n+1} g(\up) \|_\infty ^2}{\thpbracket{\up}^{2n+2}}
    \end{equation*}

    \noindent yields $\thpbracket{\up}^s g(\up) \in L^2(\R^n)$, that is $g(\ux) \in H^s(\R^n)$, from which it follows that $f(\ux) \in \thpbracket{\ux}^r H^s(\R^n)$.
\end{proof}

\begin{theorem}[Schwartz representation - union form]
    \begin{equation*}
        \cS'(\R^n) = \bigcup\limits_{s,r \in \R} \thpbracket{\ux}^r H^s(\R^n)
    \end{equation*}
\end{theorem}

\begin{proof}
    We proceed by double inclusion.

    $\boxed{\supset}$ By definition of Sobolev spaces, it is already acquired that
    \begin{equation*}
        \forall s \in \R, \: H^s(\R^n) \subset \cS'(\R^n)    
    \end{equation*}

    \noindent As the $\thpbracket{\ux}^r$ are isomorphisms of $\cS'(\R^n)$, it follows that
    \begin{equation*}
        \forall (r,s) \in \R^2, \: \thpbracket{\ux}^r H^s(\R^n) \subset \cS'(\R^n)
    \end{equation*}

    \noindent Hence
    \begin{equation*}
        \bigcup\limits_{s,r \in \R} \thpbracket{\ux}^r H^s(\R^n) \subset \cS'(\R^n)
    \end{equation*}

    $\boxed{\subset}$ Let $u \in \cS'(\R^n)$.

    \noindent Recall that by proposition \ref{prop:estimateuf}, 
    \begin{equation*}
        \exists \tilde{C}_u \in \R_+^*, \exists k \in \N, \forall f \in \cS(\R^n), \: |u(f)| \le \tilde{C}_u \| f \|^\cS_k
    \end{equation*}
    In everything that follows, we let $q$ be an even integer such that $q \geq \frac{n}{2} + k + 2$.

    \noindent Define now $R_q = \thpbracket{\ux}^{-q} \thpbracket{\uD}^{-q}$ and $R^\dagger_q = \thpbracket{\uD}^{-q} \thpbracket{\ux}^{-q}$. Since $\thpbracket{\uD} = \cF^{-1} \thpbracket{\up} \cF$, and $\thpbracket{\ux}^{-q}$, $\thpbracket{\uD}^{-q}$, $\cF$ are all isomorphisms on $\cS'(\R^n)$, so are $R_q$ and $R^\dagger_q$.

    \noindent Note also that for $(\alpha, \beta) \in (\N^n)^2$ such that $|\alpha|+|\beta| \le k$, and any $f \in \cS(\R^n)$,
    \begin{align*}
        \| R_q f \|_{\alpha, \beta} &= \sup\limits_{\ux} \left| \ux^\alpha \partial^\beta R_q f \right| \\
        &= \sup\limits_{\ux} \left| \ux^\alpha \sum\limits_{\gamma \le \beta} \begin{pmatrix} \beta \\ \gamma \end{pmatrix} \partial^\gamma \thpbracket{\ux}^{-q} \partial^{\beta-\gamma} \thpbracket{\uD}^{-q} f \right| \\
        &\le \sum\limits_{\gamma \le \beta} \begin{pmatrix} \beta \\ \gamma \end{pmatrix} C_\gamma \sup\limits_{\ux} \left| \ux^\alpha \thpbracket{\ux}^{-q - |\gamma|} \partial^{\beta-\gamma} \thpbracket{\uD}^{-q} f  \right| \quad \textrm{using } (\ref{eq:multindexleibniz}), (\ref{eq:partialbrakestimate}) \\
        &\le \sum\limits_{\gamma \le \beta} \begin{pmatrix} \beta \\ \gamma \end{pmatrix} C_\gamma \sup\limits_{\ux} \left|\thpbracket{\ux}^{|\alpha| -q - |\gamma|} \partial^{\beta-\gamma} \thpbracket{\uD}^{-q} f  \right| \\
        &\le \sum\limits_{\gamma \le \beta} \begin{pmatrix} \beta \\ \gamma \end{pmatrix} C_\gamma \sup\limits_{\ux} \left|\partial^{\beta-\gamma} \thpbracket{\uD}^{-q} f  \right| \quad \textrm{since } |\alpha| -q - |\gamma| \le k -q - |\gamma| \le 0 \\
    \end{align*}
    and as $f \in H^0(\R^n)$, $\thpbracket{\uD}^{-q} f \in H^q(\R^n)$, which since $q > n/2 + k$ and $|\beta - \gamma| \le k$ allows for the Sobolev embedding
    \begin{align*}
        \| R_q f \|_{\alpha, \beta} &\le \sum\limits_{\gamma \le \beta} \begin{pmatrix} \beta \\ \gamma \end{pmatrix} C_\gamma C_\cS^k \left\|\thpbracket{\uD}^{-q} f  \right\|_{H^q(\R^n)} \\
        &\le \underbrace{\sum\limits_{\gamma \le \beta} \begin{pmatrix} \beta \\ \gamma \end{pmatrix} C_\gamma C_\cS^k}_{C''_\beta} \left\| f \right\|_{L^2(\R^n)} \\
    \end{align*}
    Letting $C''_k = \max\limits_{|\beta|\le k} C''_\beta$, which is finite (since there's a finite amount of $C''_\beta$'s) and independent of $f \in \cS(\R^n)$, we showed
    \begin{equation*}
        \forall f \in \cS(\R^n),  \| R_q f \|_k^\cS \le C''_k \| f \|_{L^2(\R^n)}
    \end{equation*}

    It follows that:
    \begin{align*}
        \forall f \in \cS(\R^n), \: |R_q^\dagger u(f)| &= |u(R_q f)| \\
        &\le \tilde{C}_u \| R_q f \|^\cS_k \\
        &\le \tilde{C}_u C''_k \| f \|_{L^2(\R^n)}
    \end{align*}
    Which shows that $R_q^\dagger u$ is a bounded linear functional on $\cS(\R^n)$ for the topology of $L^2(\R^n)$.

    \noindent As $\cS(\R^n)$ is dense in $L^2(\R^n)$, by the Hahn-Banach theorem, it admits an extension $v$ to the whole of $L^2(\R^n)$ such that 
    \begin{equation*}
        \forall f \in L^2(\R^n), \: |v(f)| \le \tilde{C}_u C''_k \| f \|_{L^2(\R^n)}
    \end{equation*}
    By the Riesz representation theorem, 
    \begin{equation*}
        \exists F_v \in L^2(\R^n), \forall f \in L^2(\R^n), \: v(f) = \int d^n \ux F_v(\ux) f(\ux)
    \end{equation*}
    and upon identifying $R_q^\dagger u$ to $v$ and then to $F_v$, $R_q^\dagger u \in L^2(\R^n)$.

    \noindent From there, it concludes to notice that
    \begin{equation*}
        u = \underbrace{\thpbracket{\ux}^q \underbrace{ \thpbracket{\uD}^q \underbrace{R^\dagger_q u}_{\in L^2(\R^n)} }_{\in H^{-q}(\R^n)}}_{\in \thpbracket{\ux}^q H^{-q}(\R^n)}
    \end{equation*}
\end{proof}
To get an intuitive explanation of these two theorems, note that $\thpbracket{\ux}^r H^s(\R^n)$ is decreasing (for $\subset$) as $s$ increases and as $r$ decreases. Thus, the intersection and the union should respectively be interpreted as taking (formally) $(r,s)=(-\infty, +\infty)$ and $(r,s)=(+\infty, -\infty)$; qualitatively, since $\thpbracket{\ux}^r$ controls the behavior at infinity and $H^s(\R^n)$ the smoothness, these theorems essentially say:
\begin{itemize}[noitemsep,topsep=0pt]
    \item $\cS(\R^n)$ is composed of functions that are ``infinitely nice'', both in their behavior at infinity and their regularity (in a large sense - they do not have bumps that are too spiky, even $\cC^\infty$ ones).
    \item $\cS'(\R^n)$ is composed of distributions that are ``at least a bit nice'', in the sense that there is some control over their behavior both at infinity and in their regularity.
\end{itemize}Hence, to a certain extent, the $\thpbracket{\ux}^r H^s(\R^n)$ provide a scale of the ``niceness'' of distributions.
Observe also that one theorem is, bottom line, just the dual of the other - hence, it can just be viewed as one single theorem.

\subsection{A consequence of the Schwartz representation theorem}
We finally turn to the main practical consequence of the Schwartz representation theorem that we use in the main text.

\begin{proposition}\label{prop:SchwartzUseful}
    \begin{equation*}
        \forall u \in \cS'(\R^n), \exists (r,Q) \in 2\N \times \N, \exists f \in \cC^0_b(\R^n), \: u = \thpbracket{\ux}^r \thpbracket{\uD}^{2Q} f
    \end{equation*}
\end{proposition}

\begin{proof}
    We let $u \in \cS'(\R^n)$. 

    \noindent By the Schwartz representation theorem, there exists $(r,s) \in \R^2$ such that $u \in \thpbracket{\ux}^r H^s(\R^n)$, and its proof gives $r = q$, an even natural number. From there, there exists $\tilde{s} \in -2\N$ such that $\tilde{s} < s$; since $H^s(\R^n) \subset H^{\tilde{s}}(\R^n)$, it follows that $u \in \thpbracket{\ux}^r H^{\tilde{s}}(\R^n)$.

    \noindent Therefore, there exists some $\tilde{f} \in L^2(\R^n)$ such that
    \begin{equation*}
        u = \thpbracket{\ux}^r \thpbracket{\uD}^{-\tilde{s}} \tilde{f}
    \end{equation*}
    And letting $\tilde{Q} = -\frac{\tilde{s}}{2} \in \N$, this is
    \begin{equation*}
        u = \thpbracket{\ux}^r \thpbracket{\uD}^{2\tilde{Q}} \tilde{f}
    \end{equation*}
    Now, let $K = \left\lfloor \frac{n}{4} \right\rfloor + 1$, which is such that $2K > \frac{n}{2}$. Write
    \begin{equation*}
        u = \thpbracket{\ux}^r \thpbracket{\uD}^{2\tilde{Q}} \thpbracket{\uD}^{2K} \thpbracket{\uD}^{-2K} \tilde{f}
    \end{equation*}
    Letting $Q = \tilde{Q} + K \in \N$ and $f = \thpbracket{\uD}^{-2K} \tilde{f}$, this is
    \begin{equation*}
        u = \thpbracket{\ux}^r \thpbracket{\uD}^{2Q} f
    \end{equation*}
    Note now that since $f = \thpbracket{\uD}^{-2K} \tilde{f}$ with $\tilde{f} \in L^2(\R^n)$, $f \in H^{2K}(\R^n)$. But as $2K > n/2$, by Sobolev embedding, $H^{2K}(\R^n) \subset \cC^0_b(\R^n)$, and hence $f \in \cC^0_b(\R^n)$.
\end{proof}

Note that this amounts to write that for some $(R,Q) \in \N \times \N$,
\begin{equation*}
    u(\ux) = (1+\ux^2)^R (1-\Delta)^Q f(\ux)
\end{equation*}
(where compared to the previous statement, $R=r/2$). That is the final form of this result that we will consider.

\section{The coarea formula}
\label{sec:coarea}

In this whole appendix, $N \in \N^*$.

It is well known that, owing to a change of coordinates made licit by the implicit function theorem, the following holds:

\begin{proposition}[The coarea formula]
    Let $\Omega \subset \R^N$ be an open set, $f \in \cC_c(\Omega)$ and $g \in \cC^1(\Omega)$ such that $\nabla g$ does not vanish on $\supp f$.

    \noindent One has:
    \begin{equation*}
        \int_\Omega f(x) dx = \int_\R \left( \int_{g^{-1}(s) \cap \Omega} \frac{f(x)}{|\nabla g (x)|} d \cH^{N-1}(x) \right) ds
    \end{equation*}
    where $d \cH^{N-1}(x)$ is the Hausdorff measure on the $N-1$-dimensional level set $g^{-1}(s)$.
\end{proposition}

In particular, for $\Omega = \R^N$, the coarea formula simplifies to
\begin{equation*}
    \int_{\R^N} f(x) dx = \int_\R \left( \int_{g^{-1}(s)} \frac{f(x)}{|\nabla g (x)|} d \cH^{N-1}(x) \right) ds
\end{equation*}
Note also that this formula, like the more general one, is valid even when $g$ isn't surjective onto $\R$ with the convention that when $g^{-1}(s)=\emptyset$, the associated integral vanishes.

\begin{proposition}[A distributional version of the coarea formula]
    Let $f \in \cC_c(\R^N)$ and $g \in \cC^1(\R^N)$ such that $\nabla g$ never vanishes on $\supp f$.
    \begin{equation*}
        \int_{\R^N} \delta(s-g(x))f(x)dx = \int_{g^{-1}(s)} \frac{f(x)}{|\nabla g (x)|} d \cH^{N-1}(x)
    \end{equation*}
    where $d \cH^{N-1}(x)$ is the Hausdorff measure on the $N-1$-dimensional level set $g^{-1}(s)$.
\end{proposition}

\begin{proof}
    \noindent The left-hand side is a distribution in $s$: for $\varphi \in \cD(\R)$, its pairing with $\varphi$ is by definition
    \begin{equation*}
        \int_\R ds\, \varphi(s) \int_{\R^N} \delta(s-g(x))f(x)dx \colonequals \int_{\R^N} \varphi(g(x)) f(x) dx\,.
    \end{equation*}
    Applying the coarea formula to the compactly supported continuous function $x \mapsto \varphi(g(x)) f(x)$, and then using $g(x) = s$ on $g^{-1}(s)$, gives
    \begin{align*}
        \int_{\R^N} \varphi(g(x)) f(x) dx &= \int_\R \left( \int_{g^{-1}(s)} \frac{\varphi(g(x)) f(x)}{|\nabla g (x)|} d \cH^{N-1}(x) \right) ds \\
        &= \int_\R ds\, \varphi(s) \int_{g^{-1}(s)} \frac{f(x)}{|\nabla g (x)|} d \cH^{N-1}(x)\,,
    \end{align*}
    which is the claim.
\end{proof}

\begin{lemma}[A delta-distributional chain rule]
     Let $g \in \cC^1(\R^N)$. In the sense of distributions,\footnote{This result can be extended to the sense of tempered distributions with a similar proof under the extra assumption that $g(x)$ and all its derivatives are polynomially bounded.}
     \begin{equation*}
        \nabla \delta (s-g(x)) = - \delta'(s-g(x)) \nabla g(x)
     \end{equation*}
\end{lemma}

\begin{proof}
    The identity is proven by checking its application to test functions.

    \noindent Let $f(s,x) \in \cD(\R^{1+N})$ be a test function.

    \noindent Define now $F : \ux \in \R^N \mapsto f(g(x),x)$. For a given $k \in \{1, \dots, N\}$, the chain rule reads
    \begin{equation*}
        \partial_{x_k} F(x) = \partial_s f(g(x), x) \partial_{x_k}(g(x)) + \partial_{x_k} f (g(x),x)
    \end{equation*}
    Since $f$ has compact support, so does $F$; thus, upon integrating the previous equality over the whole of $\R^N$, 
    \begin{equation*}
        0 = \int d^N x \partial_s f(g(x),x) \partial_{x_k} g(x) + \int d^N x \partial_{x_k} f(g(x),x)
    \end{equation*} 
    Now
    \begin{equation*}
        \< \partial_{x_k} \delta (s-g(x)) , f(s,x) \> = - \< \delta (s-g(x)), \partial_{x_k} f(s,x) \> = - \int d^N x \partial_{x_k} f(g(x),x)
    \end{equation*}
    and
    \begin{align*}
        \< \delta'(s-g(x)) \partial_{x_k} g(x), f(s,x) \> &= \< \partial_s \delta(s-g(x)), f(s,x) \partial_{x_k} g(x) \> \\
        &= - \int d^N x \partial_s f(g(x),x) \partial_{x_k} g(x)
    \end{align*}
    which concludes.
\end{proof}

\begin{proposition}[A coarea formula-inspired integration by parts]
    Let $g \in \cC^2(\R^N)$ such that $\nabla g$ never vanishes and $f \in \cC_c^1(\R^N)$.
    \begin{equation*}
        \int \delta'(s-g(x)) f(x) dx = \int \delta(s-g(x)) \underbrace{\nabla \cdot \left[ \frac{\nabla g(x)}{|\nabla g(x)|^2} f(x) \right]}_{D_g[f](x)}dx
    \end{equation*}
\end{proposition}

\begin{proof}
    Consider:
    \begin{equation*}
        \nabla \cdot \left[ \delta(s-g(x)) f(x) \frac{\nabla g(x)}{|\nabla g(x)|^2} \right] = [\nabla \delta(s-g(x))] \cdot f(x) \frac{\nabla g(x)}{|\nabla g(x)|^2} + \delta(s-g(x)) \nabla \cdot \left[\frac{\nabla g(x)}{|\nabla g(x)|^2}f(x) \right]
    \end{equation*}
    Using the delta-distributional chain rule lemma, this is:
    \begin{equation*}
         \nabla \cdot \left[ \delta(s-g(x)) f(x) \frac{\nabla g(x)}{|\nabla g(x)|^2} \right] = -\delta'(s-g(x)) f(x) + \delta(s-g(x)) \nabla \cdot \left[\frac{\nabla g(x)}{|\nabla g(x)|^2}f(x) \right]
    \end{equation*}
    By the divergence theorem, the integral of the lhs vanishes (since $f$ is compactly supported); and hence, integrating the previous equation leads to
    \begin{equation*}
        \int \delta'(s-g(x)) f(x) dx = \int \delta(s-g(x)) \underbrace{\nabla \cdot \left[ \frac{\nabla g(x)}{|\nabla g(x)|^2} f(x) \right]}_{D_g[f](x)}dx
    \end{equation*}
\end{proof} 

\begin{proposition}[An iterated version of coarea formula-inspired integration by parts]
    The previous proposition can be iterated: for $k \in \N^*$, $g \in \cC^{k+1}(\R^N)$ such that $\nabla g$ never vanishes and $f \in \cC_c^k(\R^N)$.
    \begin{equation*}
        \int \delta^{(k)}(s-g(x)) f(x) dx = \int \delta(s-g(x))D_g^{(k)}[f](x)dx
    \end{equation*}
\end{proposition}

\begin{proof}
    Reproducing the previous steps (including the delta-distributional chain rule lemma) with $\delta^{(k)}$ instead of $\delta$, it appears that
    \begin{equation*}
        \int \delta^{(k)}(s-g(x)) f(x) dx = \int \delta^{(k-1)}(s-g(x)) \underbrace{\nabla \cdot \left[ \frac{\nabla g(x)}{|\nabla g(x)|^2} f(x) \right]}_{D_g[f](x)}dx
    \end{equation*}
    And thus, the result follows by induction.
\end{proof}

\bibliography{biblio}

\providecommand{\href}[2]{#2}\begingroup\raggedright\begin{thebibliography}{10}

\bibitem{Haag:1958vt}
R.~Haag, ``{Quantum field theories with composite particles and asymptotic
  conditions},'' \href{http://dx.doi.org/10.1103/PhysRev.112.669}{{\em Phys.
  Rev.} {\bfseries 112} (1958) 669--673}.

\bibitem{ruelle1962asymptotic}
D.~Ruelle, ``On asymptotic condition in quantum field theory,'' {\em Helvetica
  Physica Acta} {\bfseries 35} no.~3, (1962) 147.

\bibitem{LSZPAPER}
A.~Hebbar, T.~Pochart, and B.~C. van Rees, ``{How to compute S: the LSZ
  prescriptions}.'' (to appear).

\bibitem{streater2000pct}
R.~F. Streater and A.~S. Wightman, {\em PCT, spin and statistics, and all
  that}, vol.~30.
\newblock Princeton University Press, 2000.

\bibitem{Lehmann:1954rq}
H.~Lehmann, K.~Symanzik, and W.~Zimmermann, ``{On the formulation of quantized
  field theories},'' \href{http://dx.doi.org/10.1007/BF02731765}{{\em Nuovo
  Cim.} {\bfseries 1} (1955) 205--225}.

\bibitem{lehmann1957formulation}
H.~Lehmann, K.~Symanzik, and W.~Zimmermann, ``On the formulation of quantized
  field theories—ii,'' {\em Il Nuovo Cimento (1955-1965)} {\bfseries 6}
  no.~2, (1957) 319--333.

\bibitem{hepp1965connection}
K.~Hepp, ``On the connection between the lsz and wightman quantum field
  theory,'' {\em Communications in Mathematical Physics} {\bfseries 1} no.~2,
  (1965) 95--111.

\bibitem{Jost:1965yxu}
R.~Jost, {\em {The general theory of quantized fields}}, vol.~4.
\newblock American Mathematical Society, Providence, RI, 1965.

\bibitem{Reed:1979Scattering}
M.~Reed and B.~Simon, {\em {Methods of Modern Mathematical Physics. III:
  Scattering Theory}}.
\newblock Academic Press, New York, 1979.

\bibitem{araki1999mathematical}
H.~Araki, {\em Mathematical theory of quantum fields}.
\newblock OUP Oxford, 1999.

\bibitem{bogolubov2012general}
N.~N. Bogol{\^u}bov, A.~A. Logunov, A.~Oksak, and I.~Todorov, {\em General
  principles of quantum field theory}, vol.~10.
\newblock Springer Science \& Business Media, 2012.

\bibitem{glimm2012quantum}
J.~Glimm and A.~Jaffe, {\em Quantum physics: a functional integral point of
  view}.
\newblock Springer Science \& Business Media, 2012.

\bibitem{peskin1995introduction}
M.~E. Peskin and D.~V. Schroeder,
  \href{http://dx.doi.org/10.1201/9780429503559}{{\em {An Introduction to
  quantum field theory}}}.
\newblock Addison-Wesley, Reading, USA, 1995.

\bibitem{reed1980functional}
M.~Reed and B.~Simon, {\em Methods of Modern Mathematical Physics, Vol. 1:
  Functional Analysis}.
\newblock Academic Press, New York, revised and enlarged~ed., 1980.

\bibitem{Hormander1960}
L.~H{\"o}rmander, ``Estimates for translation invariant operators in $l^p$
  spaces,'' \href{http://dx.doi.org/10.1007/BF02547187}{{\em Acta Mathematica}
  {\bfseries 104} no.~1-2, (1960) 93--140}.

\bibitem{grafakos2014classical}
L.~Grafakos, \href{http://dx.doi.org/10.1007/978-1-4939-1194-3}{{\em Classical
  Fourier Analysis}}, vol.~249 of {\em Graduate Texts in Mathematics}.
\newblock Springer New York, 3rd~ed., 2014.

\bibitem{Martin:1966zz}
A.~Martin, ``{INABILITY OF FIELD THEORY TO EXPLOIT THE FULL UNITARITY
  CONDITION},''.

\bibitem{Lechner:2006kb}
G.~Lechner, ``{Construction of Quantum Field Theories with Factorizing
  S-Matrices},'' \href{http://dx.doi.org/10.1007/s00220-007-0381-5}{{\em
  Commun. Math. Phys.} {\bfseries 277} (2008) 821--860},
  \href{http://arxiv.org/abs/math-ph/0601022}{{\ttfamily
  arXiv:math-ph/0601022}}.

\bibitem{Araki:1962zhd}
H.~Araki, K.~Hepp, and D.~Ruelle, ``{On the asymptotic behaviour of Wightman
  functions in space-like directions},''
  \href{http://dx.doi.org/10.5169/seals-113273}{{\em Helv. Phys. Acta}
  {\bfseries 35} no.~III, (1962) 164--174}.

\bibitem{dellantonio_asymptotic_1959}
G.~F. Dell’Antonio and P.~Gulmanelli, ``Asymptotic conditions in quantum
  field theories,'' \href{http://dx.doi.org/10.1007/BF02747805}{{\em Il Nuovo
  Cimento (1955-1965)} {\bfseries 12} no.~1, (Mar., 1959) 38--53}.
  \url{https://doi.org/10.1007/BF02747805}.

\bibitem{ARAKI1960260}
H.~Araki, ``On asymptotic behavior of vacuum expectation values at large
  space-like separation,''
  \href{http://dx.doi.org/https://doi.org/10.1016/0003-4916(60)90135-4}{{\em
  Annals of Physics} {\bfseries 11} no.~2, (1960) 260--274}.
  \url{https://www.sciencedirect.com/science/article/pii/0003491660901354}.

\bibitem{jostandhepp}
R.~Jost and K.~Hepp, ``{Uber die Matrixelemente des Translationsoperators},''
  \href{http://dx.doi.org/10.5169/seals-113266}{{\em Helv. Phys. Acta}
  {\bfseries 35} no.~I, (1962) 34--46}.

\bibitem{Jaffe:1967nb}
A.~M. Jaffe, ``{HIGH-ENERGY BEHAVIOR IN QUANTUM FIELD THEORY. I. STRICTLY
  LOCALIZABLE FIELDS},'' \href{http://dx.doi.org/10.1103/PhysRev.158.1454}{{\em
  Phys. Rev.} {\bfseries 158} (1967) 1454--1461}.

\bibitem{Kapustin:1999ci}
A.~Kapustin, ``{On the universality class of little string theories},''
  \href{http://dx.doi.org/10.1103/PhysRevD.63.086005}{{\em Phys. Rev. D}
  {\bfseries 63} (2001) 086005},
  \href{http://arxiv.org/abs/hep-th/9912044}{{\ttfamily arXiv:hep-th/9912044}}.

\bibitem{Keltner:2015xda}
L.~Keltner and A.~J. Tolley, ``{UV properties of Galileons: Spectral
  Densities},'' \href{http://arxiv.org/abs/1502.05706}{{\ttfamily
  arXiv:1502.05706 [hep-th]}}.

\bibitem{Steinmann:1970cm}
O.~Steinmann, ``{Scattering formalism for non-localizable fields},''
  \href{http://dx.doi.org/10.1007/BF01649431}{{\em Commun. Math. Phys.}
  {\bfseries 18} (1970) 179--194}.

\bibitem{araki_collision_1967}
H.~Araki and R.~Haag, ``Collision cross sections in terms of local
  observables,'' \href{http://dx.doi.org/10.1007/BF01645754}{{\em
  Communications in Mathematical Physics} {\bfseries 4} no.~2, (Apr., 1967)
  77--91}. \url{http://link.springer.com/10.1007/BF01645754}.

\bibitem{Frohlich:1990zg}
J.~Frohlich and P.~A. Marchetti, ``{Spin statistics theorem and scattering in
  planar quantum field theories with braid statistics},''
  \href{http://dx.doi.org/10.1016/0550-3213(91)90378-B}{{\em Nucl. Phys. B}
  {\bfseries 356} (1991) 533--573}.

\bibitem{Mund:2008hn}
J.~Mund, ``{The Spin-Statistics Theorem for Anyons and Plektons in d=2+1},''
  \href{http://dx.doi.org/10.1007/s00220-008-0628-9}{{\em Commun. Math. Phys.}
  {\bfseries 286} (2009) 1159--1180},
  \href{http://arxiv.org/abs/0801.3621}{{\ttfamily arXiv:0801.3621 [math-ph]}}.

\bibitem{Buchholz:1981fj}
D.~Buchholz and K.~Fredenhagen, ``{Locality and the Structure of Particle
  States},'' \href{http://dx.doi.org/10.1007/BF01208370}{{\em Commun. Math.
  Phys.} {\bfseries 84} (1982) 1}.

\bibitem{Fredenhagen:1994dk}
K.~Fredenhagen, M.~R. Gaberdiel, and S.~M. Ruger, ``{Scattering states of
  plektons (particles with braid group statistics) in (2+1)-dimensional quantum
  field theory},'' \href{http://dx.doi.org/10.1007/BF02102411}{{\em Commun.
  Math. Phys.} {\bfseries 175} (1996) 319--336},
  \href{http://arxiv.org/abs/hep-th/9410115}{{\ttfamily arXiv:hep-th/9410115}}.

\bibitem{Jain:2014nza}
S.~Jain, M.~Mandlik, S.~Minwalla, T.~Takimi, S.~R. Wadia, and S.~Yokoyama,
  ``{Unitarity, Crossing Symmetry and Duality of the S-matrix in large N
  Chern-Simons theories with fundamental matter},''
  \href{http://dx.doi.org/10.1007/JHEP04(2015)129}{{\em JHEP} {\bfseries 04}
  (2015) 129}, \href{http://arxiv.org/abs/1404.6373}{{\ttfamily arXiv:1404.6373
  [hep-th]}}.

\bibitem{Inbasekar:2015tsa}
K.~Inbasekar, S.~Jain, S.~Mazumdar, S.~Minwalla, V.~Umesh, and S.~Yokoyama,
  ``{Unitarity, crossing symmetry and duality in the scattering of $
  \mathcal{N}=1 $ susy matter Chern-Simons theories},''
  \href{http://dx.doi.org/10.1007/JHEP10(2015)176}{{\em JHEP} {\bfseries 10}
  (2015) 176}, \href{http://arxiv.org/abs/1505.06571}{{\ttfamily
  arXiv:1505.06571 [hep-th]}}.

\bibitem{Gabai:2022snc}
B.~Gabai, J.~Sandor, and X.~Yin, ``{Anyon scattering from lightcone
  Hamiltonian: the singlet channel},''
  \href{http://dx.doi.org/10.1007/JHEP09(2022)145}{{\em JHEP} {\bfseries 09}
  (2022) 145}, \href{http://arxiv.org/abs/2205.09144}{{\ttfamily
  arXiv:2205.09144 [hep-th]}}.

\bibitem{Mehta:2022lgq}
U.~Mehta, S.~Minwalla, C.~Patel, S.~Prakash, and K.~Sharma, ``{Crossing
  symmetry in matter Chern{\textendash}Simons theories at finite $N$ and
  $k$},'' \href{http://dx.doi.org/10.4310/ATMP.2023.v27.n1.a5}{{\em Adv. Theor.
  Math. Phys.} {\bfseries 27} no.~1, (2023) 193--310},
  \href{http://arxiv.org/abs/2210.07272}{{\ttfamily arXiv:2210.07272
  [hep-th]}}.

\bibitem{Buchholz:1974je}
D.~Buchholz, ``{Collision Theory for Massless Fermions},''
  \href{http://dx.doi.org/10.1007/BF01608977}{{\em Commun. Math. Phys.}
  {\bfseries 42} (1975) 269--279}.

\bibitem{Buchholz:1976rs}
D.~Buchholz, ``{Collision Theory for Massless Bosons},''
  \href{http://dx.doi.org/10.1007/BF01625781}{{\em Commun. Math. Phys.}
  {\bfseries 52} (1977) 147--173}.

\bibitem{dybalski2005haag}
W.~Dybalski, ``{Haag-Ruelle scattering theory in presence of massless
  particles},'' \href{http://dx.doi.org/10.1007/s11005-005-2294-6}{{\em Lett.
  Math. Phys.} {\bfseries 72} (2005) 27--38},
  \href{http://arxiv.org/abs/hep-th/0412226}{{\ttfamily hep-th/0412226}}.

\bibitem{duch2015massless}
P.~Duch and A.~Herdegen, ``{Massless asymptotic fields and Haag-Ruelle
  theory},'' \href{http://dx.doi.org/10.1007/s11005-014-0733-y}{{\em Lett.
  Math. Phys.} {\bfseries 105} (2015) 245--277},
  \href{http://arxiv.org/abs/1410.7665}{{\ttfamily arXiv:1410.7665 [math-ph]}}.

\bibitem{Schroer:1963gw}
B.~Schroer, ``{Infraparticles in quantum field theory},'' {\em Fortsch. Phys.}
  {\bfseries 11} (1963) 1--31.

\bibitem{Kulish:1970ut}
P.~P. Kulish and L.~D. Faddeev, ``{Asymptotic conditions and infrared
  divergences in quantum electrodynamics},''
  \href{http://dx.doi.org/10.1007/BF01066485}{{\em Theor. Math. Phys.}
  {\bfseries 4} (1970) 745}.

\bibitem{Caron-Huot:2023vxl}
S.~Caron-Huot, M.~Giroux, H.~S. Hannesdottir, and S.~Mizera, ``{What can be
  measured asymptotically?},''
  \href{http://dx.doi.org/10.1007/JHEP01(2024)139}{{\em JHEP} {\bfseries 01}
  (2024) 139}, \href{http://arxiv.org/abs/2308.02125}{{\ttfamily
  arXiv:2308.02125 [hep-th]}}.

\bibitem{Caron-Huot:2025ymc}
S.~Caron-Huot, M.~Giroux, H.~S. Hannesdottir, S.~Mizera, and C.~Pasiecznik,
  ``{Records from the S-Matrix Marathon: Asymptotic Observables},''
\newblock 2, 2025.
\newblock \href{http://arxiv.org/abs/2502.13021}{{\ttfamily arXiv:2502.13021
  [hep-th]}}.

\bibitem{Caron-Huot:2023ikn}
S.~Caron-Huot, M.~Giroux, H.~S. Hannesdottir, and S.~Mizera, ``{Crossing beyond
  scattering amplitudes},''
  \href{http://dx.doi.org/10.1007/JHEP04(2024)060}{{\em JHEP} {\bfseries 04}
  (2024) 060}, \href{http://arxiv.org/abs/2310.12199}{{\ttfamily
  arXiv:2310.12199 [hep-th]}}.

\bibitem{Barata:1990rn}
J.~C.~A. Barata and K.~Fredenhagen, ``{Particle scattering in Euclidean lattice
  field theories},'' \href{http://dx.doi.org/10.1007/BF02102039}{{\em Commun.
  Math. Phys.} {\bfseries 138} (1991) 507--520}.

\bibitem{Patella:2024cto}
A.~Patella and N.~Tantalo, ``{Scattering amplitudes from Euclidean correlators:
  Haag-Ruelle theory and approximation formulae},''
  \href{http://dx.doi.org/10.1007/JHEP01(2025)091}{{\em JHEP} {\bfseries 01}
  (2025) 091}, \href{http://arxiv.org/abs/2407.02069}{{\ttfamily
  arXiv:2407.02069 [hep-lat]}}.

\bibitem{Luscher:1986pf}
M.~Luscher, ``{Volume Dependence of the Energy Spectrum in Massive Quantum
  Field Theories. 2. Scattering States},''
  \href{http://dx.doi.org/10.1007/BF01211097}{{\em Commun. Math. Phys.}
  {\bfseries 105} (1986) 153--188}.

\bibitem{Luscher:1990ux}
M.~Luscher, ``{Two particle states on a torus and their relation to the
  scattering matrix},''
  \href{http://dx.doi.org/10.1016/0550-3213(91)90366-6}{{\em Nucl. Phys. B}
  {\bfseries 354} (1991) 531--578}.

\bibitem{Luscher:1990ck}
M.~Luscher and U.~Wolff, ``{How to Calculate the Elastic Scattering Matrix in
  Two-dimensional Quantum Field Theories by Numerical Simulation},''
  \href{http://dx.doi.org/10.1016/0550-3213(90)90540-T}{{\em Nucl. Phys. B}
  {\bfseries 339} (1990) 222--252}.

\bibitem{Kruczenski:2022lot}
M.~Kruczenski, J.~Penedones, and B.~C. van Rees, ``{Snowmass White Paper:
  S-matrix Bootstrap},'' \href{http://arxiv.org/abs/2203.02421}{{\ttfamily
  arXiv:2203.02421 [hep-th]}}.

\bibitem{Sommer:1970mr}
G.~Sommer, ``{Present state of rigorous analytic properties of scattering
  amplitudes},'' \href{http://dx.doi.org/10.1002/prop.19700181102}{{\em
  Fortsch. Phys.} {\bfseries 18} (1970) 577--688}.

\end{thebibliography}\endgroup
\bibliographystyle{utphys}

\end{document}